\documentclass[aps,amsmath,amssymb,superscriptaddress,prb,reprint,twocolumn,longbibliography]{revtex4-2}
\usepackage{dcolumn}% Align table columns on decimal point
\usepackage{braket,bm}% bold math

\usepackage{mathptmx}
\DeclareMathAlphabet{\mathcal}{OMS}{cmsy}{m}{n}
\usepackage[dvipdfmx]{graphicx}

\DeclareMathOperator{\sign}{sign}
\begin{document}
	\preprint{APS/123-QED}
	\title{Approximating  Maximally Localized Wannier Functions with Position Scaling-Eigenfunctions }% Force line breaks with \\
	\author{Yuji Hamai}
	\author{Katsunori Wakabayashi}%
	\affiliation{%
	Department of Nanotechnology for Sustainable Energy,
	School of Science and Technology, Kwansei Gakuin University,
	Gakuen-Uegahara 1, Sanda, 669-1330, Japan
	}%
	\date{\today}% It is always \today, today,
	\begin{abstract}
  Position scaling-eigenfunctions are generated by transforming compactly
supported orthonormal scaling functions and utilized for faster alternatives
to maximally localized Wannier functions (MLWFs). The position
scaling-eigenfunctions are first applied  to numerical procedures solving
Schr\"odinger and Maxwell's equations, and the solutions well agree with
preceding results. Subsequently, by projecting the position scaling-eigenfunctions
onto the space spanned by the Bloch functions, approximated MLWFs are obtained.
They show good agreements with preceding results using MLWFs. In addition,
analytical explanations of the agreements and an estimate of  the error
associated with the approximation are provided.
\end{abstract}
	\maketitle
\section{Introduction}
The concept of Wannier functions (WFs) dates back to the work done by Wannier~
\cite{PhysRev.52.191}, and it has been widely applied to electric polarization
\cite{PhysRevB.48.4442,RevModPhys.66.899,PhysRevLett.97.107602,
    PhysRevB.47.1651,PhysRevB.48.4442, PhysRevB.92.165134,RevModPhys.84.1419,
    PhysRevB.95.075114, PhysRevB.89.155114,PhysRevB.47.1651,PhysRevB.48.4442},
chemical bonding\cite{PhysRev.129.554,PhysRevB.89.155114, PhysRevB.92.165134,
    PhysRevB.95.075114, RevModPhys.84.1419, first-guess, PhysRevB.64.245108},  the
orbital  magnetization\cite{PhysRevB.85.014435} and  the photonic confinement
\cite{Kurt2003,Busch2011,PhysRevB.88.075201} and in turn it has lead to the
development of topological electronic\cite{Liu2022} and photonic devices
\cite{Pham2016,PhysRevB.74.195116}.

The key to the successful application of WFs  is composition of maximally
localized Wannier functions (MLWFs). Following analytical investigations of
exponentially localized Wannier functions (ELWFs) and their existence
\cite{PhysRev.115.809,PhysRev.135.A685,PhysRevB.8.2485,PhysRevB.47.10112}, a
specific procedure to obtain MLWFs, the Marzari-Vanderbilt (MV) method, is
developed\cite{Marzari1997}. The MLWFs are defined in the procedure, in effect,
as the WFs with the minimum spread, and it is equivalent of the Kivelson's
definition of WF as the eigenstate of the position operator projected onto the
relevant energy bands\cite{kivelson1982}. Taking advantage of the gauge
freedom of the Bloch functions (BFs), the MV method finds MLWFs by optimizing
 the gauge. Since then various approaches to obtain MLWFs and to
improve the MV method have been proposed. Many of them focus on how to lead the
MV method to physically correct MLWFs by providing  initial localized orbits
\cite{ct500985f,damle2016scdmk, Vitale2020,PhysRevB.92.165134,
	Pizzi2020} or by making the gauge of Bloch functions (BFs)
continuous\cite{PhysRevB.95.075114}. Other attempts include use of  the
full-potential linearized augmented plane-wave
method\cite{freimuth2008publisher}, group theory\cite{PhysRevB.88.075201} and
solving eigenvalue problem  associated with the position eigenvector in 3D
space\cite{PhysRevB.103.075125}.
\begin{figure}[h]
  \centering
  \includegraphics[width=1.0\linewidth]{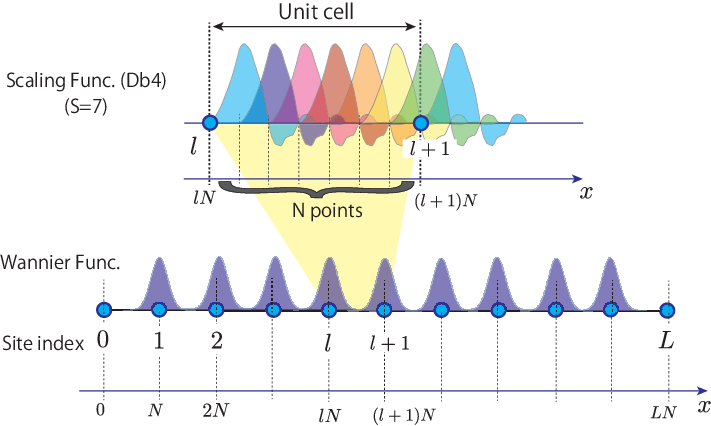}
  \caption{
    Schematic diagram of an entire crystal depicting an application of
compactly supported orthonormal scaling functions to a 1D crystal with
periodic boundary condition (lower) and the detail of one unit cell
(upper). Cyan circles represent the left edges of unit cells, with which
schematic pictures of Wannier functions drawn in grayish blue are
associated (lower). Each unit cell is divided in $N$ to assign
translationally identical scaling functions colored from blue to green
(upper), thereby $LN-1$ of them are deployed  over the entire crystal.
The scaling functions and  the Bloch functions span  the identical state
space, and hence the Wannier functions shown in the lower figure, as
well as the Bloch functions are expressed as linear combinations of the
scaling functions.
  }\label{fig:series_of_scaling_functions}
\end{figure}

Meanwhile, the wavelets and scaling functions (SFs)\cite{Daub10} have
continuously been developed and applied mostly to signal and image processing
\cite{Daub10}. In other areas related to quantum mechanics, Battle applies
wavelets to the renormalization group\cite{battle1999wavelets}, while  Evenbly
and White built approximations to the ground state of an Ising model by
establishing  a precise connection between discrete wavelet transforms and
entanglement renormalization\cite{PhysRevLett.116.140403}.

In the area of solid state physics related to Wannier functions,
Parsen\cite{parzen1953}, for example, had shown  Shannon SF\cite{hong2004-9},
$\mathrm{sinc}(x)$, as a WF in a 1D crystal even before the concept of SF
existed. Clow\cite{clow2003} composed a WF from a multi-SF\cite{keinert2003} in a
similar system. Other than electronic system, phononic\cite{PhysRevB.74.224303}
and  photonic\cite{CHECOURY2006360} wave functions composed  of 2D wavelets are
used to calculate band gaps in crystals.
This implies SFs are capable of becoming products and ingredients of BFs.

 Figure~\ref{fig:series_of_scaling_functions} depicts a schematic
application of  a series of orthonormal SFs\cite{Daub10-6} to a 1D crystal
system. The WFs and the SFs in Fig.~\ref{fig:series_of_scaling_functions}
resemble each other; they  both are translationally orthonormal and localized.
Although SFs have a clear mathematical definition\cite{Daub10-5}, they are
not eigenfunctions of any observables. This may make SFs as mere means to
calculate the results and their physical meaning is hard to conceive.   In the
paper, the authors attempt to generate position scaling-eigenfunctions, discrete
versions of the Dirac's $\delta$-function, which  are SFs and at the same time
eigenvectors of  an observable. The efficacy of the position eigenvectors
projected onto  composite  band systems,  as
alternatives to MLWFs, are  subsequently  examined.
\section{Composing Position Scaling-Eigenfunction and Computing Matrix Elements of Operators}
In this section, position eigenfunctions in a discrete system are generated from
known SFs by taking advantage of the two-scale relation, and their properties are
studied.  Subsequently, the matrix elements of the kinetic energy operator are
calculated with the generated position eigenfunctions, since the matrix is used in
Sec.~\ref{sec:Numerical} to solve Schr\"odinger and Maxwell's equations.

\subsection{Overview of Scaling Function}
The lower half of
Fig.~\ref{fig:series_of_scaling_functions} shows $N$ compactly supported SFs
deployed in one unit cell to compose BFs and WFs. The SFs in the series  have the
identical shape, and they are translationally orthonormal, i.e.,
\begin{equation}\label{eq:scaling_orthonormal }
  \begin{array}{c}
	\int \phi(x-n_1)\phi(x-n_2)dx= \delta_{n_1,n_2}\quad (n_1, n_2 \in \mathbb Z). \\
\end{array}
\end{equation}
 The particular SFs shown in the Fig.~\ref{fig:series_of_scaling_functions} are
Daubechies-4s (Db4s). Each of them has nonzero value only inside the continuous
region whose width is $7$. Since the continuous region, the support, limits the
range of calculations such as integration, compactly supported SFs are
particularly useful basis functions  when composing and decomposing a function.

While many of the orthonormal sets used in solid state physics are
eigenfunctions of Hermitian operators, an SF does not have its origin in
physics. They are the solutions of the following recursive algebraic equation
called two-scale relation\cite{Daub10-5}:
\begin{equation}\label{eq:two-scale }
\begin{split}
	\phi(x)=\sqrt 2\sum_{l=0}^S h_{l}\phi(2x-l),\\
			\{\ h_l: 0 \le l \le S,l \in \mathbb Z,  h_l \in \mathbb R\},
  \end{split}
\end{equation}
with
\begin{equation}\label{eq:ph-integral }
  \begin{split}
    \int \phi(x-n)dx= 1  \ (n \in \mathbb Z).\\
  \end{split}
\end{equation}
Once the series $\{h_l\}$ is given, Eq.~(\ref{eq:two-scale }), the two-scale
equation, is recursively solved point-by-point (see  Ref.~\cite{Daub10-6}  for
the specific values of $ \{h_{l}\}$ and more sophisticated and detailed
treatment of the two-scale relation). Thus, the series $\{h_l\}$ determines all
characteristics of the corresponding SF.

\begin{figure}[h]
	\includegraphics[width=1\linewidth]{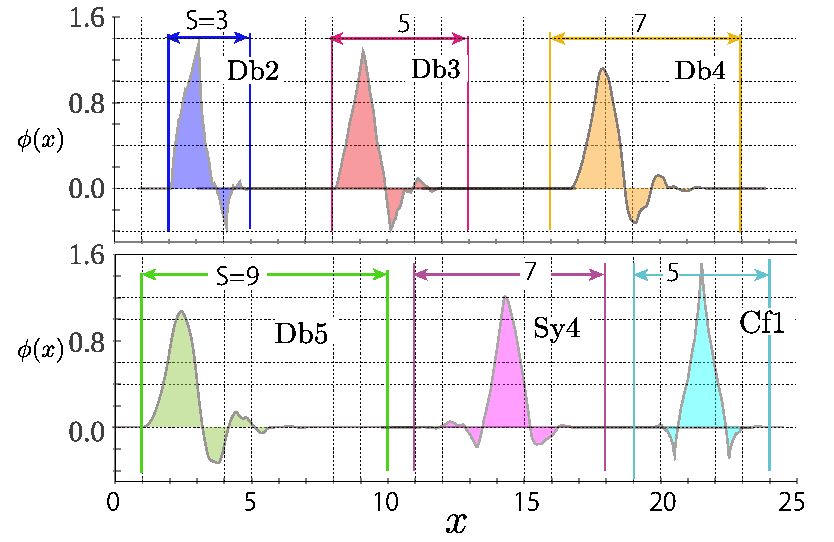}
	\caption{
    Examples of compactly supported scaling functions and their support lengths
$S$. Each compactly supported  scaling function has its own support, outside
which the value of the scaling function is completely zero.  In general, the
longer  the support is, the smoother the SF becomes.
}
	\label{fig:scaling-functions}
\end{figure}

Figure~\ref{fig:scaling-functions} shows examples of  orthonormal  compactly
supported scaling functions\cite{Daub10-6}, Daubechies-2 to -5  (Db2 to Db5),
Symlet-4(Sy4) and Coiflet-1(Cf1), which are created from different sets of $
\{h_{l}\}$.
\subsection{Composing Position Eigenfunction}\label{sec:composing-position-eigenvectors}

Position eigenvectors in a discrete system not only have to be translationally
orthonormal, but also they are desired to be strongly localized and at least
twice differentiable. If the space is spanned by a limited number of broad and
smooth basis vectors, such as plane waves, strongly localized orbits/position
eigenvectors with an exponential decay may not be composed.

The compactly supported SFs, such as Db4s, readily posse some of  the rare
characters
mentioned above.  In fact, a compactly supported SF is a good first
approximation of the constructed position eigenvector as shown later in Table
\ref{tab:xi_n} in Sec.~\ref{sec:position-eigenvectors}. Furthermore, most of
them are found to be \textit{practically twice differentiable} in a sense shown
in Appendix \ref{app:mom_op} and as demonstrated by the numerical results.
Thus, the compactly supported SFs are chosen to be the building blocks of
position eigenfunctions.
\subsubsection{Strategy}\label{sec:strategy} To switch to
the Dirac notation, we would like to have the state space representation of an
SF such as Db2-Db5, Sy4 and Cf1. Let $\vert n \rangle$ be the state
vector pertaining to  an SF, $\phi(x-n)$. Then $\vert n\rangle$ is formally
composed by the position eigenvector $\vert x \rangle $:
\begin{equation}
	\vert n \rangle = \int dx \phi(x-n)\vert x \rangle \quad(\;\mathrm{or}\;\;  \phi(x- n)= \langle x \vert n \rangle \;),
\end{equation}
where $n$ corresponds to the position of a sub-cell  seen in Fig.~(\ref{fig:series_of_scaling_functions}).
Therefore, the scaling-kets span the following space:
\begin{equation}
	\mathcal{H}_{S} ={\mathrm{Span}} \{\vert n \rangle: n \in \mathbb{Z} \}.
\end{equation}

By defining the position operator $\tilde x$  in $\mathcal{H}_{S} $:
\renewcommand{\arraystretch}{1.5}
\begin{equation}
\begin{array}{c}
\tilde x =\hat {P}_S  \hat x \hat {P}_S, \\
\hat {P}_S = \sum_n \vert n \rangle \langle n \vert,
\end{array}
\end{equation}
the eigenvectors of the position operator,  $\{\vert x_p \rangle \} $, are given as solutions of the following
eigenvalue equation:
\begin{equation}\label{eq:x-eigen}
\tilde x \vert x_p \rangle = x_p \vert x_p \rangle,
	 \ (\vert x_p \rangle \in \mathcal{H}_{S}, p \in \mathbb{Z}).
\end{equation}
$\vert x_p \rangle$ is expanded by the compactly supported scaling kets:
\begin{equation}\label{eq:x-eigen-decompose}
	\begin{split}
	\vert x_p \rangle &=\sum_n \vert  n \rangle \langle n \vert x_p \rangle \\
		 &=\sum_n \xi_{n,p} \vert  n \rangle,
	\end{split}
\end{equation}
where
\begin{equation}\label{eq:x-eigen-decompose_xi}
	\begin{split}
	\xi_{n,p} = \langle n \vert x_p \rangle.
	\end{split}
\end{equation}
Thus, the unknown to solve is now shifted from the position eigenvectors to the
matrix  $\{\xi_{n,p}\}$.
By multiplying $\langle m \vert$ from the left to the both sides of Eq.~(\ref{eq:x-eigen}):
\begin{equation}\label{eq:Xmp_1}
\begin{split}
	\langle m\vert\tilde{x}\vert x_{p}\rangle  &=\sum_{n}X_{m,n}\xi_{n,p},\\
	\langle m\vert x_p \vert x_{p}\rangle &=  x_p \xi_{m,p},  \\
\end{split}
	\end{equation}
where
\begin{equation}\label{eq:Xmp_2}
		X_{m,n}  =\langle m\vert\tilde x \vert n\rangle.
\end{equation}
And hence the equation for $\xi_{m,n}$ is obtained:
\begin{equation}\label{eq:Xmn}
		\sum_{n}X_{m,n}\xi_{n,p} = \xi_{m,p} x_p.
\end{equation}
\subsubsection{Position Operators and Matrix Elements}\label{sec:Xn}
$\{X_{m,n}\}$ can be obtained from the following integral:
\begin{equation}\label{eq:two-scale-integration-1}
\begin{split}
X_{m,n} =\int \phi(x-m)x\phi(x-n)dx.\\
\end{split}
\end{equation}
Owing to  the two-scale relation, Eq.~(\ref{eq:two-scale }), however,
the above is recursively expressed by  $\{X_{m,n}\}$ without actual integration:
 \begin{equation}\label{eq:two-scale-integration-2}
 \begin{split}
\int &\phi(x-m)x\phi(x-n)dx\\
 & =\sum_{l,k} 2\int  h_{l} h_{k}\phi(2x-2m -l)x\phi(2x-2n-k)dx\\
 &=\frac{1}{2}\sum_{l,k} h_{l} h_{k}X_{2m+l, 2n +k}.
 \end{split}
 \end{equation}
 By combining Eqs.~(\ref{eq:two-scale-integration-1}) and (\ref{eq:two-scale-integration-2}), we have
solvable algebraic  equations for $\{X_{m,n}\}$ with given coefficients
  $\{h_l\}$(see Appendix~\ref{app:x_op} for the general strategy to calculate matrix elements).
Since the support is compact, the integration becomes $0$ for
$ \vert m - n \vert \ge S$. This makes matrix $\{X_{m,n} \}$ sparse
and easy to handle.
From  the translational symmetry,  the diagonal elements are
simplified  to be:
\begin{equation}\label{eq:XmmX00}
\begin{split}
	X_{m,m}&=\int \phi(x-m)x \phi(x-m) dx \\
		&=\int \phi(y)(y+m)\phi(y) dy \quad(y=x-m)\\
		&= X_{0,0} +m.
\end{split}
\end{equation}
By defining the following series  $\{X_{r}\}$:
\begin{equation}\label{eq:Xseries}
X_r=
\begin{cases}
		X_{m+r,m} & (\mathrm{for\ } r \ne 0)\\
		X_{0,0}  & (\mathrm{for\ } r= 0),
\end{cases}
\end{equation}
$X_{m,n}$ is described by $X_r$ as follows:
\begin{equation}\label{eq:Xseries-1}
X_{m,n} = X_{r} + m\delta_{m,n} \quad(r = m-n).
\end{equation}

By substituting this into Eqs.~(\ref{eq:two-scale-integration-1}) and (\ref{eq:two-scale-integration-2}), the following equations
for $\{X_r\}$ are obtained:
\begin{equation}\label{eq:Xn-def}
	\begin{array}{l}
		  \boldsymbol{X}=\mathbf{H}\boldsymbol{X}+\boldsymbol{h},\\
		\left[ \mathbf{H} \right ]_{r,q}= \frac{1}{2}\sum_{k} h_{k} h_{k+2r-q},\\
		\left[ \boldsymbol{X} \right ]_{q}=X_q,\\
		\left[ \boldsymbol{h} \right ]_{r}= \frac{1}{2}\sum_{k}k h_{k-2r} h_{k}.
	\end{array}
\end{equation}
Since the above equations are linear algebraic and $\{h_l\}$ is given, it does
not take any actual  integration to obtain the matrix elements of the position
operator, $\{X_r\}$.

Table~\ref{tab:Xn} in Appendix~\ref{app:p_op} shows  $\{X_r\}$ calculated with
the listed SFs by numerically solving Eq.~(\ref{eq:Xn-def}).
\subsubsection{Position eigenvector}\label{sec:position-eigenvectors}
Since  the shape of the SFs and the position eigenfunctions are translationally invariant,
the matrix  $\{\xi_{m,n}\}$ is also reduced to a series  $\{\xi_{r}\}$:
\begin{equation}
 \xi_{m+r,m} \rightarrow \xi_r,
\end{equation}
as detailed in
Appendix~\ref{app:Position_Eigenvector_Equation}, this simplifies the Eq.~(\ref{eq:Xmn})
to the following equations for $\{\xi_n\}$,
\begin{equation}\label{eq:xi_eigen_eq}
\begin{split}
\sum_{n}\{X_{m-n}\xi_{n-p}+m\delta_{m,n}\xi_{n-p} \}=x_p \xi_{m-p}.
\end{split}
\end{equation}
 As described in Appendix~\ref{app:xi_solution}, Eq.~(\ref{eq:xi_eigen_eq})
 can be solved both analytically and numerically.
 The central part of the series, $\{\xi_{r}\}$, is shown in Table \ref{tab:xi_n}.
\begin{table}[h]
  \caption{Central part of the numerical values of  of $\{\xi_p\}$ corresponding to the SF indicated in the first column. The values are obtained
  by numerically solving Eq.~(\ref{eq:xi_eigen_eq}) with GSL library\cite{gsl}.}
  \label{tab:xi_n}
  \centering
  \begin{tabular}{l c c c c c c }
    \hline
    \hline
   &&$\xi_{-2}$  &  $\xi_{-1}$  &  $\xi_{0}$  &  $\xi_{1}$  &  $\xi_{2}$  \\  \hline
    \ 1. SF  		&& &  & & &   \\
    \hspace{3mm}a. Sy4  && -0.016\footnotemark[1] & -0.023 & 1.000 &  0.023 &  0.017 \\
    \hspace{3mm}b. Db2 	&& 0.004  & -0.070 & 0.995 &  0.071 &  0.001  \\
    \hspace{3mm}c. Db3 	&& 0.017  & -0.119 & 0.985 &  0.121 &  -0.002  \\
    \hspace{3mm}d. Db4 	&& 0.032 & -0.159 & 0.973 &  0.165 &  -0.006   \\
    \hline \hline
  \end{tabular}
  \footnotetext[1]{All figures are rounded off to the third decimal place.}
\end{table}
By using Eq.~(\ref{eq:x-eigen-decompose}) and  $\{\xi_p\}$, the coordinate representation of the position eigenvector
 $\xi_p(x)$ is obtained:
\begin{equation}\label{eq:xi-composed}
	\begin{split}
		 \xi_p(x)&= \langle x \vert x_p \rangle \\
		 &=\sum_n \xi_{n,p}\langle x  \vert  n \rangle \\
		 &=\sum_n \xi_{n-p}\phi(x-n), \\
	\end{split}
\end{equation}
where $\phi(x)$ is the given compactly supported SF, such as Db2-Db5 and Sy4.
\begin{figure}[b]
	\includegraphics[width=0.9\linewidth]{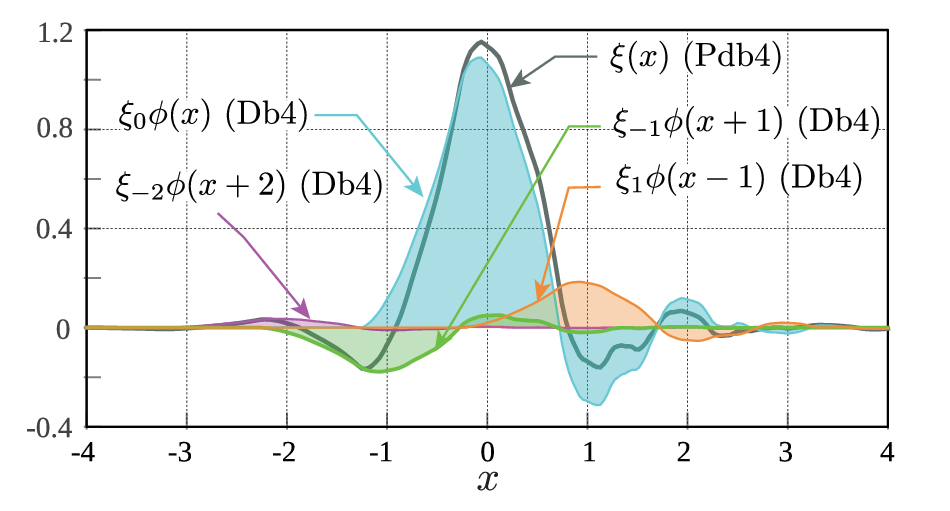}
\caption{Position eigenfunction $\langle x \vert \xi_{p=0} \rangle = \xi(x)$ (Pdb4) composed of Db4s.
A Pdb4 is drawn with a gray thick  line, and
 its components, Db4s multiplied by $\xi_r$,  are drawn with colored lines.
 $\xi_{-1}$ in green adds negative leading part to the central
 cyan-colored Db4 and the $\xi_1$ in pink lifts up the trailing negative part of the central Db4, thereby making the
 position eigenfunctions more symmetric than
 the Db4 colored in cyan,
 which is the main component of the Pdb4.
}\label{fig:synthesizingpositionscalingfunctions}
\end{figure}
Fig.~\ref{fig:synthesizingpositionscalingfunctions} shows  a position eigenfunction Pdb4
  with its component Db4s. It is seen the Pdb4 appears  more symmetric than Db4, and the
 Pdb4 decays exponentially (see Appendix~\ref{ap:decay}). The support appears to be
 virtually compact, extending approximately  from $-3$ to $3$.

\begin{figure}[h]
	\includegraphics[width=1.0\linewidth]{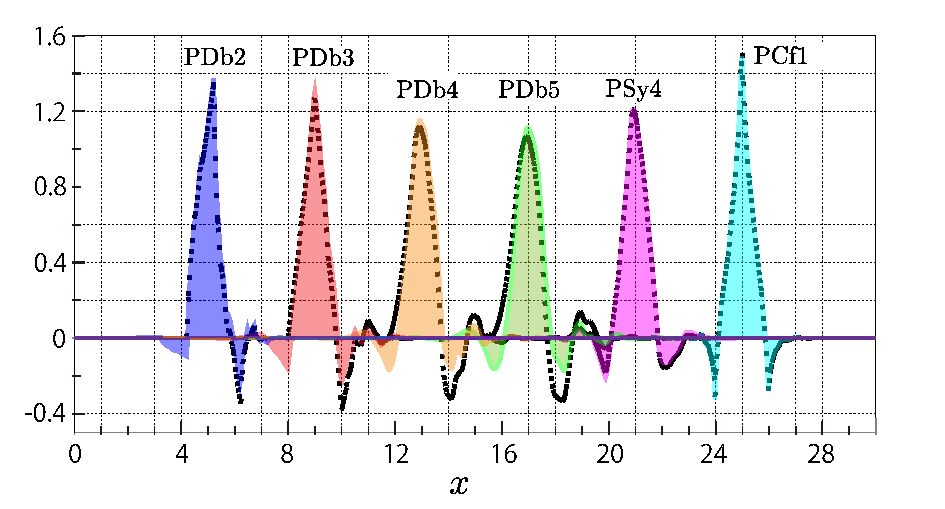}
	\caption{Position eigenfunctions.
	From left to right, Pdb2, Pdb3, Pdb4, Pdb5, Psy4 and Pcf1. The corresponding SFs are drawn  with  dashed black lines.}.
	\label{fig:eigenvectors}
\end{figure}

Figure~\ref{fig:eigenvectors} shows position eigenfunctions composed of Db2-Db5 (Pdb2-Pdb5), Sy4 (Psy4) and Cf1 (Pcf1).
As does the Pdb4 in Fig.~\ref{fig:synthesizingpositionscalingfunctions},
Pdb2, Pdb3 and Pdb5 appear to be more symmetric than the corresponding compactly supported SFs.
On the other hand, Psy4 and Sy4, Pcf1 and Cf1 look nearly the same,
since Sy4 and Cf1 are designed to be more symmetric than Dbns are\cite{Daub10-6,Daub10-8}.
Note that  the number of basis vectors in $\{\vert n\rangle \}$ and $\{\vert x_n \rangle
\}$ are the same and both sets are orthonormal, and hence they span the identical
space:
\begin{equation}
\begin{split}
&\mathcal{H}_{X} = {\mathrm{Span} }\{\vert x_p \rangle: p \in \mathbb{Z} \},\\
&\mathcal{H}_{S} =\mathcal{H}_{X}.
\end{split}
\end{equation}
\subsubsection{Shifting Eigenfunction and Eigenvalue}
The $0$-th eigenvector has $X_0$ as its eigenvalue (see Table \ref{tab:Xn} in Appendix~\ref{app:p_op}).
It is however more convenient to have a direct association of  the eigenvector number and the eigenvalue.
By  shifting the eigenfunctions  by  $X_0$:
\begin{equation}
	\xi_p (x) \rightarrow \xi_p (x+X_0),
\end{equation}
the transformation from an SF to the position eigenfunction changes as follows:
\begin{equation}\label{eq:position-shift}
\xi_p (x)  =\sum_n \xi_{n-p} \phi((x+X_0) -n).
 \end{equation}
Consequently  the eigenvalue is adjusted to be:
\begin{equation}
\tilde x \vert x_p \rangle = p \vert x_p \rangle.
\end{equation}
\subsubsection{Position Eigenvector as Scaling Function}
As detailed in Appendix~\ref{app:xi_solution}, the translational invariance of
$\{\xi_{n-m} \}$ leads to the conclusion that  the position eigenvectors are
also categorized to be  SFs. Namely, they themselves satisfy the following
two-scale relation:
\begin{equation}
\xi(x)=\sqrt{2}\sum_{l}\eta_{l}\xi(2x-l),\\
		\quad\eta_{l}=\sum_{p,q,l}\xi_{p}\xi_{q}h_{q+l-2p}.
\end{equation}
The position eigenvectors therefore have their own wavelets, which are
shown in Appendix~\ref{app:pos-eigen}. They may be utilized
to solve or analyze quantum mechanical systems where
multi-resolution capability are important.
\subsection{Computing Matrix Elements of Kinetic Energy Operator}\label{app:kinetic-q}
 The matrix elements of the kinetic energy operator, $\{P_r\}$, are first
calculated with a set of compact support SFs without any actual  integration or
differentiation. Subsequently, $\{P_r\}$ are converted to the matrix elements
pertaining to  the position scaling-eigenvectors, $\{Q_r\}$, by using  $\{\xi_r \}$.

As detailed in Appendix~\ref{app:mom_op}, we have the equations for the matrix
elements of
$\hat p^n $:
\begin{equation}
	\begin{array}{l}
		  \mathbf{K}\cdot\boldsymbol{P}=\rho \boldsymbol{P},\\
\left[   \mathbf{K} \right]_{r,q}=\sum_{k} h_{k} h_{k+2r-q}, \\
\begin{aligned}
\left[   \boldsymbol{P}  \right]_r&=P_r\\
 &=
 \langle r\vert \hat p^{n} \vert 0\rangle,\\
\end{aligned}\\
n = \log_2 \rho.
	\end{array}
\end{equation}
The specific values of $\{P_r\}$ obtained from the above equations are listed on
Tables \ref{tab:Pn} and \ref{tab:P2n}. Note that when the eigenvalue $\rho=2$,
$\{P_r\}$ are the matrix elements of the momentum operator. When
$\rho=4$,  $\{P_r\}$  are the elements of the kinetic energy operator. The
reason and the details of solution procedure are detailed in Appendix
\ref{app:mom_op}.

By using $\{\xi_p\}$,  we get the matrix elements of the kinetic energy
operator associated with the position scaling-eigenvectors:
\begin{equation}\label{eq:ktoq}
\begin{split}
Q_r &= \langle x_{r} \vert \hat{p}^2 /2 \vert x_ 0 \rangle \\
 &= \sum_{k,l} \{\xi_{k-r} \langle k \vert \} \hat{p}^2 /2 \{\xi_{l} \vert l \rangle \}\\
&=\sum_{k,l} \xi_{k-r} \xi_{l} P_{k,l}\\
&=\sum_{k,q} \xi_{k-r} \xi_{k-q} P_{q}.\\
\end{split}
\end{equation}
Table \ref{tab:Q2n} shows the numerically obtained  $\{Q_r\}$.

A measure of differentiability, the H\"older exponents\cite{Daub10-7} are shown
in Table \ref{tab:P_Holder} together with the support lengths. Roughly speaking,
for instance, Db4 is $1.6179$ times differentiable and other H\"older exponents
are also smaller than $2$. The matrix elements of the kinetic energy are
nevertheless obtained, and they produce as good results as the preceding  ones
as shown in Sec.~\ref{sec:results}. The reason Pdb2 does not have adequate
values of the matrix elements are discussed in  Appendix~\ref{app:mom_res}.
\begin{table}[h]
	\caption{\label{tab:Q2n}
		Matrix elements of $\hat p^2/2$, $Q_r= \langle x_r \vert \ \hat p^2/2 \vert x_0 \rangle $ obtained with different  position scaling-eigenfunctions.
	}
	\centering
	\begin{tabular}{l c c c c c c }
  \hline
  \hline
		 && $-Q_{0}$ &  $-Q_{1}$  &  $-Q_2$  &  $-Q_{3}$  &  $-Q_{4}$\footnotemark[1] \\
		\hline
   \ 1. Position eigenfunction  		&& &  & & &   \\
		\hspace{3mm}a. Psy4  && -4.008\footnotemark[2]  &  2.540  & -0.671 & 0.145 & -0.010 \\
		\hspace{3mm}b. Pdb2 	&& \footnotemark[3] & \footnotemark[3]  & \footnotemark[3] & \footnotemark[3]  & \footnotemark[3] \\
		\hspace{3mm}c. Pdb3 	&& -5.265 & 3.389 & -0.876  & 0.114  & 0.005 \\
		\hspace{3mm}d. Pdb4 	&& -3.991 & 2.529 & -0.668 & 0.144 & -0.010 \\
    \hline \hline
	\end{tabular}
	\footnotetext[1]{$Q_{-n}=Q_{n}$ when $\rho=4$. }
	\footnotetext[2]{All figures are rounded off to the third decimal place.}
	\footnotetext[3]{Not available. See  Appendix~\ref{app:mom_res} for the reason. }
\end{table}
\begin{table}[h]
	\caption{\label{tab:P_Holder}
		Support lengths and H\"older exponents of SFs. }
	\centering
	\begin{tabular}{l c c l}
		\hline \hline
		 & Support && H\"older exponent $\alpha$ \\
		\hline
    1. SF	  & && \\
		\hspace{3mm}a. Db2		& 3 && 0.5500\footnotemark[1] \\
		\hspace{3mm}b. Db3		& 5 && 1.0878\footnotemark[1] \\
		\hspace{3mm}c. Db4		& 7 && 1.6179\footnotemark[1] \\
		\hspace{3mm}d. Db5		& 9 && 1.596\footnotemark[2] \\
		\hspace{3mm}e. Sy4  	& 7 &&\footnotemark[3]	 \\
		\hspace{3mm}f. Cf4		& 5 && \footnotemark[3]  \\
    \hline \hline
	\end{tabular}
	\footnotetext[1]{Reference\cite{Daub-Lag-1}.}
	\footnotetext[2]{Reference\cite{Daub1988}.}
	\footnotetext[3]{Not available.}
\end{table}
\subsection{Adapting Eigenvectors to Crystal System}\label{adaptation}
 To apply the position eigenfunctions to 1D electronic and photonic crystals with continuous real coordinates, the following assumptions are made: (1) The periodic boundary condition is imposed at the both ends of the crystal.
(2) The unit cell length is set at 1. (3) In electronic systems, the following
units are used: $m=1,\; \hbar = 1$. (4) In optical systems, the speed of light
$c$ is set at 1. (5) The unit cells are divided in $N$ and $\Delta x = 1/N$.
These assumptions  result in  the following adjustment to the functions,
operators and variables as described in  Appendix~\ref{app:coordinate-trans}:
\begin{equation}\label{eq:disc-transform}
\begin{array}{c}
p \rightarrow x_p=l+ m\Delta x, \\
Q_r \rightarrow  N^{2}Q_r \quad (\rho=4),\\
\end{array}
\end{equation}
with
\begin{equation}\label{eq: p_mod_N}
  \begin{array}{c}
    \Delta x = 1/N,\\
    l = p\!\!\!\! \mod N,\\
    m=p - lN,\\
  \end{array}
\end{equation}
and
\begin{equation}\label{eq:def_xi}
\begin{split}
  \langle x\vert x_{p}\rangle
  &=\sqrt N \xi(N(x-x_p)),\\
\end{split}
\end{equation}
\begin{equation}\label{eq:xi_integral}
\begin{split}
  \int dx \langle x \vert x_p \rangle &=\sqrt N\int dx \xi(N(x-x_p))\\
  &=\sqrt N\int \xi(y-p)\frac{dy}{N}\\
  &=1/\sqrt{N}.\\
\end{split}
\end{equation}
\subsection{Single-Band Bloch Functions}
A BF pertaining to the $n$-th band with its wave number $k$ has the next form:
\begin{equation}\label{key}
\begin{split}
\vert \psi_k^n \rangle &= \frac{1}{\sqrt L }e^{ik\tilde x} \vert u_k^n\rangle. \\
\end{split}
\end{equation}
The cell-periodic part $\vert u_k^n\rangle$ is expanded by $\{\vert x_p \rangle\}$:
\begin{equation}
	\begin{split}
		\vert u_k^n \rangle &=\left\{\sum_{p=0}^{NL-1} \vert x_p \rangle \langle x_p \vert \right\} \vert u_k^n \rangle  \\
&=\sum_{p=0}^{NL-1} c_{p,k}^{n}\vert x_p \rangle,\\
	\end{split}
\end{equation}
where $\{c^n_{p, k} \}$ is the $N$-periodic (cell-periodic) coefficients of expansion:
\begin{equation}\label{key}
	c^n_{p, k}=c_{p+N,k}^{n}=\langle x_p \vert u_k^n\rangle,
\end{equation}
and hence by utilizing Eq.~(\ref{eq: p_mod_N}):
\begin{equation}
	\begin{split}
		\vert u_k^n \rangle
		&=\sum_{l=0}^{L-1} \sum_{m=0}^{N-1} c_{m,k}^{n} \vert x_{Nl+m}\rangle \\
		&=\sum_{l=0}^{L-1} \sum_{m=0}^{N-1} c_{m,k}^{n} \vert x_{p}\rangle. \\
	\end{split}
\end{equation}
The BF is thus decomposed with $\{\vert x_p \rangle\}$:
\begin{equation}\label{eq:bloch-x}
	\begin{split}
		\vert\psi^n_{k}\rangle &=\frac{1}{\sqrt L }\sum_{l=0}^{L-1} \sum_{m=0}^{N-1} e^{ik\tilde x} c_{m,k}^{n}\vert x_{p}\rangle  \\
		&=\frac{1}{\sqrt L }\sum_{l=0}^{L-1} \sum_{m=0}^{N-1} e^{ikl}e^{ikm\Delta x }c_{m,k}^{n}\vert x_{p}\rangle,
	\end{split}
\end{equation}
where Eq.~(\ref{eqap:fn_phi_n}) in Appendix~\ref{sec:interpolation} is applied.
\subsection{State Vectors and Spatial Resolution}\label{sec:vector-space}
In many cases, to solve the Schr\"odinger equation, it takes discretization of
the entire state space $\mathcal H$ by a set of a finite number of basis
vectors, $\{\vert \eta_m \rangle\}$, which may be plain waves, Gaussian
functions, atomic orbits and so on. The position eigenvectors of the discretized
space:
\begin{equation}\label{key}
	\mathcal H_\eta = \mathrm{Span} \{\vert \eta_m\rangle\:,0 \le m < N \},
\end{equation}
are obtained by solving the following equation:
\begin{equation}\label{eq:eta-eigen}
	\begin{split}
		\tilde x \vert x^\eta_p \rangle = x^\eta_p \vert x^\eta_p \rangle, \
		\left(\vert x^\eta_p \rangle  = \sum_m u^p_m \vert \eta_m\rangle \right).
	\end{split}
\end{equation}
Then $\xi^\eta_p (x)=\sqrt N \langle x \vert x^\eta_p \rangle $ is the
discretized version of $\delta$-function in $\mathcal H_\eta$, and it provides
the information on the spatial resolution constrained by  $\{\vert \eta_m
\rangle\}$.

In this paper, the basis vectors themselves are position eigenvectors and the spatial
resolution of the solutions such as BFs and WFs are clear from the time of
formulation, as described in Appendix~\ref{app:interpolatin}.
\section{MLWF and Position Scaling-Eigenvector in State Space}\label{sec:MLWF_SF}
The section focuses on the properties of the position eigenvectors
projected onto a composite energy band space,
$\mathcal H_C$, consisting of $n_c$ energy bands defined as follows:
\begin{equation}\label{eq:HB}
    \begin{split}
        &\mathcal{H}_{C}=\sum_n \mathcal{H}_{B_n}, \\
        &\mathcal{H}_{B_n}=
        \mathrm{Span}\{\vert\psi_{l k_0}^{n}\rangle\:\  0\le l < L,  k_0 = 2\pi/L \},
    \end{split}
\end{equation}
where  $L$ is the number of the unit cells in the system and energy bands
included in $\mathcal H_C$ are arbitrary and $n$  does not have to begin from
$0$ and the band indexes do not have to be successive numbers.

Subsequently, the relationship between the projected position eigenvectors and MLWFs in $\mathcal H_C$ leading to approximations of MLWFs by the projected eigenvectors is discussed.
\subsection{MLWF}
In the section, important properties of MLWFs utilized in the succeeding
sections are reviewed in preparation for  discussing approximation utilizing the
position scaling-eigenvectors,
\subsubsection{Eigenvalue Equations for MLWFs}\label{sec:eq-MLWF}
MLWFs satisfy the following equation\cite{Marzari1997,kivelson1982}:
 \begin{equation}\label{eq:btow_lambda}
\begin{split}
\tilde{x}^C  \vert W \rangle &= x \vert W \rangle, \\
\end{split}
\end{equation}
where $\tilde x^C $ is the position operator  in $\mathcal{H}_C$:
\begin{equation}\label{eq:PB}
	\begin{split}
		\tilde {x}^C&= \hat {P}_C \hat {x} \hat {P}_C,\\
		\hat {P}_C &= \sum_{k,n} \vert \psi_k^n \rangle \langle \psi_k^n \vert.
	\end{split}
\end{equation}
Because of  the number of basis vectors and the translational invariance of the
system, the solution has to have the following form (see
Appendix~\ref{app:wannier-eigeen}):
\begin{equation}\label{eq:btow_s_l}
	\begin{split}
		\tilde{x}^C  \vert W^s_l \rangle &= (x_{0s}+l) \vert W^s_l  \rangle, \\
	\end{split}
\end{equation}
with
\begin{equation}
\tilde x^C \vert W^s_0 \rangle = x_{0s}\vert W^s_0 \rangle,
\end{equation}
 where $s$ and $l$  denote the series and the cell number, respectively.
 The number of series is $n_c$, for the number of energy bands involved is $n_c$.

The MLWF as a function of position is given by the following projection
 (see Appendix~\ref{app:suppl_0}):
\begin{equation}\label{eq:MLWF(x)}
\begin{split}
\langle x_p\vert W^s_{l}\rangle
&=N^{-1/2}W^s_0(x_p-l)\\
&=N^{-1/2}W^s_l(x_p).
\end{split}
\end{equation}
 \subsubsection{Orthonormality of MLWFs}\label{sec:Orthonormmality of MLWFs}
Because MLWFs have to be orthonormal:
\begin{equation}\label{eq:MLFs-should-orthonormal}
    \begin{split}
        \langle  W^{s_1}_{l_1}  \vert  W^{s_2}_{l_2} \rangle =\delta_{s_1,s_2}\delta_{l_1, l_2}.\\
    \end{split}
\end{equation}
The numbers of MLWFs and  BFs are the same, and they span the same space, $\mathcal H_C$ and hence:
\begin{equation}\label{key}
        \hat {P}_C  = \sum_{l,s} \vert W_l^s \rangle \langle W_l^s \vert
       = \sum_{k,n} \vert \psi_k^n \rangle \langle \psi_k^n \vert.
\end{equation}
Since MLWFs are expanded as described in Appendix~\ref{app:wannier-eigeen}:
\begin{equation}\label{eq:wls-expansion}
    \vert W_{l}^{s}\rangle=\sum_{k}e^{-ikl}w_{k}^{s,n}\vert\psi_{k}^{n}\rangle,
\end{equation}
the following holds:
\begin{equation}\label{key}
    \langle W_{l_{1}}^{s_{1}}\vert W_{l_{2}}^{s_{2}}\rangle
    =\sum_{k}e^{-ik(l_{2}-l_{1})}\sum_n \left( \overline{w}_{k}^{s_{1},n}w_{k}^{s_{2},n} \right).
\end{equation}
Throughout the paper \textit{overline} on a variable, such as $\overline
\square$, denotes the complex conjugate of the variable underneath.

By the requirement given by Eq.~(\ref{eq:MLFs-should-orthonormal}) for any $l_1,
l_2 \in \mathbb Z$, we have:
\begin{equation}\label{eq:MLWF-orthonormal-delta}
    \sum_n  \overline{w}_{k}^{s_{1},n}w_{k}^{s_{2},n}=\frac{1}{L}\delta_{s_1,s_2}.
\end{equation}
\subsection{Position Eigenvectors Projected onto Composite Band Space}\label{sec:ape-general}
The section focuses on the properties of the position eigenvectors when they are projected
onto composite bands.
\subsubsection{Definition}
MLWFs are the eigenvectors of the operator that is the position operator  projected onto $\mathcal H_C$,
as described in Eq.~(\ref{eq:PB}).  In contrast to this, the  projected position
eigenvectors (PPE)
are  defined as the position eigenvectors, $\vert x_p \rangle \in \mathcal{H}_{X}$, projected onto $\mathcal H_C$:
\begin{equation}\label{eq:ppe-definition}
    \begin{split}
        \vert  x^C_p \rangle&= \hat {P}_C \vert x_{p}\rangle  \\
            &= \sum_{k,n} \!\vert\psi^n_{k}\rangle\langle\psi^n_{k}\vert x_{p}\rangle  \\
        &=\frac{1}{\sqrt{L}}\sum_{k,n} e^{-ikl}e^{-ikm\Delta x}\bar{c}^n_{m,k}\vert\psi^n_{k}\rangle,
    \end{split}
\end{equation}
where the superscript $C$ denotes that the vector is not a position  eigenvector in $\mathcal H_X$,
but in $\mathcal H_C$.
\subsubsection{Properties of PPE}
Since any vector in $\mathcal H_C$ is expanded by $\vert \psi^n_k \rangle \in \mathcal H_C$:
\begin{equation}\label{eq:ape-functionsqrt N value}
    \vert f \rangle = \sum_{n,k} f^n_k \vert \psi^n_k \rangle.
\end{equation}
The projection of $\vert f \rangle $ onto a PPE becomes:
\begin{equation}\label{eq:ape-character_1}
    \begin{split}
        \langle  x^C_p \vert f \rangle
        &=\left\{ \langle x_{p}\vert \sum_{n_1,k_1} \vert\psi^{n_1}_{k_1}\rangle\langle\psi^{n_1}_{k_1} \vert \right\}  \sum_{n,k} f^n_k \vert \psi^n_k \rangle \\
        &=\sum_{n,k}\langle x_{p} \vert \psi^{n}_{k}\rangle f^n_k  \\
        &=\langle  x_p \vert f \rangle,
    \end{split}
\end{equation}
and hence PPEs behave as if they were the position eigenvectors in
$\mathcal H_C$.

It is also noted that the inner product is a measure of
closeness of the two vectors and by definition:
\begin{equation}\label{eq:closeness}
    \cos \theta_{x_p,f} = \langle  x^C_p \vert f \rangle
    / \sqrt{\langle  x^C_p \vert x^C_p \rangle \langle  f \vert f \rangle}.
\end{equation}
From Eq.~(\ref{eqap:fn_phi_xp}) in Appendix~\ref{sec:approx_BF},
\begin{equation}\label{eq:ape-character_2}
        \langle  x_p \vert f \rangle = \frac{1}{\sqrt N } (f(x_p) + O(\sigma_0^2/N^2)),
\end{equation}
the value of a function at $x=x_p$ is a measure of the closeness between the vector and
the corresponding PPE.
\subsubsection{Approximation of PPE}
By utilizing Eq.~(\ref{eq:ape-character_2}),
\begin{equation}\label{key}
        \langle\psi^n_{k}\vert x_{p}\rangle \simeq \frac{\overline \psi^n_k(x_p)}{\sqrt N}.
\end{equation}
PPE defined by Eq.~(\ref{eq:ppe-definition}) is therefore approximated with the values of
the BFs pointwise:
\begin{equation}\label{eq:pwe-definition}
        \vert \tilde{x}^C_p \rangle=\sum_{k,n} \frac{\overline \psi^n_k(x_p)}{\sqrt N}  \vert \psi_k^n \rangle,
\end{equation}
which is calculated from the values of the BFs rather than $\{c^n_{pk} \}$.
From Eq.~(\ref{eqap:fn_phi_xp}), the error is given as follows:
\begin{equation}\label{eq:pape-definition}
        \langle  x \vert x^C_p \rangle = \langle x_p \vert \tilde{x}^C_p \rangle + O(\sigma_0^2/N^2).
\end{equation}
To distinguish $\vert \tilde x^C_p \rangle $ from PPE, it is termed pointwise projected position eigenvector (PWE).

\subsection{Approximation of MLWF with PPE}\label{sec:PPE-approx}
The section describes the following properties of PPEs and PWEs in relation to
MLWFs to examine the possibility of PPEs and PWEs being used as alternatives or
supplements to MLWFs: (1) Closeness in terms of shape and position. (2)
Closeness in terms of their characters in the relevant space, i.e., intra- and
inter-series orthonormality.

The results of the analysis justify the intuitive closeness of PPEs and PWEs to
MLWFs. The similarities and differences are summarized in Appendix
\ref{contrast}. Furthermore, it is shown that a relatively light calculation,
searching the maximum point of a function, $F(m)= \sum_{k}\left|c_{m,k}\right|^2
$, gives the maximum point of the MLWF.

The specific estimate of error associated with the approximation of an MLWF by
the PPE will be presented later in Sec.~\ref{sec:ape-error}, albeit, for single
band cases.
\subsubsection{Maximum Point of PPE and MLWF}\label{sec:Closest PPE to MLWF}
When $\vert x_p^C \rangle$ is closest to $\vert W^s_{l}\rangle $
among:
\begin{equation}\label{key}
    \{\vert x^C_q \rangle: 0 \le q < NL \},
\end{equation}
$x_p$ is the maximum point of $\left| W_{l}(x)  \right|$ from Eqs.~(\ref{eq:MLWF(x)}) and (\ref{eq:closeness}).
Combining the above and the following equation
obtained in Appendix \ref{app:why x_p is maximum p}:
\begin{equation}\label{eq:ape-max-is-MLWF-max}
    \begin{split}
        \xi_p^C(x)=\langle x \vert x_p^C\rangle &\simeq N^{-1/2}W^s_{l}(x)\overline{W}^s_{l}(x_p),
    \end{split}
\end{equation}
we notice the position $x_p$ is also the maximum point of $\xi^C_p(x)$. Thus,
finding the maximum point of an MLWF is achieved by searching the maximum point
of the tallest PPE. It should be however noted, in cases PPEs do not have a
clear maximum, as will be shown later in Sec.~\ref{sec:num-photonic}
(e.g.,Fig.~\ref{fig:bush-E-1}(d)), the actual $\sigma^2_x$ (spread) has to be
calculated to identify the maximally localized PPE.

To signify the difference between PPE and PWE, and to show the procedure of
finding the approximate position of an MLWF to a \textit{programmable} degree,
the section focuses on the single band case. By restricting the band index to
$n$,  a PPE projected onto itself is calculated as follows:
\begin{equation}\label{eq:MLWF-max-and-ape}
    \begin{split}
        \xi_p^n(x_{p}) & =\langle x^n_{p}\vert x_{p}^n \rangle\\
        & =\sum_k \langle x_{p}\vert\psi^n_{k}\rangle\langle\psi^n_{k}\vert x_{p}\rangle\\
        & =\sum_{k,p_{1},p_{2}}\langle x_{p}\vert\left\{\frac{1}{\sqrt{L}}e^{ik x_{p_1}}c^n_{p_{1},k}\vert x_{p_{1}}\rangle\right\} \\
        &\quad\quad\quad\quad\quad \left\{\frac{1}{\sqrt{L}}\langle x_{p_{2}}\vert e^{-ik x_{p_2}}\overline{c}^n_{p_{2},k}\right\} \vert x_{p}\rangle\\
        & =\sum_{k}\left|c^n_{p,k}\right|^{2},
    \end{split}
\end{equation}
with
\begin{equation}\label{-}
    \begin{split}
        m = p\ \mathrm{mod}\ N, x_p = p\Delta x.
    \end{split}
\end{equation}
On the other hand, a PWE projected onto itself is calculated as follows:
\begin{equation}\label{eq:pwe-projection}
    \begin{split}
        &\ \langle\tilde{x}_{p}^{n}\vert\tilde{x}_{p}^{n}\rangle
        =\frac{1}{N}\sum_{k_1,k}\left\{\langle\psi_{k_1}^{n}\vert\psi^{n}_{k_1}(x_{p})\right\} \left\{\overline{\psi}_{k}^{n}(x_{p})\vert\psi^{n}_{k}\rangle\right\} \\
        &\quad\quad\quad=\frac{1}{N}\sum_{k}\psi_{k}^{n}(x_{p})\overline{\psi}_{k}^{n}(x_{p})\\
       &\ \quad\quad \left(=\sum_{k}\left|c_{p,k}^{n}\right|^{2}+O\left(\sigma^{2}/N^{2}\right)   \right) ,
    \end{split}
\end{equation}
where Eq.~(\ref{eqap:fn_phi_xp}) in Appendix \ref{sec:interpolation} is
utilized.

From the above, the following points are easily understood: (1) Finding the
approximate position of the MLWF in a unit cell  is reduced to searching  the
maximum point of $F(m) =\sum_{k}\left|c_{m,k}\right|^{2} $. This is
expected to be a relatively light work on computer.  (2) PWEs are potentially
useful, for it can be computed from known values of the BFs calculated in some
other ways.

However, it has to be noted: (a) A PPE and the corresponding PWE are not
identical. (b)  The spatial resolution is
limited by the number of the independent bases vectors used in the calculation of BFs.
(c)  The distinctiveness of the peak of the PWE depends on the nature of the
basis as is mentioned in the opening of Sec.~\ref{sec:composing-position-eigenvectors}.
\subsubsection{Maximum Values of PPE and MLWF}\label{sec:Value PPE to MLWF}
By normalizing the PPE closest to an MLWF, we have the following approximation:
\begin{equation}\label{eq:w=x_p}
    \vert W_l^s\rangle \simeq  \frac{1}{\sqrt{\left| \langle x^C_p \vert x^C_p \rangle \right| }} \vert x^C_p \rangle.
\end{equation}
By combining the above and  Eq.~(\ref{eq:MLWF(x)}), we have:
\begin{equation}\label{key}
    \vert {W}^s_{l}(x_p) \vert=\sqrt{N\left| \langle x^C_p \vert x^C_p \rangle \right| }.
\end{equation}
Thus, the maximum value of an MLWF is obtained from that of the tallest PPE.

    \subsubsection{Intra-Series Orthonormalization of PPEs }\label{sec:Orthonormalization of PPEs }
Since the PPEs defined and described so far are not translationally orthonormal within a series, i.e.:
\begin{equation}\label{key}
    \langle  x^C \vert  (x+l)^C \rangle =
    N^{-1}\sum_{k,n} \psi^n_k(x)  \overline \psi^n_k(x+l).
\end{equation}

To use PPEs as an alternative to MLWFs, they have to be orthonormalized (Appendix~\ref{app:orthcond}),
and hence PPEs is redefined as follows:
\begin{equation}\label{eq:orthonormal-ape}
    \begin{split}
        \vert  x^{\perp C}_p \rangle &= \sum_{k,n} \frac{1}{\Phi_k} \left\{
        \vert\psi^n_{k}\rangle \langle \psi^n_k \vert x_p \rangle
        \right\},\\
        \Phi_k &= \sqrt {\sum_n \vert \langle \psi^n_k \vert x_p \rangle \vert^2 }.
    \end{split}
\end{equation}
The above definition differs from that of Eq.~(\ref{eq:ppe-definition}), for $ \Phi_k $ is
at the denominator.
However, when $\vert  x^C_p \rangle$ is closest to an MLWF,
 from Eq.~(\ref{eqap:xcxc}) in Appendix \ref{app:intra}, the following holds:
 \begin{equation}\label{eq:phi=sqrt xp2}
     \begin{split}
         \Phi_k          \simeq \sqrt{\left| \langle x^C_p \vert x^C_p \rangle \right| }.
     \end{split}
 \end{equation}
And hence:
\begin{equation}\label{eq:ppe-factor}
    \begin{split}
        \vert  x^{\perp C}_p \rangle &
        \simeq  \frac{1}{\sqrt{\left| \langle x^C_p \vert x^C_p \rangle \right| }} \vert x^C_p \rangle.
    \end{split}
\end{equation}
Therefore, the orthonormalized PPEs are equal to the approximated MLWFS seen in Eq.~(\ref{eq:w=x_p}).

Similarly, orthonormalized PWEs are defined as follows:
\begin{equation}\label{eq:orthonormal-pwe}
    \begin{split}
        \vert  \tilde x^{\perp C}_p \rangle &= \sum_{k,n} \frac{\psi^n_k(x_p)}{\tilde \Phi_k}
        \vert\psi^n_{k}\rangle,\\
        \tilde \Phi_k &= \sqrt {\sum_n \vert \psi^n_k(x_p) \vert^2 }.
    \end{split}
\end{equation}
\subsubsection{Inter-Series Orthonormalization of PPEs }\label{sec:Inter-Orthonormalization of PPEs }
When there are two or more series of MLWFs, the same number of PPE series have to exist, and
one series  has to be orthogonal to others. Let $\vert  x^C_{p_1} \rangle$ be the closest PPE to an MLWF in
series $s_1$,
and
  $x_{p_0}$ be the zero of $ \xi^C_{p_1} (x)= N^{-1/2}\langle x \vert x^C_{p_1} \rangle $
then
\begin{equation}\label{eq:w_s_w_s_1is0}
	\begin{split}
	\vert \xi_{p_1} (x_{p_0}) \vert
		&= N^{-1/2}\vert \langle x^C_{p_0} \vert x^C_{p_1}  \rangle \vert \\
		&= N^{-1/2}\vert \langle  x^C_{p_1} \vert x^C_{p_0}  \rangle \vert
		= \vert \xi_{p_0} (x_{p_1}) \vert
		=0.
	\end{split}
\end{equation}
We find  $\vert  x^C_{p_1} \rangle$ and $\vert  x^C_{p_0} \rangle$ are
orthogonal. Thus, an approximation of another series is obtained
(see 	Fig.~\ref{fig:wang_4} for example). One unresolved
problem is it does not guarantee that $ \vert  x^C_{p_0}  \rangle$ is the most
localized PPE in the vicinity of $ x_{p_0}$. Likewise, if $x_{p_0}$ is chosen so
that $\vert x^C_{p_0}  \rangle$ is most localized in the vicinity, the
inter-series orthogonality is not guaranteed. Nonetheless, the numerical results
shown in Sec.~\ref{sec:results} well agree with MLWFs at least on the figures.
\subsection{Density Matrix and Initial-Guess Orbit}\label{sec:scdm}
Position representation of $\hat P_C$, the density matrix\cite{Vitale2020},
\begin{equation}\label{eq:dm}
    \begin{split}
		&\rho_{x_2}(x_1)=\langle x_1 \vert \hat P_C \vert x_2\rangle\\
        &\left(
        \vert x_n \rangle \in \mathcal H,\ \langle x \vert x_n \rangle = \delta(x-x_n)\ ,n = 1,\ 2
        \right),
    \end{split}
\end{equation}
is  proved to fall off exponentially as a function of
$x_1$~\cite{PhysRev.129.554, PhysRev.135.A685, PhysRev.135.A698, Prodan2005,
  Benzi2013}, and it is used as an initial-guess of the MV method either as it is
\cite{Pizzi2020,Vitale2020,ct500985f,damle2016scdmk} or  by replacing the
$\vert x_2 \rangle$ with a localized wave function such as an atomic orbit
expected to be close to the MLWF\cite{freimuth2008publisher,PhysRevB.95.075114,
  PhysRevB.103.075125,PhysRevB.74.195116,Vitale2020,PhysRevB.92.165134,Pizzi2020}.
The position representation of PPE defined by Eq.~(\ref{eq:ppe-definition}) and
$\langle  x^C_p \vert f \rangle$ in Eq.~(\ref{eq:ape-character_2}) resemble the
density matrix. In the continuous limit, they are identical, however  $\vert
x_p\rangle$ and $\vert f \rangle$ in the paper are vectors in a discretized space
 $\mathcal H_X$.

When the state space is spanned by a finite number of orthonormal vectors
 $\{\vert \eta_m \rangle\}$, as in the case of numerical simulations or
mathematical modeling, the state space is discretized as discussed in
Sec.~\ref{sec:vector-space}:
\begin{equation}\label{eq:cont2discrete}
	\begin{split}
		\mathcal H  &  \rightarrow \mathcal H_\eta, \\
		\int dx \vert x \rangle \langle x \vert &  \rightarrow  \sum_m \vert \eta_m \rangle  \langle \eta_m \vert\\
		&  = \sum_p \vert x^\eta_p \rangle \langle  x^\eta_p \vert, \\
	\end{split}
\end{equation}
where $\{\vert x^\eta_p \rangle\}$ is the set of position eigenvectors in $\mathcal H_\eta$
and spans the same space:
\begin{equation}
    \mathrm{Span} \{\vert x^\eta_p \rangle \} = \mathrm{Span} \{\vert \eta_m \rangle \}=\mathcal H_\eta.
\end{equation}
And hence the density matrix becomes:
\begin{equation}\label{eq:dm-continuous}
	\begin{split}
		\langle x \vert \sum_p \vert x^\eta_p \rangle \langle  x^\eta_p \vert x_2  \rangle
		=& N^{-1/2}  \langle x \vert \sum_p\vert x^\eta_p  \rangle  \bar \xi^\eta_p (x_2) \\
		=& N^{-1}\sum_p  \xi^\eta_p (x) \bar \xi^\eta_p (x_2).
  	\end{split}
\end{equation}
The right-hand side of Eq.~(\ref{eq:dm-continuous}) involves more than one
position eigenvectors, thereby making  the density matrix potentially broader
than the inherent spread of $\vert x^\eta_p\rangle$ (one example is a delta
function overlapping all molecular orbits).

On the other hand, the PPE for $\vert x_{p_2} \rangle \in \mathcal H_x$ is:
\begin{equation}\label{eq:dm-discrete}
	\begin{split}
		\langle x \vert \sum_p \vert x_p \rangle \langle  x_p \vert x_{p_2}\rangle
		=& \langle x \vert \sum_p \vert x_p \rangle \delta_{p,p_{p_2}}\\
		=& N^{-1/2} \xi_{p_2}(x).
	\end{split}
\end{equation}
The right-hand side of the above equation consists of only one position
eigenfunction.

In the case of the position scaling-eigenfunction, Pdb3, Pdb4, Pdb5 and Psy4, the spreads
increase by 10 to 50 \% by replacing $\vert x_p\rangle \in \mathcal H_X$ with
$\vert x_{2}\rangle \in \mathcal H$.
\section{MLWF and PPE in k-Space}\label{sec:k-space}
Partial differential equations for MLWFs in k-space give insight into properties of MLWFs and their relatives.
The focus of the section is on a description of the degree of the closeness of MLWFs and PPEs, namely
Eq.~(\ref{eq:w=x_p}).

\subsection{Partial Differential Equations for MLWFs}
By expanding  MLWFs with the BFs, another solution procedure for MLWFs  is obtained:
\begin{equation}\label{eq:btow}
\begin{split}
\vert W \rangle = \sum_{k} \sum_{n} w_{k}^n \vert \psi_k^n \rangle,
\end{split}
\end{equation}
where the series number $s$ is dropped and the home unit cell index $l$ is assumed to be $0$.
After projecting the Eq.~(\ref{eq:btow_lambda}) onto the bra of a BF,
we have a set of partial differential equations for $w_k$ (see Appendix~\ref{app:supp} for the detail):
\begin{equation}\label{eq:wannier-k-eq}
\frac{\partial w_{k}^{n_{1}}}{\partial k}=\sum_{m,n_{2}}\frac{\partial\bar{c}_{m,k}^{n_{1}}}{\partial k}c_{m,k}^{n_{2}}w_{k}^{n_{2}}-ix_0 w_{k}^{n_{1}},\\
\end{equation}
where $x_0$ has to be chosen to have maximum localization.

When all energy bands are included in the calculation of MLWFs, the following holds:
\begin{equation}
	\mathcal H_C = \mathcal H_B=\sum_{n=0}^{N-1} \mathcal{H}_{B_n}\\ = \mathcal H_X.
\end{equation}
Therefore, position eigenvectors $\{\vert x_p\rangle\}$, MLWFs and PPEs are identical.
In fact, the position eigenvectors can conversely be expanded by the BFs:
 \begin{equation}\label{eq:x-bloch}
 \begin{split}
	 \vert x_{p}\rangle &=\sum_{k} \sum_{n=0}^{N-1} \vert \psi_k^n \rangle \langle \psi_k^n
	 \vert x_{p}\rangle \\
	 &=\frac{1}{\sqrt L }\sum_{k} \sum_{n=0}^{N-1} e^{-ikp\Delta x }\overline{c}_m^{n,k} \vert \psi_k^n \rangle,\\
\end{split}
\end{equation}
 and it is easily checked the following satisfies Eq.~(\ref{eq:wannier-k-eq}):
\begin{equation}
	\begin{split}
 	 &w_{k,p}= e^{-ik x_p }\overline{c}_m^{n,k},\\
   &x_0= p\Delta x=x_p.
 	 \end{split}
\end{equation}
 \subsection{MLWF and PPE in Single-Band System}
  \subsubsection{Phase of MLWF}
  When an MLWF's home unit cell is $[0:1)$,
Eq.~(\ref{eq:wannier-k-eq}) for single-band system is  simplified:
\begin{equation}\label{eq:wannier-k-eq-single}
	\begin{split}
		\frac{\partial w_{k}}{\partial k}
		&=\sum_{m}\frac{\partial\bar{c}_{m,k}}{\partial k}c_{m,k}w_{k}-ix_0 w_{k}.\\
	\end{split}
\end{equation}
By defining $r_{m,k}$ and $\theta_{m,k}$ in the following way:
\begin{equation}
	c_{m,k}=r_{m,k} e^{i \theta_{m,k}}, \;r_{m,k},  \in \mathbb{R}^{+},  \;  \theta_{m,k}  \in  \mathbb{R},
\end{equation}
the first term on the right-hand side of Eq.~(\ref{eq:wannier-k-eq-single})
 turns out to be the cell expectation value of the change in $\theta_{m,k}$:
 \begin{equation}\label{eq:av-theta}
 \begin{split}
\sum_{m=0}^{N-1}\frac{\partial\bar{c}_{m,k}}{\partial k}c_{m,k}
&=\sum_{m=0}^{N-1}\frac{\partial r_{m,k}e^{-i\theta_{m,k}}}{\partial k}r_{m,k}e^{i\theta_{m,k}}  \\
&=-i\sum_{m=0}^{N-1}\frac{\partial\theta_{m,k}}{\partial k}r_{m,k}^{2}. \\
\end{split}
 \end{equation}

By defining:
 \begin{equation}\label{eq:theta=bxw}
\bar{\theta}_k=\int^{k}dk_1 \langle \frac{\partial\theta_{k_1}}{\partial k_1}  \rangle_{cell}=
\int^{k}dk_1 \sum_{m=0}^{N-1}\frac{\partial\theta_{m,k_1}}{\partial k_1}r_{m,k_1}^{2},
\end{equation}
and choosing  $x_0$ to make the WF most localized, the MLWF is solved as follows:
\begin{equation}\label{eq:MLWF-single}
	\begin{split}
		\vert W_l \rangle &=
    \sum_k e^{-i\bar{\theta}_k}e^{-ikx_0} e^{-ikl}\vert \psi_k \rangle. \\
	\end{split}
\end{equation}
In the above equation, the term $e^{-i\bar{\theta}_k}$ works to reshape  the resulting WF sharper,
by making the expectation value of the phase zero at each wave number.
 \subsubsection{The Most Localized PPE and Its Error}\label{sec:ape-error}
A PPE in $\mathcal{H}_{B_1}$  is:
\begin{equation}\label{ape-psi-expantion}
\begin{split}
\sum_{k}\!\!\vert\psi_{k}\rangle\langle\psi_{k}\vert x_{p}\rangle  &\!=\!\frac{1}{\sqrt{L}}\sum e^{-ikl}e^{-ikm\Delta x}\bar{c}_{m,k}\vert\psi_{k}\rangle.\\
\end{split}
\end{equation}
By applying the procedure described in  Sec.~\ref{sec:Orthonormalization of PPEs }, we have orthonormalized PPE:
\begin{equation}\label{eq:bnko}
\begin{split}
 \vert  x^{\perp}_p \rangle
 &= \frac{1}{\sqrt{L}} \sum_k  e^{-ikl}e^{-ikm\Delta x}
 \frac{\bar{c}_{m,k}}{\vert c_{m,k}  \vert }
 \vert\psi_{k}\rangle \\
&= \frac{1}{\sqrt{L}} \sum_k  e^{-ikl}e^{-ikm\Delta x}
e^{-i\theta_{m,k}}
\vert\psi_{k}\rangle \\
&=\sum_k b_{p,k}  \vert\psi_{k}\rangle.
\end{split}
\end{equation}
Thus, the series $\{b_{p,k} \}$ satisfies the following equation:
\begin{equation}\label{eq:wannier-k-a}
\frac{\partial b_{p,k}}{\partial k} =-i\frac{\partial\theta_{m,k}}{\partial k}b_{p,k}-i(l+m\Delta x)b_{p,k}.
\end{equation}
$\bar{c}_{m,k}$ in Eq.~(\ref{eq:bnko})  offsets the phase of each BF $k$-by-$k$ at $x= p \Delta x =l+m\Delta x $,
so that all BFs are aligned at this position.

One of the possible first approximation of Eq.~(\ref{eq:theta=bxw}),
is to replace the expectation values of the phases with the phases at maximum point of  the
following \textit{probability density}:
\begin{equation}\label{eq:MLWF-max-and-ape_theta}
    \sum_{k}\left|c^n_{m,k}\right|^{2}
    =\sum_{k}\left|r_{m,k}\right|^{2}.
\end{equation}
And hence,
\begin{equation}\label{eq:theta-bar}
	\begin{split}
\int^{k}dk_1\sum_{m=0}^{N-1}\frac{\partial\theta_{m,k_1}}{\partial k_1}r_{m,k_1}^{2} &\simeq \sum_{m=0}^{N-1}\theta_{m,k}r_{m,k}^{2} \\
		&\sim \theta_{m_0,k},\\
	\end{split}
\end{equation}
where $m_0$ is the maximum point of Eq.~(\ref{eq:MLWF-max-and-ape_theta}).

For the reason, PPEs and MLWFs are close to each other when they are highly localized.
And the measure of the error $\delta_k$ is:
\begin{equation}\label{eq:single-band-error}
\delta_k = \vert \overline \theta_k - \theta_{m_0,k} \vert.
\end{equation}
The above analysis  does not  directly apply in the case of composite band systems, and still the analysis needs more development. In the paper,
the numerical results supporting the validity of the PPE
 will be shown later in  Sec.~\ref{sec:results}.
\section{Numerical Scheme}\label{sec:Numerical}
The section describes solution procedures of  the 1D Schr\"odinger and Maxwell's equations. Especially, the capability of the position scaling-eigenfunctions
to diagonalize the potential energy and the electric susceptibility
is explained and utilized.
\subsection{Diagonalization of Potential Energy and Electric Susceptibility}
Since the position scaling-eigenfunctions are chosen as the basis vectors, the
position operator $\tilde x \in \mathcal H_X$ is diagonal by definition.
And if $V(x)$ is piecewise quadratic over a few calculation cells,
as described in detail in Appendix~\ref{sec:potential-elem},
the following holds:
\begin{equation}\label{apeq:v-diagonal-0}
  \begin{split}
    \langle x_{p_1} \vert V(\hat x) \vert  x_{p_2} \rangle =
    V(x_{p_2}) \delta_{p_1, {p_2}}+O(\sigma^2_0/N^2),
  \end{split}
\end{equation}
where
\begin{equation}\label{key}
  \begin{array}{c}
   \sigma^2_0=\langle x_0 \vert \hat x^2 \vert x_0 \rangle,\\
  p_1, p_2 \in \mathbb Z.
  \end{array}
\end{equation}
Since $\sigma^2_0$ is of the order of $0.1$, the error above is estimated to be
$\sim 0.1/N$ and practically negligible. Thus, the potential energy and the
electric susceptibility  become diagonal with respect to the position
scaling-eigenfunctions.

When  $V(x)$ is given as a $\delta$-function such as:
\begin{equation}\label{key}
  \begin{split}
    V (x,a) &= V_0\delta (x-a),
  \end{split}
\end{equation}
the corresponding potential energy  operator is discretized to be
(see Appendix~\ref{sec:potential-elem} for detail):
\begin{equation}\label{ape: disc-potential-0}
  \begin{split}
    \hat V(a) = NV_0 \vert x_p \rangle \langle x_p \vert
    \quad (x_p = a, p= a/\Delta x).
  \end{split}
\end{equation}
Thus, the numerical expression of the matrix is diagonal:
\begin{equation}\label{ape: disc-potential-1}
  \begin{split}
    \langle x_{p_1} \vert\hat V(a)  \vert x_{p_2} \rangle =  V_0 N\delta_{p_1, p_2}.
  \end{split}
\end{equation}
	\subsection{Electronic System}\label{sec:num-electronic}
Owing to Eq.~(\ref{apeq:v-diagonal}), the operators of the discretized
Hamiltonian in $\mathcal{H}_{X}$ becomes:
\begin{equation}
\begin{split}
\hat {H} (\hat p^2, \hat x) \rightarrow \tilde {H} (\tilde p^2, \tilde x).
\end{split}
\end{equation}
By this, the potential energy matrix  becomes diagonal if the operator  is composed of $\hat x$ alone.
Thus, the Schr\"odinger equation becomes:
\begin{equation}\label{eq:Hamiltonian0}
\tilde H \vert\psi^n_{k}\rangle=
	E_k^n  \vert\psi^n_{k}\rangle.
\end{equation}
By multiplying the ket of a position eigenvector and  substituting Eq.~(\ref{eq:bloch-x}) into Eq.~(\ref{eq:Hamiltonian0}), we have:
\begin{equation}
\begin{split}
\sum_{m_{2}=0}^{N-1}\langle x_{m_{1}}\vert H\vert x_{m_{2}}\rangle e^{ik(m_{2}-m_{1})\Delta x}c^{n,k}_{m_2}=E_{k}^{n}c^{n,k}_{m_1},\\
\langle x_{m_{1}}\vert H\vert x_{m_{2}}\rangle=\frac{1}{2}\langle x_{m_1}\vert\hat{p}^{2}\vert x_{m_2}\rangle+v(x_{m_{1}})\delta_{m_{1,}m_{2}}.
\end{split}
\end{equation}
With the kinetic energy matrix elements, $\{Q_r\}$, listed on Table \ref{tab:Q2n},
the final form of the eigenvalue equation to determine the series $\{c_{m}^{n,k}\}$ is obtained as follows:
\begin{equation}\label{eq:matrix_hamiltonian}
\begin{split}
\sum_{m_{2}=0}^{N-1} \left[  Q_{m_{1}-m_{2}}  + \delta_{m_{1},m_{2}} V(m_{2}\Delta x) \right]&
e^{ik(m_{2}-m_{1})\Delta x} c_{m_{2}}^{n,k} \\
=&E_{k}^{n}c_{m_{1}}^{n,k}.
\end{split}
\end{equation}
%
%It is noted that (1) the total number of energy bands, (2) the total number of
%position eigenvectors (orbits) in one unit cell and (3) the division in one
%unit cell  are the same in this scheme.
\subsection{Optical System}\label{sec:num-photonic}
The equations for the electric field $f(x)$ and magnetic field $h(x)$ satisfy the following equations\cite{Gupta2022}:
\begin{equation}\label{key}
	\begin{split}-\frac{\partial^{2}}{\partial x^{2}}f(x) & =\left(\frac{\omega}{c}\right)^{2}\epsilon(x)f(x),\\
		h(x) & =-\frac{c}{\omega}\frac{\partial}{\partial x}f(x),
	\end{split}
\end{equation}
where $\epsilon(x) \ge 0$ is the dielectric susceptibility function.

By using the same strategy described in Sec.~\ref{sec:num-electronic}, the discretized equation for
the electric field  becomes:
\begin{equation}\label{eq:matrix_maxwell}
\begin{split}
\sum_{m_{2}=0}^{N-1}Q_{m_{1}-m_{2}}& e^{ik(m_{2}-m_{1})\Delta x}c_{m_{2}}^{n,k}\\
&=\omega^{2}_k \epsilon(m_{1}\Delta x)c_{m_{1}}^{n,k}.
\end{split}
\end{equation}
The set of eigenvalue equations are solved with a math library such as  the GSL
\cite{gsl}.
\subsection{Basis Vectors and BFs}\label{sec:basis}
The solutions of Eqs.~(\ref{eq:matrix_hamiltonian}) and
(\ref{eq:matrix_maxwell}), $\{c_{k,m}^n\}$ are elements of a matrix $\boldsymbol
c_k $ transforming the position scaling-eigenvectors $\{\vert x_m \rangle \}$ to
the BFs $\{\vert \psi^k_n \rangle \}$ and the matrix is inherently square with
respect to $m$ (position) and $n$ (energy band). It is noted even if we replace
the position scaling-eigenvectors with other  basis set, the number of the
independent basis vectors in a unit cell and the number of the computable bands
are identical.
\section{Results and Discussions}\label{sec:results}
The section compares
orthonormalized PPEs defined by Eq.~(\ref{eq:orthonormal-ape}), orthonormalized
PWEs defined by Eq.~(\ref{eq:orthonormal-pwe}) and MLWFs  obtained from
Eq.~(\ref{eqap:btow_w_U}) with preceding results.

As the position scaling-eigenfunction, Psy4 is used. Replacing Psy4 with Pdb3,
Pdb4 and  Pdb5  results in no discernible difference except minor numerical
deviation.

Throughout the section, the units of the vertical and horizontal axis are
adjusted to the figures in the references, for convenience in comparison, and hence units
differ from figure to figure.
 \subsection{Numerical Verification of Energy Eigenvalues}
 To verify the obtained position eigenvectors and the numerical scheme developed
in the paper, the electronic and optical energy eigenvalues calculated by Eqs.
(\ref{eq:matrix_hamiltonian}) and (\ref{eq:matrix_maxwell}) are compared with
the preceding results. The agreements are quite good, and hence MLWF, PPE and
PWE produced by the scheme deserve to be discussed. \subsubsection{Electronic
    System} 
\begin{figure}[h]
	\centering
	\includegraphics[width=1.0\linewidth]{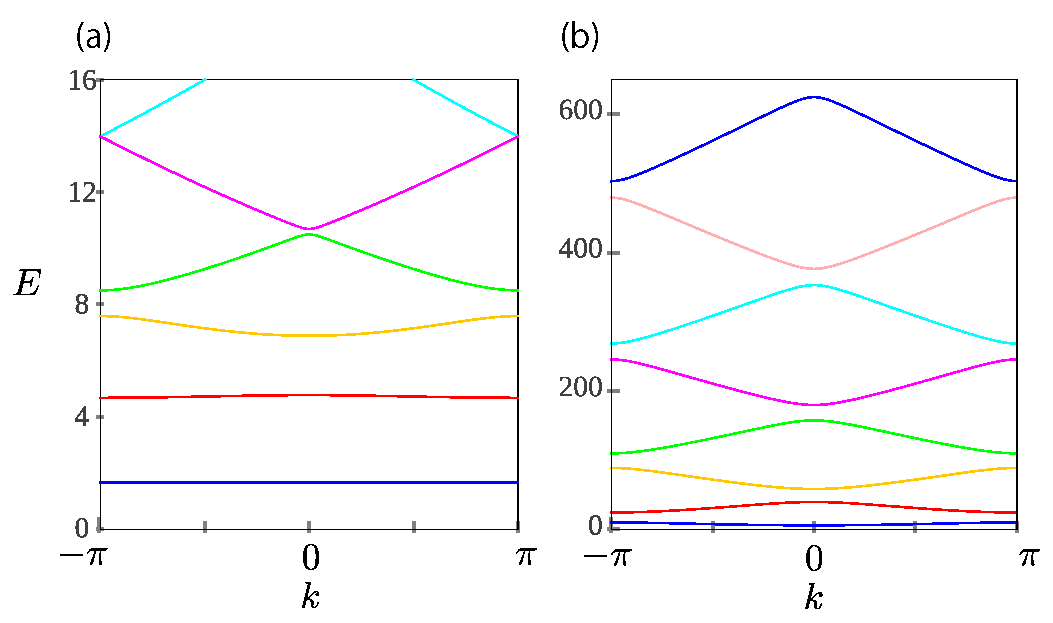}
		\caption{Calculated electronic band structures, where $N =
 40$, $L=40$. (a) $V(x)=V_0 \left[1-\cos({2\pi x}/{a})\right]$~\cite{Grosso2000} (b) $V(x)=\sum_n V_0\delta(x
 -na)$~\cite{johnston2019}. 		%	\label{fig:electronic-bands}
	}%
	\label{fig:electronic-bands}
\end{figure}

Figure~\ref{fig:electronic-bands} shows two energy band diagrams
produced from the method described in Sec.~\ref{sec:num-electronic}. The
calculation conditions are listed in Table \ref{tab:Electronic-Conditions}.
Both show good agreements with the preceding results.
\begin{table}[h]
	\caption{\label{tab:Electronic-Conditions}  Units and potential energy employed in the calculation
    of energy band diagrams
    shown in Figs.~\ref{fig:electronic-bands}(a) and (b).
  }
	\centering
	\begin{tabular}{l c  c c}
		\hline
    \hline
		  &&  Fig.~\ref{fig:electronic-bands}(a) \footnotemark[1]  & Fig.~\ref{fig:electronic-bands}(b) \footnotemark[2]
		\\ \hline
    1. Units  	  &&	    &  \\
		\hspace{3mm}a. Energy 	&&	 $\hbar^2/(2ma_B^2)$ 	     &  $\hbar^2/(2ma^2)$  \\
		\hspace{3mm}b. Length 	&&	$a=5 a_B$ 		   &  $a=$Unit Cell Size  							\\
    2. Potential energy  	  &&	    &  \\
		\hspace{3mm}a. $V(x)$    	&& 	 $V_0 \left[ 1-\cos\frac{2\pi x}{a} \right]$	  &  $\sum_n V_0\delta(x -na)$  \\
		\hspace{3mm}b. $V_0$ 	    && 	 $10(\hbar^2/2m)(\pi/a)^2$	   &  $6\hbar^2/(2ma)$  \\
    \hline \hline
	\end{tabular}
\footnotetext[1]{Reference\cite{Grosso2000}.}
\footnotetext[2]{Reference\cite{johnston2019}.}
\end{table}
\begin{figure}[h]
	\centering
	\includegraphics[width=0.8\linewidth]{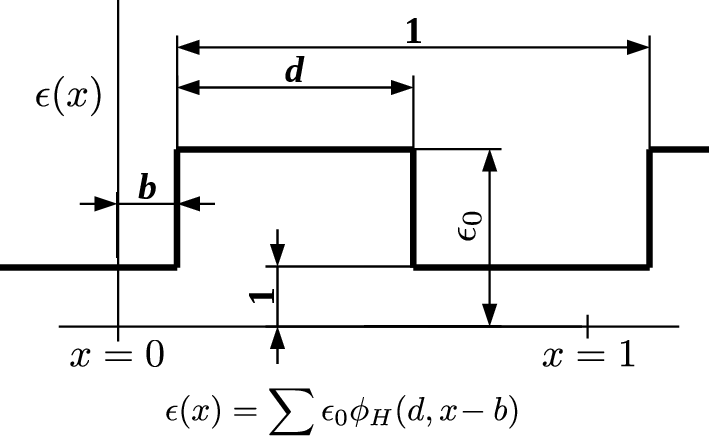}
	\caption{Parameters of dielectric susceptibility  $\epsilon(x)$.}
	\label{fig:epsilone}
\end{figure}
\subsubsection{Photonic Systems}
\begin{figure}[h]
	\centering
	\includegraphics[width=1.0\linewidth]{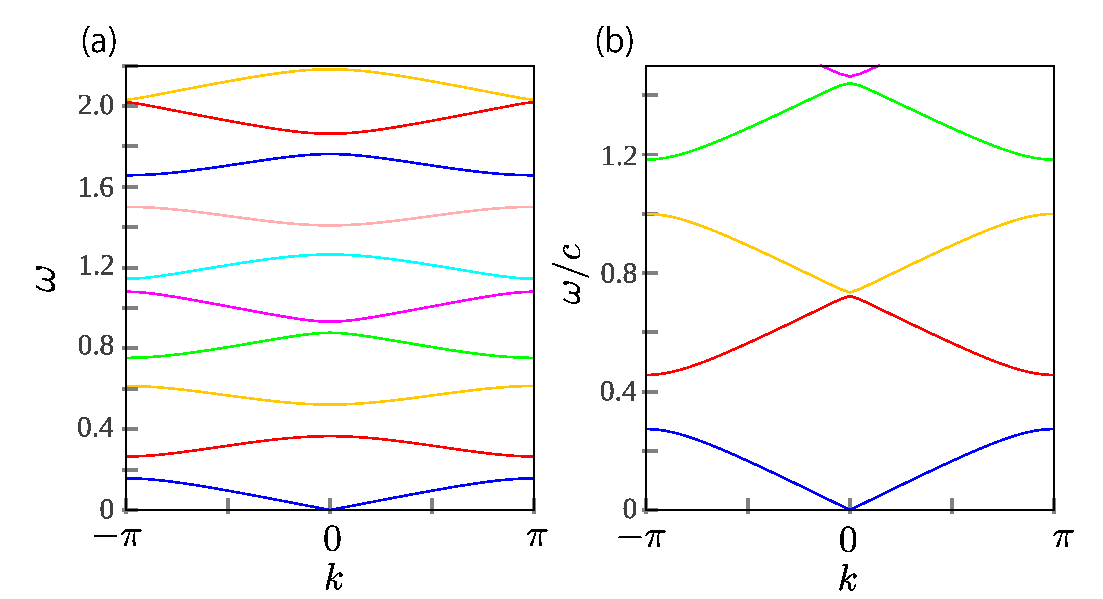}
	\caption{\label{fig: band exapmples}%
		Calculated photonic band structures, where $N = 40$, $L=40$.
(a) $\epsilon(x)=\sum 12\phi_H(1/2, x-n)$~\cite{Busch2011}. (b)
 $\epsilon(x)=\sum 5\phi_H(0.3, x-0.35-n)$~\cite{Gupta2022}.}
		\label{fig:dispersion }
	\end{figure}

The two dispersion diagrams shown in Fig.~\ref{fig:dispersion } are calculated
from the method described in Sec.~\ref{sec:num-photonic}. The dielectric
susceptibility  $\epsilon (x)$, in both cases, is a sort of step functions that
is  shown in the Fig.~\ref{fig:epsilone}.  The calculation conditions and
parameters are shown in Table \ref{tab:Photonic-Conditions}. Both figures well
agree with the preceding results.
\begin{table}[h]
	\caption{\label{tab:Photonic-Conditions}
  Units and parameters for electric susceptibility $\epsilon(x)$ employed in the calculation of the dispersion diagrams
 and Wannier functions shown in Figs.~\ref{fig:dispersion },  \ref{fig:gupta_3_7_E}, \ref{fig:bush-E-1} and \ref{fig:bush-H-1}.
   The parameters $\epsilon_0$, $b$ and $d$ are defined in the Fig.~(\ref{fig:epsilone}).
}
	\centering
	\begin{tabular}{l c c c c}
		\hline
		  &\ \ \ &
    Figs.~\ref{fig:dispersion }(a), \ref{fig:bush-E-1} and \ref{fig:bush-H-1}\footnotemark[1]
 &&
 Figs.~\ref{fig:dispersion }(b) and  \ref{fig:gupta_3_7_E}\footnotemark[2]
		\\ \hline  \hline
    1. Units && & \\
		\hspace{3mm}a. Frequency 	  	&&	 $2\pi c  $ 	&&  $c$  \\
		\hspace{3mm}b. Length 	 &&	$a=$Unit Cell Size 		&&  $a=$Unit Cell Size \\
    2. Parameters  && &  \\
		\hspace{3mm}a. $\epsilon_0$	   && 	 12 && 5  \\
		\hspace{3mm}b. $b$ 	&& 	 $0$	&&  $0.35$ \\
		\hspace{3mm}c. $d$ 	 && 	 $0.5$	&&  $0.3$  \\
    \hline \hline
	\end{tabular}
\footnotetext[1]{Reference\cite{Busch2011}.}
\footnotetext[2]{Reference\cite{Gupta2022}.}
\end{table}
\subsection{WFs in Single Band Systems }
The MLWFs, PPEs and PWEs pertaining to $n_b$-th energy band are qualitatively and quantitatively
compared with preceding results.
\subsubsection{Electronic Systems }
\begin{figure}[h]
  \centering
  \includegraphics[width=1.0\linewidth]{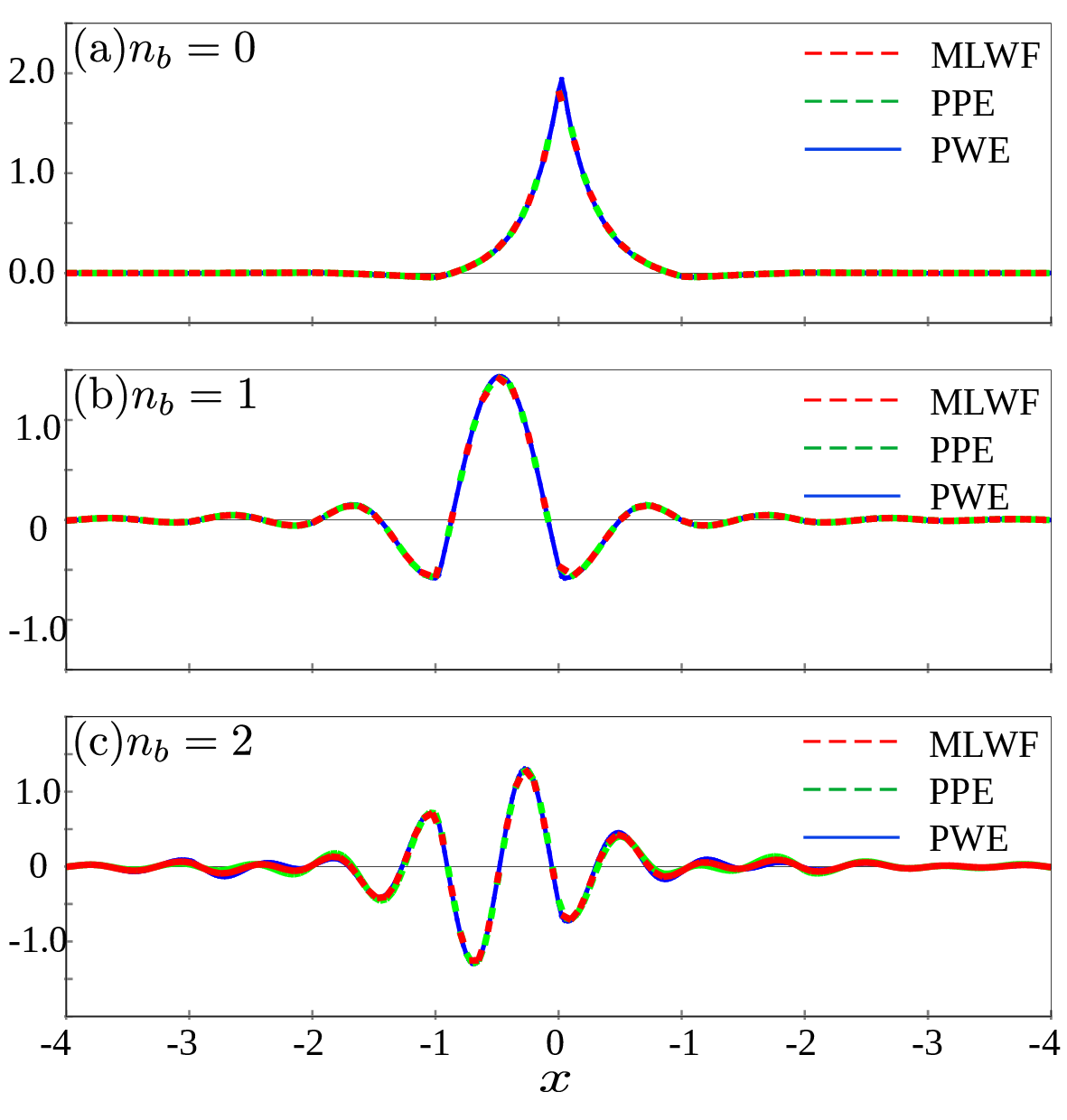}
	\caption{Profiles of MLWF, PPE and PWE. $n_b$ at the left upper corner of each panel denotes
    the band index. In all panels, $N=40$, $L=200$ and $V(x)=\sum V_0 \delta(x-n)$ with $V_0=-0.2(2\pi^2)$~\cite{VELLASCO2020}.}
\label{fig:vellasco}
\end{figure}

The potential energy employed in the cases of Fig.~\ref{fig:vellasco}, is of Kr\"onig-Penny type as noted in the captions and the MLWF, PPE and PWE well agree with the preceding results\cite{VELLASCO2020}
in each energy band.

\begin{figure}[h]
\centering
\includegraphics[width=1.0\linewidth]{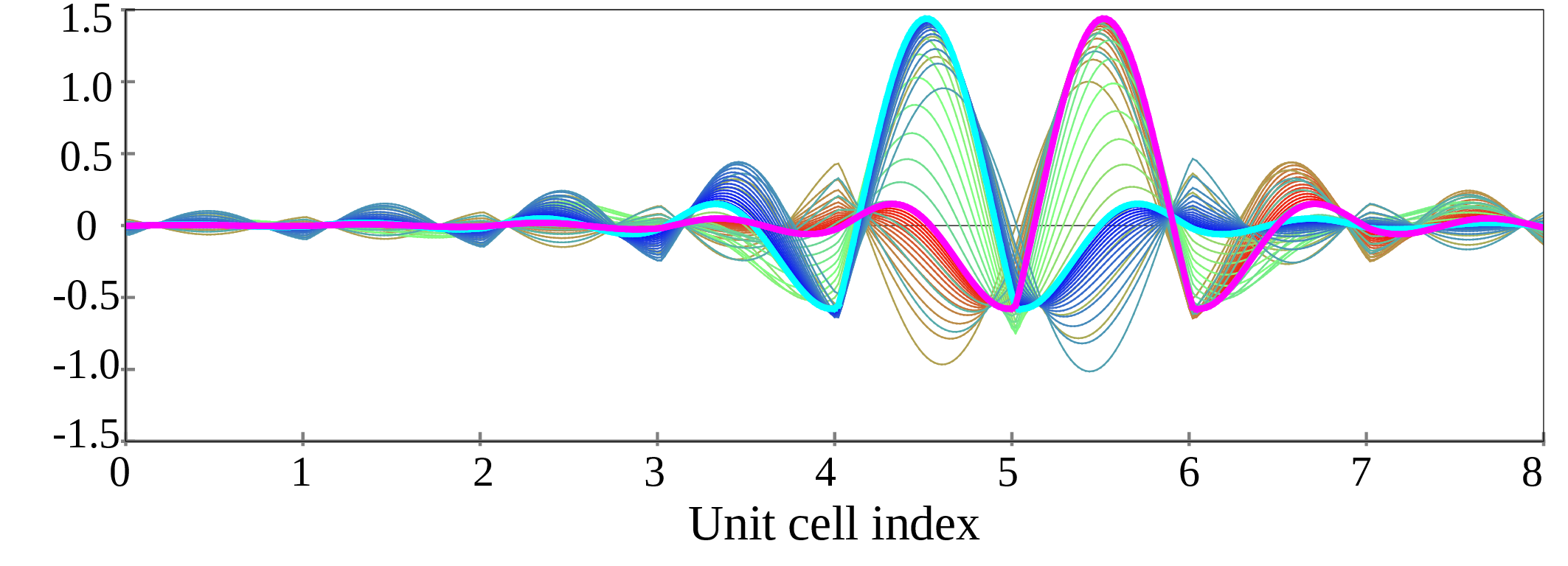}
\caption{Profiles of MLWF(thick lines colored (red or blue) and
 shifting PPE (thin lines colored gradually from red to blue),
$V(x)=\sum V_0 \delta(x-n)\ (V_0=-0.2(2\pi^2))$, $N=40$, $L=200$.
The band index $n_b=1$.}
\label{fig:vellasco-shift}
\end{figure}

Fig.~\ref{fig:vellasco-shift} is a graphical representation of
Eq.~(\ref{eq:ape-max-is-MLWF-max}) and  Eq.~(\ref{eq:MLWF-max-and-ape}), and it
shows shifting PPEs under the same condition employed in
Fig.~\ref{fig:vellasco}(b). The MLWFs and PPEs are shown in thick and thin
lines, respectively, and the color of PPEs ($\xi^{\perp C}_{p}(x)= \langle x
\vert x^{\perp C}_p \rangle $) changes from blue to red as the parameter $x_p$
varies from 0 to 1.

Table \ref{tab:Vellasco}  compares the localization indicator $\sigma_x^2$ calculated with each method.
The PPEs and PWEs
show as good localization as the known results.
When the cost of calculation is taken into account, the methods using PPE and PWE are attractive
alternatives.
\begin{table}[h]
  \caption{Comparison of spreads, $\sigma_x^2 =\langle W\vert  \{\hat{x} - \langle W\vert \hat{x} \vert W\rangle \}^2\vert W \rangle $, obtained from  different methods.}
  \label{tab:Vellasco}
  \centering
  \begin{tabular}{c c c c c c c}
    \hline
   \hline
   \multicolumn{2}{c}{Calculation condition} && \multicolumn{4}{c}{$\sigma_x^2$}\\
   \cline {1-2} \cline{4-7}
    $V_0$ & Band &&  Ref.~\cite{VELLASCO2020}   & MLWF   & PPE   & PWE \\ \hline
    $-0.2\times 2\pi^2$ &0 && 0.03  & 0.03 & 0.03 &  0.03 \\
    $-0.2\times 2\pi^2$ &1 && 0.12  & 0.12 & 0.12 &  0.12  \\
    $-0.2\times 2\pi^2$ &2 && 0.24  & 0.24 & 0.24 &  0.26   \\
    \hline \hline
  \end{tabular}
\end{table}
\subsubsection{Photonic Systems }
\begin{figure}[h]
	\centering
	\includegraphics[width=1.0\linewidth]{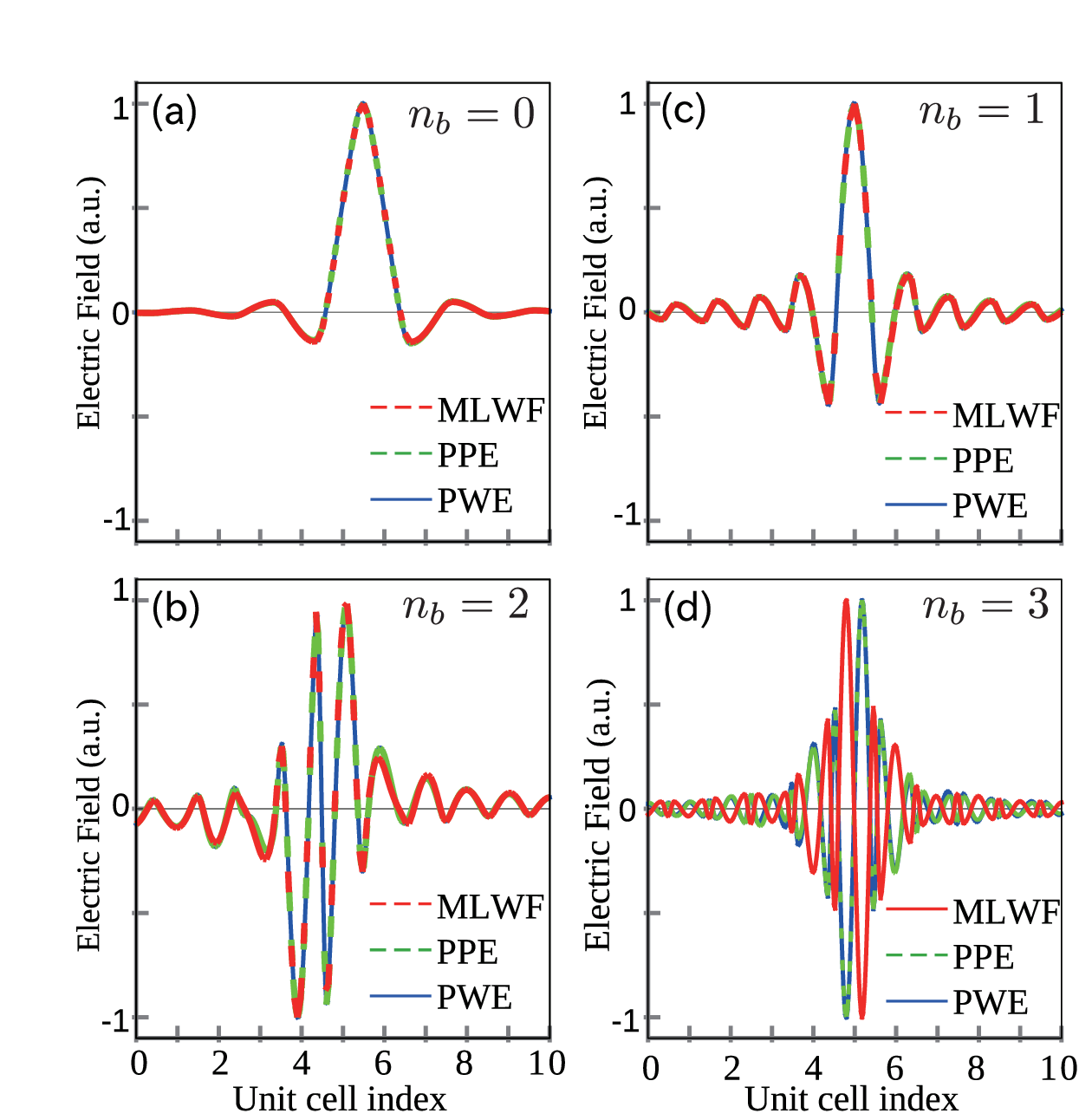}
	\caption{Profiles of MLWF, PPE and PWE. $n_b$ at the left upper corner of each panel denotes
    the band index. In all panels,  $N=40$, $L=40$ and  $\epsilon(x)=\sum 12\phi_H(0.3,x\!-0.35-\!n)$~\cite{Gupta2022}.}
	\label{fig:gupta_3_7_E}
\end{figure}
\begin{figure}[h]
	\centering
	\includegraphics[width=1.0\linewidth]{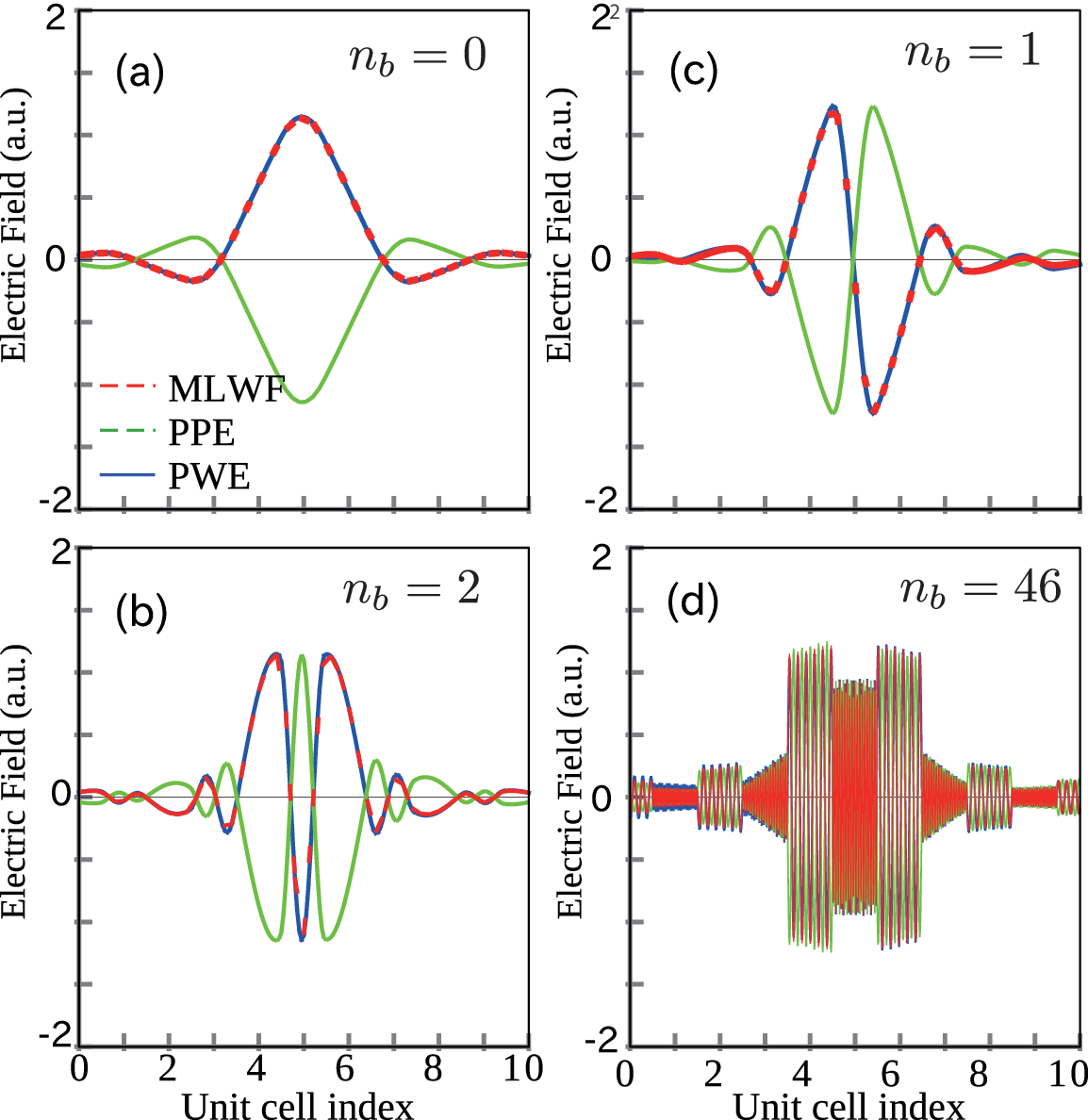}
	\caption{
    Profiles of electric field  for MLWF, PPE and PWE. $n_b$ at the left upper corner of each panel denotes
    the band index.  In (a)-(c) $N=40,\ L=40$. In (d) $N=192,\ L=96$. In all panels, $\epsilon(x)=\sum 5\phi_H(1/2, x\!-\!na),\ a$: cell length\cite{Busch2011}.}
	\label{fig:bush-E-1}
\end{figure}
\begin{figure}[h]
	\centering
	\includegraphics[width=1.0\linewidth]{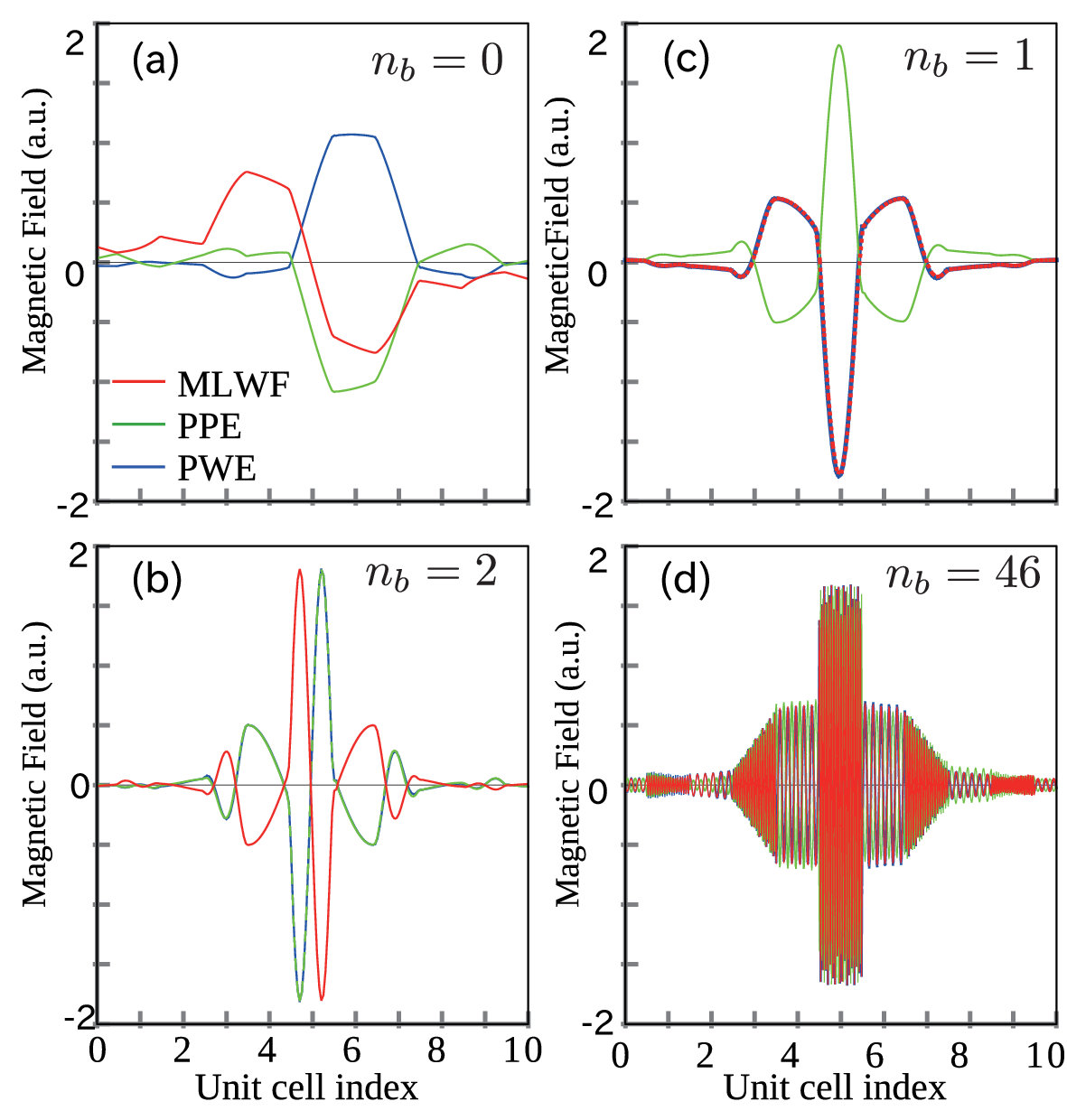}
	\caption{Profiles of magnetic field  for MLWF, PPE and PWE.  $n_b$ at the left upper corner of each panel denotes
    the band index.  In (a)-(c) $N=40,\ L=40$. In (d) $N=192,\ L=96$. In all panels, $\epsilon(x)=\sum 5\phi_H(1/2, x\!-\!na),\ a$: cell length\cite{Busch2011}. }
	\label{fig:bush-H-1}
\end{figure}
This section focuses on comparing MLWFs, PPEs and PWEs  with the preceding results
done for photonic crystals. The dielectric  susceptibility functions used in
Figs.~\ref{fig:gupta_3_7_E} to \ref{fig:bush-H-1}, are listed in Table
\ref{tab:Photonic-Conditions}, and all MLWFs,  PPEs and PWEs
% in Figs.~\ref{fig:gupta_3_7_E} to \ref{fig:bush-H-1}
  show good agreements, in shape and
position, with the preceding results\cite{Busch2011,Gupta2022}. The disagreement
with MLWF shown in Fig.~\ref{fig:bush-H-1}(a)  will be discussed later in
Sec.~\ref{sec:note}.

Even in  the 46th band of Figs.~\ref{fig:bush-E-1} and \ref{fig:bush-H-1} with
very high wave number, all WFs calculated in this paper show  good agreements
with the preceding results calculated with $(N,\ L)=(512,\ 99)$.
% and $L$  set at 512 and 99,respectively.
 As is noted in Secs.~\ref{sec:Closest PPE to MLWF}, $\sigma^2_x$ is actually
calculated to find out the maximally localized PPE and PWE in Figs.~\ref{fig:bush-E-1}(d) and \ref{fig:bush-H-1}(d).

\subsubsection{Note on 0th band MLWF in Fig.~\ref{fig:bush-H-1}(a) }\label{sec:note}
The MLWF drawn in red in Fig.~\ref{fig:bush-H-1}(a) disagrees with the PPE, PWE
and the result shown in Ref.~\cite{Busch2011}.
The WF drawn in red in the present calculation is actually the  eigenfunction of
the position operator but not the maximally localized WF. As described in
Ref.~\cite{Romano2010}, the magnitude of
the WFs do not decrease exponentially, thereby inappropriately influenced by the
boundary condition through the eigenvalue equation, Eq.~(\ref{eqap:btow_w_U}).

On the other hand, PPEs and PWEs  show good agreements with
results seen in Refs.~\cite{Romano2010,Busch2011}, since they are calculated locally with
little influence from  the boundary.
\subsection{WFs in Composite Band Systems}\label{sec:composite-band}
The section focuses on an electronic composite band
 system and the comparisons of MLWFs, PPEs and PWEs with the results
shown in Ref.~\cite{Wang2014}.
\begin{figure}[h]
	\centering
    	\includegraphics[width=1.0\linewidth]{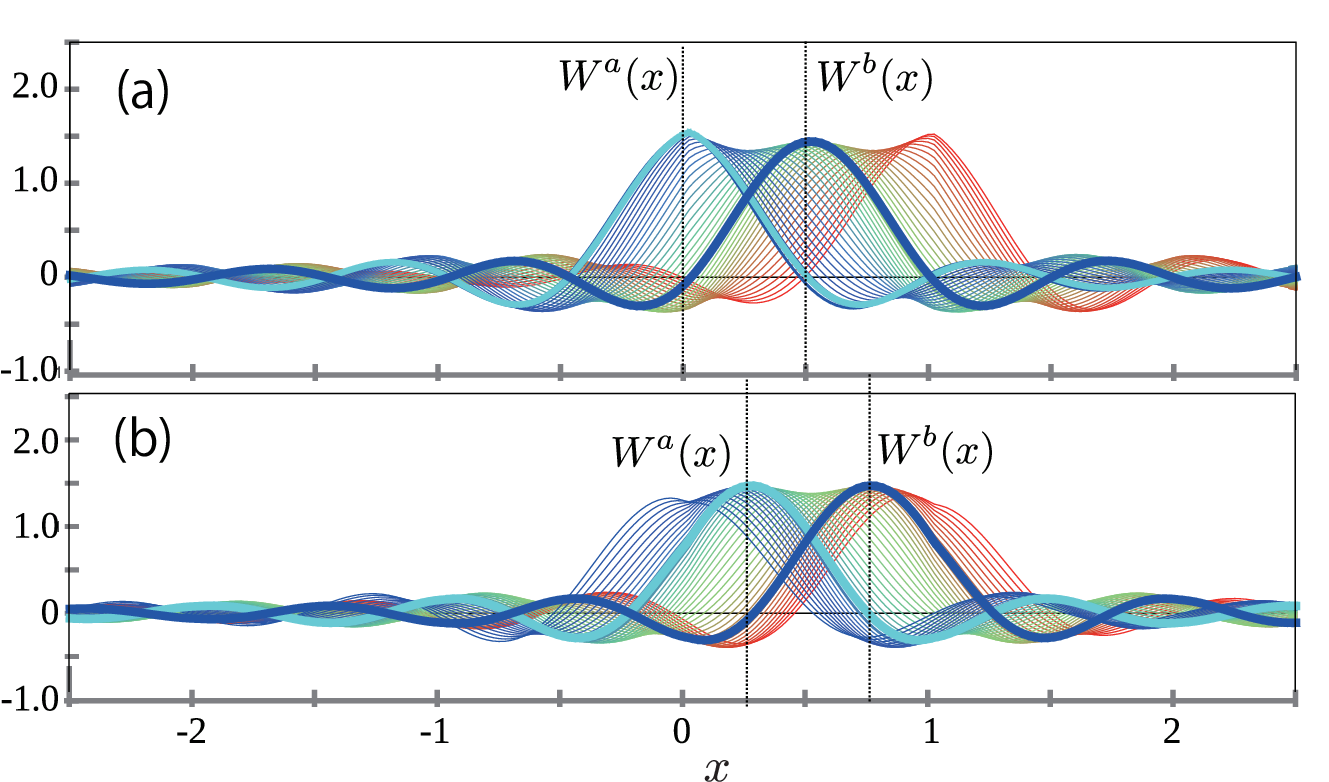}
	\caption{Searching maximally localized position with PPE in two-band systems consisting of $0$th and
    first energy bands.
   The superscript $a$ and $b$ on  $W(x)$ denote the series.
   In (a) and (b) $N=40, L=100$ and  $V(x)=\sum V_0 \delta(x -n)$,
		in (a) with  $V_0=-0.661$ and in (b) with $V_0=+0.661$~\cite{Wang2014}.
It shows the maximum point of one series coincides with the zero of the other.
 }\label{fig:wang_shifting-2}
\end{figure}
\begin{figure}[h]
    	\includegraphics[width=1.0\linewidth]{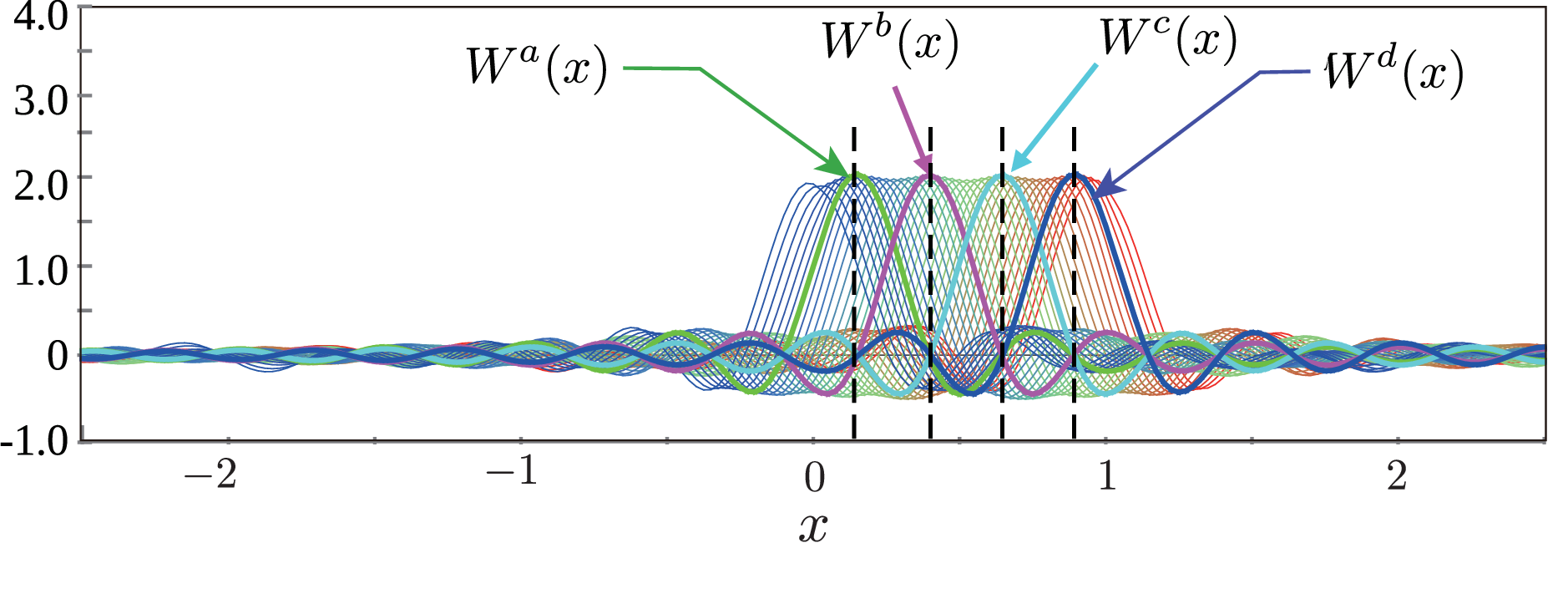}
	\caption{Searching maximally localized position with PPE in four-band system consisting of
    $0$th to third energy bands.
    The superscript $a$ to $d$ on  $W(x)$ denote the series.
    In the panel, $N=40, L=100$ and $V(x)=\sum V_0 \delta(x -n)$, with $V_0=0.661$~\cite{Wang2014}.
  It shows the maximum point of one series coincides with the zero of others. }
	\label{fig:wang_4}
\end{figure}
\begin{table}[h]
    \caption{Comparison of spreads, $\sigma_x^2 =\langle W\vert  \{\hat{x} - \langle W\vert \hat{x} \vert W\rangle \}^2\vert W \rangle $, obtained from  different methods.}
	%\label{tab:wang_sqrt}
	\label{tab:wang}
	\centering
  \begin{tabular}{c c c c c c c}
  \hline
  \hline
  \multicolumn{2}{c}{Calculation condition} && \multicolumn{4}{c}{$\sigma_x^2$}\\
  \cline {1-2} \cline{4-7}
  $V_0$ & Bands &&  Ref.~\cite{Wang2014}   & MLWF   & PPE   & PWE \\ \hline
		0.661 & 0 to 3 && 0.29    & 0.29   &  0.46 & 0.46  \\
		-0.661  & 0 to 1 && 0.27   & 0.27   & 0.37  & 0.37   \\
    \hline \hline
	\end{tabular}
\end{table}

\subsubsection{Comparison of the Results } By using
Eq.~(\ref{eq:orthonormal-ape}), Fig.~\ref{fig:wang_shifting-2}(a) shows how
shifting  PPEs coincide with  MLWFs in the same way Fig.~\ref{fig:vellasco-shift}
are drawn. In addition to 	\textit{the maximally localized } WF near the origin
of the Fig.~\ref{fig:wang_shifting-2}(a), \textit{the second  maximally
  localized}  WF is observed near $x_p=0.5$, which is not seen in
Ref.~\cite{Wang2014}. This existence of two MLWFs in one unit cell is  simply a
reflection of the completeness of the MLWFs in $\mathcal{H}_{B_{12}}=\mathcal
H_{B_1}+\mathcal H_{B_2}$; it takes two eigenfunctions per unit cell to
span$\mathcal{H}_{B_{12}}$.

By setting  $V_0=+0.661$, Fig.~\ref{fig:wang_shifting-2}(b) is drawn in the same
manner (which does not appear in Ref.~\cite{Wang2014}). The figure also shows
two peaks on the envelope in a unit cell and the MLWFs and PPEs coincide.

By defining four-band PPEs in the same manner, Fig.~\ref{fig:wang_4} demonstrates the
PPEs almost coincide with \textit{four} four-band MLWFs, which are also shown in
Ref.~\cite{Wang2014}.  
Table \ref{tab:wang} compares the localization measure
obtained by MLWF, PPE, PWE with the preceding results. The PPE and PWE again
show decent agreement with the others.

Most importantly, by reflecting Eq.~(\ref{eq:ape-MLWF-orthogonal}) and the
accuracy of PPEs as an alternative to MLWF, at the maximum position of an
MLWF and PPE, the other MLWFs and PPEs have their zeroes.
\section{Summary}\label{sec:conclusion}
\subsection{Position Eigenfunctions}
Position scaling-eigenfunctions with practically compact supports and relatively
good interpolation capability are generated from compactly supported
orthonormal SFs.

The matrix elements of the operators, especially those of the kinetic energy
operator, are obtained without performing differentiation. This relaxes the
condition on differentiability of the original compact support SFs.
 \subsection{MLWF, PPE and PWE}
When the  MLWFs are composed of all BFs of all energy bands, the  MLWFs and the
position scaling-eigenfunctions
are identical.

In  single band cases, the difference between PPEs and MLWFs is the choice of
\textit{anti-phases} to cancel the phases of the BFs. In case the error
estimated by Eq.~(\ref{eq:single-band-error}) is small enough, it is sufficient to
use PPEs as an alternative to MLWFs.

The relationship between the zeros and maxima of MLWFs and the corresponding
PPEs  are explained in relation to the inter- and intra-series orthogonality of
MLWFs.

One of the advantages of using PPE is that it  does not take additional
localized orbits/bases other than the position scaling-eigenfunctions.
Especially, in cases the centers of the MLWFs do not coincide with the centers
of the atoms\cite{freimuth2008publisher} and photonic crystal cases, in which
there is no inherently localized orbits\cite{PhysRevB.74.195116}, the PPEs seem
to be particularly  useful.

The other advantage followed from the investigation of the PPE and PWE is the
variable to be maximized has a simple form:
\begin{equation}\label{eq:computationl-cost-in-summary}
    \begin{split}
        &F(m)=\sum_{k}\left|c_{m,k}\right|^2 %\left|\psi_{k}(x_{p})\right|^2
        \ \left(m = p\ \mathrm{mod}\ N\right).
        \\
%        &(m = p\ \mathrm{mod}\ N, x_p = p\Delta x).
    \end{split}
\end{equation}
And hence it is expected to be computationally inexpensive.
\subsection{Numerical Scheme}
The results well agree with the preceding results, in some cases, with less
calculation cells in a unit cell.

The number of the bases, i.e., the position scaling-eigenfunctions in one unit
cell and the spatial resolution are equivalent, and hence all spatial
information on the system are readily stored in $\{c^n_{k,m}\}$
position-by-position.

The particular advantage of the numerical scheme developed is to diagonalize the
potential energy, when it is composed only of the position operator. Because of
this, once the discretization of the Hamiltonian is done, no actual integration
is needed. In addition, the kinetic energy part is band diagonal with constant
matrix elements and hence the eigenvalue and vectors are easy to obtain.
\subsection{Future Direction} On the mathematical side, the mathematical
mechanism of  giving  compactly supported SFs  a higher degree of
differentiability than the H\"older exponent  through Eq.~(\ref{eq:Pnm}) will
have to be rigorously investigated. A method to estimate the error of  PPEs
with respect to  the MLWFs  in composite band cases has also to be developed to the
degree performed in single band cases.

One of the most important steps to take is to extend the entire scheme to 2 and
3D. For the purpose, the multidimensional scaling functions may have to be
constructed in the way described in Ref.~\cite{Cohen1993}, so that the symmetry
of the system is inherent to the basis set. If the proposed scheme is not
seriously restricted by the extension,  it allows actually to calculate the
measurable physical properties of materials, such as electric polarization,
chemical bonding known to be closely related to the configuration of MLWFs
\cite{PhysRevB.48.4442,RevModPhys.66.899,PhysRevLett.97.107602,
    PhysRevB.47.1651,PhysRevB.48.4442, PhysRevB.92.165134,RevModPhys.84.1419,
    PhysRevB.95.075114, PhysRevB.89.155114,PhysRevB.47.1651,PhysRevB.48.4442,
    PhysRev.129.554,PhysRevB.89.155114, PhysRevB.92.165134, PhysRevB.95.075114,
    RevModPhys.84.1419, first-guess, PhysRevB.64.245108}. Concrete estimates of the
efficacy of the method, such as the accuracy, speed and portability, in relation
to the  existing codes can be performed at this future stage.

Since the obtained position eigenfunctions are also scaling functions and
satisfy the two-scale relation, the computational speed of the numerical
procedures such as the maximum point searching,  solving the eigenvalue problems
associated with  Schr\"odinger equations, can be improved by combining the
multiscale nature, i.e., iterative methods and mesh refinement, in the future
step.
\begin{acknowledgments}
K.W. acknowledges the financial support by JSPS KAKENHI
(Grant No. 22H05473, 21H01019 and JP18H01154), and JST CREST (Grant No. JPMJCR19T1).
\end{acknowledgments}

\appendix
\section{Matrix Elements of  Operators  }\label{app:ops} \subsection{General
	Strategy }\label{app:x_op} If an operator $\hat A$ satisfies the following
dilation relation:
\begin{equation}
(x \rightarrow 2x) \longmapsto  (\hat A \rightarrow 2^n \hat A).
\end{equation}
The two-scale relation becomes linear equations for the matrix elements
of the operator:
\begin{equation}\label{eq:X_two_scale}
\begin{split}A_{m,n} & =\langle m\vert\hat{A}\vert n\rangle\\
	& =\int\phi(x-m)A(x,\frac{1}{i}\frac{\partial}{\partial x})\phi(x-n)dx\\
	& =2\sum_{l,k}h_{l}h_{k}\int\phi(2x-2m-l)\\
	& \quad\quad A(x,\frac{1}{i}\frac{\partial}{\partial x})\phi(2x-2n-k))dx\\
	& =2\sum_{l,k}h_{l}h_{k}\int\phi(y-2m-l)\\
	& \quad\quad A(2^{-1}y,\frac{1}{i}\frac{\partial}{\partial\left(2^{-1}y\right)})\phi(y-2n-k)d\left(\frac{y}{2}\right)\\
	& =2\sum_{l,k}h_{l}h_{k}\int\phi(y-2m-l)\\
	& \quad\quad2^{-n}A(y,\frac{1}{i}\frac{\partial}{\partial y})\phi(y-2n-k)d\left(\frac{y}{2}\right)\\
	& =2^{-n}\sum_{l,k}h_{l}h_{k}A_{2m+l,2n+k}.
\end{split}
\end{equation}
It only takes linear algebraic calculations to obtain the
matrix elements, since the final form of the equations involves neither integration
nor differentiation.
\subsection{Position Operator $\tilde x$ }\label{app:p_op}
\subsubsection{Matrix Elements of $\tilde x$}\label{sec:Xelem}
Since the following holds:
\begin{equation}
(x \rightarrow 2x) \longmapsto (\hat x^n
\rightarrow 2^{n} \hat x^n ),
\end{equation}
Eq.~(\ref{eq:two-scale-integration-1}) is a special case of Eq.~(\ref{eq:X_two_scale}).
The solution, $X_r=\langle r \vert \hat{x} \vert 0 \rangle$, is
listed on Table \ref{tab:Xn}. From the table,  $X_{m,n}$ is calculated
as follows:
\begin{equation}\label{key}
	X_{m,n}= m + X_{r} \quad(r = m-n).
\end{equation}
\begin{table}[h]
	\caption{\label{tab:Xn}  Matrix elements of position operator $X_r = \langle r \vert \hat{x} \vert 0 \rangle $
    obtained with different SFs. }
	\centering
	\begin{tabular}{l c c c c c c }
		\hline
    \hline
		 && $X_{0}$ &  $X_{1}$  &  $X_2$  &  $X_{3}$  &  $X_{4}$\footnotemark[1]  \\ \hline \hline
   	1. SF 	&& &  &  &  &  \\
		\hspace{3mm}a. Sy4  &&	4.126\footnotemark[2]  &  -0.023  & -0.033 & 0.000 & 0.000  \\
		\hspace{3mm}b. Db2 	&&  0.771 & -0.071  & -0.002 &  0.000 &  0.000  \\
		\hspace{3mm}c. Db3  && 1.022 & -0.121 & 0.019 & -0.001  & 0.000 \\
		\hspace{3mm}d. Db4 	&&  1.266 & -0.164 & 0.039 & -0.005 & 0.000 \\
    \hline \hline
	\end{tabular}
	\footnotetext[1]{$X_{-n}=X_{n}$. }
	\footnotetext[2]{All figures are rounded off to the third decimal place.}
\end{table}
\subsubsection{Matrix Elements of $\tilde x^2$}\label{sec:X2elem}
Table \ref{tab:SX2n} shows the matrix elements:
\[\sigma^2_r=\langle r \vert \hat x^2  \vert 0 \rangle. \]
They are revisited in Appendix~\ref{eq:P-integration} to discuss
the interpolation capability of the position scaling-eigenfunctions.
\begin{table}[h]
	\caption{\label{tab:SX2n}  Matrix elements of $\hat x^2$,  $\sigma^2_r=\langle r \vert \hat x^2  \vert 0 \rangle $, obtained with
    different SFs. }
	\centering
	\begin{tabular}{l c c c c c c }
		\hline
    \hline
		 && $\sigma^2_{0}$ &  $\sigma^2_{1}$  &  $\sigma^2_2$  &  $\sigma^2_{3}$  &  $\sigma^2_{4}$\footnotemark[1] \\
		\hline
    1. SF && & & & &\\
		\hspace{3mm}a. Sy4  && 0.103\footnotemark[2]  &  -0.285  & -0.314 & 0.004 & 0.000  \\
		\hspace{3mm}b. Db2 	&& 0.077 & -0.206  & -0.003 &  0.000 &  0.000  \\
		\hspace{3mm}c. Db3  && 0.094 & -0.414 & 0.087 & -0.001  & 0.000 \\
		\hspace{3mm}d. Db4 	&& 0.119 & -0.647 & 0.207 & -0.031 & 0.000 \\
    \hline \hline
	\end{tabular}
	\footnotetext[1]{$\sigma^2_{-n}=\sigma^2_{n}$. }
	\footnotetext[2]{All figures are rounded off to the third decimal place.}
\end{table}
\subsection{Momentum Operator $\tilde p$ }\label{app:mom_op}
\subsubsection{Computation}\label{app:mom_cmp}
Since:
\begin{equation}
(x \rightarrow 2x) \longmapsto (\hat p^n \rightarrow 2^{-n} \hat p^n),
\end{equation}
the following equations for the matrix elements of $\hat p^n$ are obtained  from Eq.~(\ref{eq:X_two_scale}):
\begin{equation}\label{eq:Pnm}
\begin{split}
P_{m,n} & =\frac{1}{2}\langle m\vert \hat p^n \vert n\rangle\\
&=2^n \sum_{l,k} h_{l} h_{k} P_{2m+l,2n +k}.
\end{split}
\end{equation}
As in  the case of the position operator,  $\{P_r\}$ is defined:
\begin{equation}\label{eq:Pseries}
P_r=P_{m+r,m}.
\end{equation}
By substituting the above to Eq.~(\ref{eq:Pnm}), we have the equation for $\{P_r\}$:
\begin{equation}\label{eqap:Pr}
P_r =2^n \sum_{k,q } h_{k} h_{k+2r-q}P_q.
\end{equation}
Unlike $\hat{x}$, $\hat {p}$ is translationally invariant, thus the equation becomes homogeneous eigenvalue equation:
\begin{equation}
\begin{array}{l}
	\mathbf{K}\cdot\boldsymbol{P}=\rho \boldsymbol{P},\\
\left[   \mathbf{K} \right]_{r,q}=\sum_{k} h_{k} h_{k+2r-q}, \\
\left[   \boldsymbol{P}  \right]_r=P_r,
\end{array}
\end{equation}
and hence the physical substance of $\{P_r\}$ depends on  $\rho $:
\[ P_r = \langle r\vert  \hat{p}^n \vert 0 \rangle, \quad n=-\log_2 \rho. \]
If $\rho=2$, it is the matrix elements of the momentum operator,
if $\rho=4$, it is those of  the kinetic energy operator.
Since Eq.~(\ref{eqap:Pr}) is homogeneous, supplemental conditions
have to be  provided to determine their magnitude:
\begin{equation}\label{eq:pr-cond-1}
\begin{split}
\sum_r r P_r = \frac{1}{i} &\quad(\mathrm{for \ }\rho = 2),\\
\sum_r r^2 P_r = -2  &\quad(\mathrm{for \  }\rho = 4).
\end{split}
\end{equation}
Thus, the series $\{P_r\}$ is uniquely determined.
\subsubsection{Results and Implication  }\label{app:mom_res}
\begin{table}[h]
	\caption{\label{tab:Pn}
		Matrix elements of $i\hat p $, $iP_r = \langle r \vert \ i\hat p \vert 0 \rangle $ obtained with different SFs
		(eigenvalue $\rho=2$).							 }
	\centering
	\begin{tabular}{l c c c c c c }
		\hline
    \hline
		 && $iP_{0}$ &  $iP_{1}$  &  $iP_2$  &  $iP_{3}$  &  $iP_{4}$\footnotemark[1] \\ \hline
    1. SF && & & & &\\
		\hspace{3mm}a. Sy4  && 0.000\footnotemark[2]  &  -0.793  & 0.192 & -0.034 & 0.002 \\
		\hspace{3mm}b. Db2 	&& 0.000 & -0.667  & 0.083 &  0.00 &  0.00  \\
		\hspace{3mm}c. Db3  && 0.000 & -0.745 & 0.145 & -0.015  & 0.000 \\
		\hspace{3mm}d. Db4 	&& 0.000 & -0.793 & 0.192 & -0.034 & 0.002 \\
    \hline \hline
	\end{tabular}
	\footnotetext[1]{$P_{-n}=-P_{n}$. }
		\footnotetext[2]{All figures are rounded off to the third decimal place.}
\end{table}
\begin{table}[h]
	\caption{\label{tab:P2n}
		Matrix elements of $\hat p^2/2$, $P_r= \langle r \vert \ \hat p^2/2 \vert 0 \rangle $ obtained with different SFs
		(eigenvalue $\rho=4$). }
	\centering
	\begin{tabular}{l c c c c c c }
		\hline
    \hline
		 && $-P_{0}$ &  $-P_{1}$  &  $-P_2$  &  $-P_{3}$  &  $-P_{4}$\footnotemark[1] \\
    \hline
    1. SF && & & & &\\
		\hspace{3mm}a. Sy4  && -4.166\footnotemark[2]  &  2.642  & -0.698 & 0.151 & -0.011 \\
		\hspace{3mm}b. Db2 	&& -6/0\footnotemark[3] & 4/0  & -1/0 &  0/0 &  0/0  \\
		\hspace{3mm}c. Db3  && -5.268 & 3.390 & -0.876  & 0.114  & 0.005 \\
		\hspace{3mm}d. Db4 	&& -4.166 & 2.642 & -0.698 & 0.151 & -0.011 \\
    \hline \hline
	\end{tabular}
	\footnotetext[1]{$P^2_{-n}=P^2_{n}$. }
		\footnotetext[2]{All figures are rounded off to the third decimal place.}
		\footnotetext[3]{The left-hand side of  Eq.~(\ref{eq:pr-cond-1}) is $0$. }
\end{table}
 The numerically obtained series $\{P_r\}$ corresponding to the momentum and the kinetic energy operators
are listed in the Table \ref{tab:Pn} and Table \ref{tab:P2n}, respectively.

The following conventional calculation to obtain $\{P_r\}$ requires the H\"older exponent
of the SFs  to be greater than $n$:
\begin{equation}\label{eq:P-integration}
\begin{split}
P_{r}&=\langle r\vert\int dy\vert y\rangle\langle y\vert\hat{p}^n\int dx\vert x\rangle\langle x\vert n=0\rangle  \\
&=\int dy\int dx\langle r\vert y\rangle\langle y\vert x\rangle \left(\frac{1}{i} \right)^n
\frac{{\partial}^n}{\partial x^n}\langle x\vert n=0\rangle  \\
&=\left(\frac{1}{i} \right)^n\int dx\phi(x-r)\frac{\partial\phi(x)}{\partial x}.
\end{split}
\end{equation}
	 On the other hand, the current method does not have explicit conditions on
	 a differentiability per se, since no actual differentiation takes place in Eq.~(\ref{eq:X_two_scale}).
	 In fact, it works well with the SFs having lower H\"older exponents
	 such as Daubechies-3, as shown in Table \ref{tab:Q2n}.

	 The reason  seems to be
 Eq.~(\ref{eq:P-integration}) forces to calculate the unwanted higher resolution components out of
	  $\mathcal{H_S}$, which are  discarded  when coming back to $\mathcal{H_S}$.
	 In Eq.~(\ref{eq:X_two_scale}), contrary, the higher resolution parts are dealt with by the two-scale relation and algebra,
	 but not by differentiation.

	 In the case of Db2 ($\mathrm{C}^{0.5}$-continuous), however,
	 the eigenvector corresponding to the kinetic energy is:
   \begin{equation}\label{key}
   \boldsymbol{P}= [1,-4,6,-4,1]^\mathrm{T},
   \end{equation}
	 and this fails to satisfy the condition given by Eq.~(\ref{eq:pr-cond-1}).

	 The differentiability in the sense of Eq.~(\ref{eqap:Pr}) will have to be
   investigated further.
\subsubsection{Commutation Relation }\label{app:mom_commu}
Since the operators $\tilde {x}$ and $\tilde {p}$ are defined in a space spanned by a finite number of basis
vectors:
\begin{equation}\label{trace}
\mathrm{Tr}\{\tilde x \tilde p \}=\mathrm{Tr}\{\tilde p \tilde x \},
\end{equation}
and hence, they cannot satisfy the usual commutation relations such as:
\begin{equation}
[\tilde{p},\tilde{x} ]  = -i.
\end{equation}
In stead, they satisfy the following commutation relation:
\begin{equation}
\begin{split}
\sum_l &P_{m,l}X_{l,n}-X_{m,l}P_{l,n} \\
=&\sum_l P_{m-l}(X_{l-n}+n\delta_{l,n})-(X_{m-l}+m\delta_{m,l})P_{l-n} \\
=&(m-n)P_{m-n}.
\end{split}
\end{equation}
\section{Eigenvectors of the Operators }\label{app:Eigen}
\subsection{Position Eigenvalue  }\label{app:Eigen_Position}
\subsubsection{From translational symmetry}\label{app:Eigen_Position_Symm}
A position eigenvector satisfies the following:
\begin{equation}
\tilde x \vert x_n \rangle = x_n \vert x_n \rangle  \label{eqa:x_eigen_eq}.
\end{equation}
By introducing the translation operator:
\begin{equation}
T(\beta) =e^{i\beta\hat{p}}  \ (\beta \in \mathbb{R}),
\end{equation}
it is obvious:
\begin{equation}
\hat{T}(1) \tilde x \vert x_n \rangle =  \hat{T}(1)\{x_n \vert x_n \rangle \}=  x_n \{\hat{T}(1)\vert x_n \rangle \}.
\end{equation}
On the other hand, if  $\hat{T}(1)$  is applied first to $\tilde x$:
\begin{equation}
\begin{split}
\hat{T}(1) \tilde x \vert x_n \rangle &= \hat{T}(1) \tilde x  \hat{T}^{\dagger}(1)\hat{T}(1)\vert x_n \rangle \\
 &= (\tilde x +1) \{\hat{T}(1)\vert x_n \rangle \}.
 \end{split}
\end{equation}
By comparing the two equations, we get:
\begin{equation}
\begin{split}
 x_n \{\hat{T}(1)\vert x_n \rangle \}&=(\tilde x +1) \{\hat{T}(1)\vert x_n \rangle \}\\
\tilde x \{\hat{T}(1)\vert x_n \rangle \}&=  (x_n -1)\{\hat{T}(1)\vert x_n \rangle \}.\\
\end{split}
\end{equation}
And hence
\begin{equation}\label{key}
	\begin{array}{c}
		\hat{T}(1)\vert x_n \rangle = \vert x_{n-1} \rangle,\\
		 x_n = x_{n-1} + 1.
		\end{array}
\end{equation}
From the above, $x_n$ has to have the following form with an unknown $C_0$:
\begin{equation}\label{eqap:xn-a}
x_n = C_0 + n.
\end{equation}
\subsubsection{With trace of the operator}\label{app:Eigen_Position_Tr}
On the other hand by using Eq.~(\ref{eq:XmmX00}), we have:
\begin{equation}
\begin{split}
\sum_n x_n &= Tr(\tilde{x})  \\
&=\sum_{n=-N}^N X_{n,n} \\
&=  (2N+1)X_0.
\end{split}
\end{equation}
By comparing the above and Eq.~(\ref{eqap:xn-a})
\begin{equation}\label{eqap:diag-a_0}
	\begin{array}{c}
	C_0=X_0, \\
x_n = n + X_0.
\end{array}
\end{equation}
\subsection{Position Eigenvector }\label{app:Eigen_Position_Vector}
\subsubsection{Equation }\label{app:Position_Eigenvector_Equation}
The position eigenvectors are expressed by  a unitary transformation and the SFs:
\begin{equation} \label{eqap:x-eigen_1}
\vert x_n \rangle = \sum_n \xi_{n,m} \vert  m\rangle.
\end{equation}
By translating an eigenvector by 1:
\begin{equation} \label{eqap:x-eigen_2}
	\begin{split}
\vert x_{n-1} \rangle &= \hat{T}(1) \vert x_n \rangle \\
&= \sum_m \xi_{n,m} \hat{T}(1) \vert  m\rangle   \\
&= \sum_m \xi_{n,m}\vert  m-1\rangle   \\
&= \sum_m \xi_{n,m+1}\vert  m\rangle.
\end{split}
\end{equation}
On the other hand, by definition:
\begin{equation}
\vert x_{n-1} \rangle = \sum_m \xi_{n-1,m}\vert  m\rangle.
\end{equation}
From this, we find:
\[\xi_{n-1,m}=\xi_{n,m+1},\]
and the matrix $\{\xi_{n,m}\}$ is simplified to a series:
\[\xi_{n,m} \rightarrow \xi_{n-m}.\]
By combining the above with  Eq.~(\ref{eq:Xmn})
and Eq.~(\ref{eqap:diag-a_0})
the equation for $\{\xi_n\}$  becomes:
\begin{equation}\label{eqap:xi_eigen_eq}
\sum_{n}\{X_{m-n}\xi_{n-p}+m\delta_{m,n}\xi_{n-p} \}=x_p \xi_{m-p}.
\end{equation}
Since the solution has the translational symmetry, it is enough to solve for
$p=0$. From Eq.~(\ref{eqap:xn-a}) regarding the eigenvalue and
Eq.~(\ref{eqap:diag-a_0}) regarding the diagonal elements,
Eq.~(\ref{eqap:xi_eigen_eq}) has $X_0$ on the both side of it. By subtracting
$X_0$ from the both side of Eq.~(\ref{eqap:xi_eigen_eq}), or equivalently
setting $X_0=0$ and $x_{p=0}=0$, we get:
\begin{equation}\label{eqap:y_eigen_eq}
\sum_{n}\{X_{m-n}\xi_{n}+m\delta_{m,n}\xi_{n} \}=0.\\
\end{equation}
\subsubsection{Analytical Solution Procedure }\label{app:xi_solution}
Although Eq.~(\ref{eqap:y_eigen_eq}) for $\{\xi_n\}$ is numerically solvable
with a math library such as  GSL\cite{gsl}, the analytical solution gives  more
insight into the position eigenvectors.

Since Eq.~(\ref{eqap:y_eigen_eq}) has a form of convolution, Fourier
transformation makes it easier  to solve:
\begin{equation}
\begin{aligned}
\sum_{m,n}X_{m-n}e^{-im\omega}e^{in\omega}\xi_n e^{-in\omega}+\sum_{m,n}m\delta_{m,n}\xi_n e^{-im\omega}&=0.\\
\end{aligned}
\end{equation}
By defining the following two functions:
\begin{equation}\label{apeq:xomega}
\begin{aligned}
X(\omega)=\sum_{m}X_{m}e^{-im\omega},\\
\xi(\omega)=\sum_{m}\xi_{m}e^{-im\omega},\\
\end{aligned}
\end{equation}
we get:
\begin{equation}\label{apeq:X_xi_omega}
\begin{aligned}
X(\omega)\xi(\omega)+i\frac{{\partial}}{\partial\omega}\xi(\omega)&=0.
\end{aligned}
\end{equation}
With $C$ as the normalization factor, the solution is thus:
\begin{equation}\label{eqap:xi-omega-xm}
  \begin{split}
    \xi(\omega)&=C\exp\left(i\sum_{m}\int^{\omega}d\Omega  X_{m}e^{im\Omega}\right)\\
    &=C\exp\left(-\sum_{m\neq0}\frac{X_{m}}{m}e^{im\omega}\right).
  \end{split}
\end{equation}
By using the Bessel function of the first kind $J_n(z)$, the following holds\cite{cuyt2008handbook}:
\begin{equation}\label{key}
  \begin{split}
    e^{-(\frac{X_{m}}{m}e^{im\omega}-\frac{X_{m}}{m}e^{-im\omega})}
    &=e^{-2iX_{m}\sin(m\omega)}\\
    &=\sum_{n=-\infty}^{+\infty}J_{n}\left(-\frac{2X_{m}}{m}\right)e^{inm\theta}.\\
  \end{split}
\end{equation}
By defining:
\begin{equation}\label{eq:YandX}
  Y_m = -\frac{X_m}{m},
\end{equation}
 $\xi(\omega)$ is thus factorized as follows:
\begin{equation}\label{eq:xi_omega_bessel}
  \begin{split}
    \xi(\omega)=C
    &\left\{\sum_{n_{1}=-\infty}^{+\infty}J_{n_{1}}(Y_{1})e^{in_{1}\omega}\right\}
    \left\{\sum_{n_{2}=-\infty}^{+\infty}J_{n_{2}}(Y_{2})e^{i2n_{2}\omega}\right\}  \\
    &...
    \left\{\sum_{n_{T}=-\infty}^{+\infty}J_{n_{M}}(Y_{M})e^{iMn_{M}\omega}\right\}. \\
  \end{split}
\end{equation}
By segregating the terms in Eq.~(\ref{eq:xi_omega_bessel}) exponent-by-exponent, we have:
\begin{equation}\label{eq:xi_n-expansion}
  \begin{split}
    \xi_n &= \frac{1}{2\pi}  \int_0^{2\pi} \xi(\omega)e^{in\omega} d\omega\\
    &=C\sum_{n_{1}+2n_{2}+...+Mn_{M}=-n}J_{n_{1}}(Y_{1})J_{n_{2}}(Y_{2})...J_{n_{M}}(Y_{M}).
  \end{split}
\end{equation}

The following is an example hand calculation of Pdb2:
\begin{equation}\label{key}
  \begin{split}
  \xi(\omega)
  \simeq&
 (J_2(Y_1)e^{-2i\omega} -J_1(Y_1)e^{-i\omega}  \\
 &\quad+  J_0(Y_1)  +  J_1(Y_1)e^{i\omega}  +  J_2(Y_2)e^{2i\omega}) \\
  &(J_2(Y_2)e^{-4i\omega}-J_1(Y_2)e^{-2i\omega}  \\
 &\quad+  J_0(Y_2)  +  J_1(Y_2) e^{2\omega} +  J_2(Y_2)e^{4i\omega}) \\
  \simeq&
   (0.0025e^{-2i\omega}-0.0708e^{-i\omega}  \\
    &\quad +  0.9950 + 0.0708e^{i\omega}+0.0025e^{2i\omega} ) \\
  &\left( -0.0001e^{-2i\omega}  +  1.0000  + 0.0001e^{2\omega}  \right) \\
  \simeq&
   0.0015e^{-2i\omega}
    -0.0709 e^{-i\omega} \\
 &+  0.9950
 +  0.0708e^{i\omega} +0.0035e^{2i\omega},
   \end{split}
\end{equation}
where the values of $\{Y_m\}$ are obtained from Table~\ref{tab:Xn} through Eq.~(\ref{eq:YandX}).
After paying attention to Eq.~(\ref{apeq:xomega}), we have  $\{\xi_n\}$
approximately equals to the corresponding series shown in Table~\ref{tab:xi_n}.
\subsubsection{Rate of Decay of $\xi_n$}\label{ap:decay}
If
\begin{equation}\label{key}
  F^m(\omega)=\left\{\sum_{n=-\infty}^{+\infty}J_{n}(Y_{m})e^{inm\omega}\right\}.
\end{equation}
is analytic, $\xi(\omega)$ is also analytic\cite{ahlfers1966complex} and hence
$\xi_n$ decays faster than any polynomial\cite{pinsky2008introduction}.

The decay of the following Fourier series is, therefore, first estimated to show
$ F^m(\omega)$ is analytic:
\begin{equation}\label{key}
 f^m_{nm} =\frac{1}{2\pi} \int F^m(\omega)e^{-inm\omega} d\omega.
\end{equation}
When $\vert n\vert $ is large enough to meet:
\begin{equation}\label{key}
  0< \vert Y_m \vert \ll  \sqrt{n+1},\\
\end{equation}
the following holds\cite{AbramowitzStegun64}:
\begin{equation}
  J_{n}(Y_{m})\sim \sign(n)^{\vert n \vert }\frac{1}{\vert n\vert !}\left(\frac{ Y_{m}  }{2}\right)^{\vert n\vert}.
\end{equation}
And hence
\begin{equation}\label{key}
  \begin{split}
    \vert f^m_{nm} \vert &=\frac{1}{\vert n\vert!} \left|\frac{ Y_{m}  }{2}\right|^{\vert n \vert}\\
   &=\frac{1}{\sqrt{2\pi \vert n\vert }} \left(\frac{ e}{\vert n \vert }\right)^{\vert n\vert}
  \left|\frac{ Y_{m}  }{2}\right|^{\vert n\vert }\\
 &=\frac{1}{\sqrt{2\pi \vert n\vert }} \left(\frac{e\left|Y_m \right|} { 2\vert n\vert }\right)^{\vert n\vert }.
  \end{split}
\end{equation}
From Table~\ref{tab:Xn}
\begin{equation}\label{key}
 \left(\frac{e\left|Y_m\right|} { 2\vert n\vert }\right)^{\vert n\vert } < 1,
\end{equation}
 and hence  positive constants $A$ and $c_0$
satisfying the following system of inequality exist:
\begin{equation}\label{key}
  \vert f^m_{nm} \vert < Ae^{-c_0\vert n \vert }\  \quad(n\in \mathbb Z).
\end{equation}
By defining $r=nm$, $c_1=c_0/m$, we have
\begin{equation}\label{key}
  \vert f^m_{r} \vert < Ae^{-c_1\vert r \vert }\  \quad(r\in \mathbb Z).
\end{equation}
Thus, $ F^m(\omega)$ is analytic\cite{pinsky2008introduction}, and hence $\xi(\omega)$ is analytic and $\xi_n$ decays faster than any polynomial of $r$ as $r$ goes to infinity.
\subsubsection{Normalization of $\xi_m$ and  Convergence of the Sum}\label{ap:xi-norm}
The normalization coefficient $C$ is determined as follows:
\begin{equation}\label{key}
	\begin{split}
		\delta_{mn}
				=&\langle m\vert\hat{1}\vert n\rangle\\
		=&\langle m\vert \lbrace \sum_{p=-\infty}^{\infty}\vert x_{p}\rangle\langle x_{p}\vert \rbrace \vert n\rangle\\
		= &  \sum_{p}\xi_{m-p}\xi_{n-p}\\
		= & \frac{C^{2}}{(2\pi)^{2}}\!\!\!\sum_{p=-\infty}^{\infty}\!\!\!\int e^{i(m-p)\omega_{1}}\xi(\omega_{1})d\omega_{1}\int e^{i(n-p)\omega_{2}}\xi(\omega_{2})d\omega_{2}\\
		= & \frac{C^{2}}{(2\pi)^{2}}\!\!\!\int \!\! d\omega_{1}\!\!\! \int\!\!\! d\omega_{2}\!\!\!\sum_{p=-\infty}^{\infty}\!\!\!e^{-ip(\omega_{1}+\omega_{2})}\xi(\omega_{1})\xi(\omega_{2})e^{in\omega_{2}}e^{im\omega_{1}}\\
		= & \frac{C^{2}}{2\pi}\int d\omega_{1}\xi(\omega_{1})\xi(-\omega_{1})e^{i(m-n)\omega_{1}}\\
		= & \frac{C^{2}}{2\pi}\int d\omega_{1}e^{i(m-n)\omega_{1}}\\
		=&  C^{2}\delta_{m,n}.\\
	\end{split}
\end{equation}
Thus,
\begin{equation}\label{key}
	\vert C \vert = 1.
\end{equation}
\subsubsection{Convergence of $\sum \xi_m$  }\label{ap:sumxi}
Since $X_m/m$ is antisymmetric with respect to $m$, summing
Eq.~(\ref{eq:xi_n-expansion}) from $-\infty$ to $\infty$ gives:
\begin{equation}
\begin{split}\sum_{m=-\infty}^{\infty}\xi_{m}
	% & =\frac{1}{2\pi}\int\sum_{m=-\infty}^{\infty}e^{-im\omega}\xi(\omega)d\omega\\
	& =\frac{1}{2\pi}\int2\pi\delta(\omega)\xi(\omega)d\omega\\
	& =\xi(0)
	 =\exp\left(-\sum_{m\neq0}\frac{X_{m}}{m}\right) =1.
\end{split}
\end{equation}
\subsection{Position Eigenvector as Scaling Function }\label{app:pos-eigen}
\begin{figure}[h]
\includegraphics[width=1.0\linewidth]{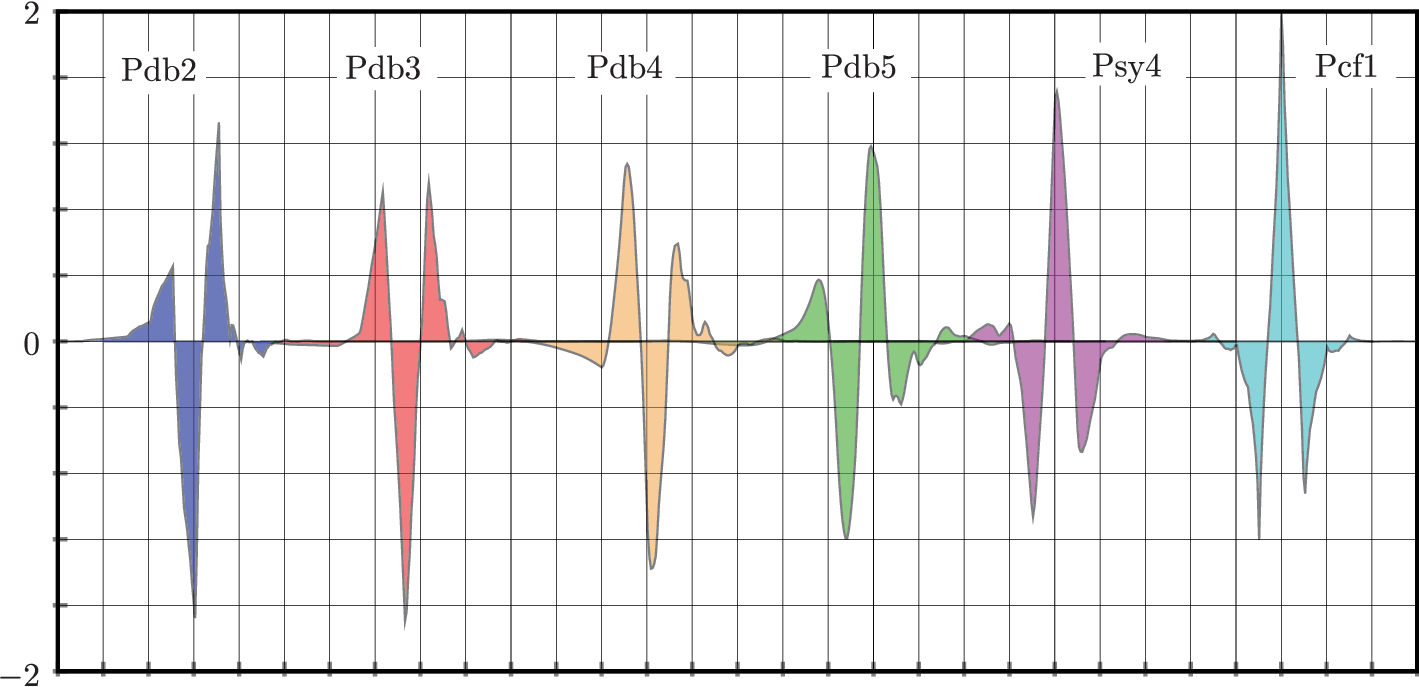}
	\caption{Position wavelets. From left to right, wavelet of Pdb2,
 Pdb3, Pdb4, Pdb5, Psy4 and Pcf1.}
\label{fig:eigenwavelts}
\end{figure}

By applying the two-scale relation to the position eigenfunctions:
\begin{equation}
\begin{split}
\xi(x)&=\sum_{p}\xi_{p}\phi(x-p)\\
&=\sqrt{2}\sum_{p}\xi_{p}\sum_{n}h_{n}\phi(2x-2p-n),
\end{split}
\end{equation}
and utilizing the following:
\begin{equation}
	\begin{split}
		\phi(x)&=\langle x \vert 0 \rangle \\
		&= \langle x  \vert \sum_p \vert x_p \rangle \langle x_p  \vert 0 \rangle\\
	&=  \sum_p \xi_p(x) \xi_{-p}, \\
	\end{split}
\end{equation}
we have:
\begin{equation}
	\begin{split}
\sqrt{2}\sum_{p}\xi_{p}&\sum_{n}h_{n}\phi(2x-2p-n)\\
		&=\sqrt{2}\sum_{p}\xi_{p}\sum_{n}h_{n}\sum_{q}\xi_{-q}\xi(2x-2p-n-q)\\
		&=\sqrt{2}\sum_{p,n,q}\xi_{p}h_{n}\xi_{q}\xi(2x-2p-n+q)\\
		&=\sqrt{2}\sum_{p,n,q}\xi_{p}h_{l-2p+q}\xi_{q}\xi(2x-l).
	\end{split}
\end{equation}
Thus, we have:
\begin{equation}
	\begin{split}
		&\xi(x)=\sqrt{2}\sum_{l}\eta_{l}\xi(2x-l),\\
	\end{split}
\end{equation}
where
\begin{equation}
	\begin{split}
\quad\eta_{l}=\sum_{p,q,l}\xi_{p}\xi_{q}h_{q+l-2p}.
	\end{split}
\end{equation}
Therefore, the position eigenfunctions turn out to be scaling functions. 
The central part of $\{\eta_l\}$ 
is listed in Table~\ref{tab:eta_l}
and the wavelets corresponding to the position scaling-eigenfunctions are shown in Fig.~\ref{fig:eigenwavelts}.
\begin{table}[h]
	\caption{\label{tab:eta_l}Central part of $\{\eta_r\} $ corresponding
to the position eigenfunction indicated in the first column.
 }
	\centering
	\begin{tabular}{l c c c c c c c }
    \hline\hline
		  &$\eta_{-3}$  & $\eta_{-2}$  &  $\eta_{-1}$  &  $\eta_{0}\footnotemark[1]$  &  $\eta_{1}$  &  $\eta_{2}$ & $\eta_{3}$  \\ \hline
	   1. SF  	& &  &  & &  &  & \\
    \hspace{3mm}a. Psy4  	& -0.026\footnotemark[2] & -0.096 & 0.316 &  0.808 & 0.480 & -0.039 &  -0.066 \\
		\hspace{3mm}b. Pdb2 	&  -0.037 & -0.023 & 0.526 &  0.822 & 0.193 & -0.083  & 0.022   \\
		\hspace{3mm}c. Pdb3 	&  -0.046  & -0.062 & 0.376 &  0.827 &  0.400 & -0.089 & -0.015  \\
		\hspace{3mm}d. Pdb4 	&  -0.038  & -0.093 & 0.244 &  0.772 &  0.572 & -0.018  & -0.074  \\
    \hline \hline
	\end{tabular}
		\footnotetext[1]{Since the series is infinite, the  number $r$ is set to $0$ where the  element is maximum.}
		\footnotetext[2]{All figures are rounded off to the third decimal place.}
\end{table}

In comparison with the original compact support SFs, the position
scaling-eigenfunctions have the following properties: (1) Infinite support
length. The position scaling-eigenfunctions in real space decay exponentially,
but the support is no longer compact. (2) Substantially small  $\sigma_0^2 $.
Despite the infinitely wide supports of the position scaling-eigenfunctions,
$\sigma_0^2 $s are nearly identical to those of the original compactly
supported SFs (see Tables~\ref{tab:SX2n} and \ref{tab:X2n}). (3) Substantially
small $\sigma_r^2 (r\ne0)$. From the tables, the position
scaling-eigenfunctions are closer to the eigenvector of $\hat x^2 $ than the
original compactly supported SFs are, since the off-diagonal parts are in
general substantially smaller.

\begin{table}[h]
	\caption{\label{tab:X2n} Matrix elements of $\hat x^2$,  $\sigma^2_r=\langle x_r \vert \hat x^2 \vert x_0 \rangle $, obtained with
    different position scaling-eigenfunctions.
    }
	\centering
	\begin{tabular}{l c c c c c c}
		\hline
    \hline
  && $\sigma^2_{0}$  &  $\sigma^2_{1}$  &  $\sigma^2_2$  &  $\sigma^2_{3}$  &  $\sigma^2_{4}$\footnotemark[1]  \\
\hline
	   1. SF  	& &  &  & &  & \\
		\hspace{3mm}a. Psy4 && 0.108\footnotemark[2]  &  -0.077  & 0.026 & 0.000 & 0.000  \\ \hline
		\hspace{3mm}b. Pdb2	&& 0.088 & -0.040  & -0.002 &  0.001 &  0.000  \\ \hline
		\hspace{3mm}c. Pdb3	&& 0.102 & -0.064 & 0.009 & 0.004  & -0.001 \\ \hline
		\hspace{3mm}d. Pdb4	&& 0.116 & -0.082 & 0.024 & 0.002 & -0.003 \\ \hline
	\end{tabular}
	\footnotetext[1]{$\sigma^2_{-n}=\sigma^2_{n}$ }
	\footnotetext[2]{All figures are rounded off to the third decimal place.}
\end{table}
\subsection{Transformation of Position Scaling-Eigenvector in Physical and Numerical Coordinate Systems }\label{app:coordinate-trans}
The formulation  can be done either in the physical or in numerical domain. In
the numerical domain, the unit cell length coincides with the number of division
in one unit cell $N$. It disagrees with the convention, but the formulation
becomes simplest. When the physical domain is chosen to keep the unit cell
length at $1$, the following  constant for conformation $c$ has to be chosen,
\begin{equation}
    \langle x \vert x_p \rangle = c\xi(N  (x-x_p)).
\end{equation}
A choice of $c$ makes some relationships simple and others a little more
complex. In the paper, $c$ is chosen so that the basis set becomes orthonormal.
The resulting  relationships between variables and functions are listed in
Table~\ref{tab:ta-transform-L } for convenience.
\begin{table}[h]
  \caption{
   Correspondence of parameters, variables, functions, inner products and
   operator  between physical and numerical domains.
  }\label{tab:ta-transform-L }
  \centering
  \begin{tabular}{c c c}
    \hline
    \hline
   & \multicolumn{2}{c}{ Domain }\\
   \cline{2-3}
    & Physical  & Numerical \\
    \hline
    Unit cell length  & $1$ & $N$ \\
    Range & $0\le x<L$ & $0\le x<NL$\\
    $\Delta x$ & $1/N$ & $1$\\
    $x_p$ &  $p\Delta x$  & $p$ \\
    $\langle x\vert x_{p}\rangle$ & $\sqrt N \xi(N(x-x_p))$ & $\xi(x-x_p)$\\
    $\langle f\vert x_{p}\rangle$ & $f_{p}$ & $f_{p}$\\
    $f(x)$ & $\sqrt N \sum_{p}f_{p}\xi(N(x-x_p))$ & $\sum_{p}f_{p}\xi(x-x_p)$\\
    $f(x_p)$
  & $\sqrt N \{f_p +O(\sigma^2_{0}/N^2)\}$
   & $ f_p+O(\sigma^2_{0})$\\
    $\int dx f(x)g(x) \simeq $  &\multicolumn{2}{c}{ $\sum_p f(x_p)g(x_p)\Delta x $ } \\
    $\hat 1 $ & $\sum_{p}\vert x_{p}\rangle\langle x_{p}\vert$ & $\sum_{p}\vert x_{p}\rangle\langle x_{p}\vert$\\
    $\langle x_p \vert x_q \rangle$  & $\delta_{p,q}$ & $\delta_{p,q}$\\
    $\int dx \langle x \vert x_p \rangle $  & $1/\sqrt{N}$ & $1$\\
    \hline \hline
  \end{tabular}
\end{table}
\subsection{Approximation of Functions and Matrix Elements Using Position eigenvectors' Properties}\label{app:interpolatin}
A function or signal can be projected onto  $\mathcal{H_S}$
spanned by a set of SFs $\{\vert \phi_p \rangle \}$:
\begin{equation}
	\begin{split}
		\tilde{f}(x)&= \sum f_p \phi (x-p),\\
		f_p &=\langle \phi_p \vert f \rangle.
	\end{split}
\end{equation}
Certain SFs called interpolation SFs, such as the Shannon SF\cite{Daub10-2}, satisfy:
\begin{equation}\label{fpfp}
		f_p =f(p).
\end{equation}
This implies the values of an incoming signal or a function on
the ticks are by themselves  expansion coefficients, thereby substantially
reducing the cost of decomposition (albeit, the Shannon SF is hard to employ in
practical cases for the slow decay). This section therefore focuses on the
interpolation capability of the position scaling-eigenfunctions and the eventual
reduction in the potential energy matrix elements.
\subsubsection{Interpolation Capability of Position Scaling-Eigenfunctions}\label{sec:interpolation}
In general case, by using a set of  general orthonormal scaling functions
$\{\phi_p(x)\}$,  $f_p$ is expressed as follows:
\begin{equation}\label{key}
\begin{split}
	f_{p} & =\int dxf(x)\phi_{p}(x)\\
	& =\int dx\langle x\vert f(\hat{x})\vert\phi_{p}\rangle.
\end{split}
\end{equation}
If $f(x)$ is piece-wise  quadratic polynomial over a few calculation cells:
\begin{equation}\label{eq:f(x)}
	f(\hat x)=a_0+a_1 \hat x + a_2 \hat x^2.
\end{equation}
And if the sampling rate or the resolution is high enough to reconstruct the
signal or the function, the operation can be done in $\mathcal{H_S}$.
Thus, following approximation is valid in the vicinity where Eq.~(\ref{eq:f(x)})
holds:
\begin{equation}\label{eqap:xhat_approx}
	\begin{split}
	f(\hat{x})\vert\phi_{p}\rangle &=a_{0}\vert\phi_{p}\rangle+a_{1}\hat{x}\vert\phi_{p}\rangle+a_{2}\hat{x}^{2}\vert\phi_{p}\rangle
	\\
		& =a_{0}\vert\phi_{p}\rangle+a_{1}(\sum_{r}X_{r}\vert\phi_{p+r}\rangle+p\vert\phi_{p}\rangle) \\
		&\  +\!\!a_{2}(\sum_{r}\sigma_{r}^{2}\vert\phi_{p+r}\rangle
		\!+\!2p\sum_{r}X_{r}\vert\phi_{p+r}\rangle \!+\! p^{2}\vert\phi_{p}\rangle)\\
		&=\lbrace f(p)+(a_{1}X_{0}+2a_{2}pX_{0} +a_{2}\sigma^2_{0})\rbrace \vert \phi_p \rangle\\
		&\  +\sum_{r\ne 0}\lbrace  (a_{1}+2a_{2}p)X_{r}+a_2 \sigma_{r}^{2} \rbrace \vert\phi_{p+r}\rangle\\
		&=f_p \vert \phi_p \rangle +\sum_{r\ne 0} f_{p}^r\vert\phi_{p+r}\rangle,
	\end{split}
\end{equation}
where
\begin{equation}\label{eq:fpr}
	\begin{array}{l}
		f_p=f(p)+ (a_{1}X_{0}+2a_{2}pX_{0} +a_{2}\sigma^2_{0}), \\
		f_p^r=(a_{1}+2a_{2}p)X_{r}+a_2 \sigma_{r}^{2},
	\end{array}
\end{equation}
and  the contribution from the higher resolution wavelets are ignored when the
operator acts on the kets.

By choosing one of the position scaling-eigenvectors as $\phi(x)$, all elements
of $\{X_r\}$ become zero. Furthermore, $\sigma_r^2 (r \ne 0)$ shown in Table \ref{tab:X2n}
becomes substantially smaller than those of the original SFs listed in Table
\ref{tab:SX2n}. And hence:
\begin{equation}\label{eqap:fn_phi_n}
	\begin{array}{l}
		f_{p}= 	f(p)+O(\sigma^2_{0}),\\
		f_p^r= 0  +O(\sigma^2_{r}).
	\end{array}
\end{equation}
In the physical domain:
\begin{equation}\label{eqap:fn_phi_xp}
	\begin{array}{l}
		f_{p}= 	N^{-1/2}f(x_p)+O(\sigma^2_{0}/N^2),\\
		f_p^r= 0  +O(\sigma^2_{r}/N^2),\\
			 x_p= p/N\ (	L_0, V_0 \in \mathbb R).
	\end{array}
\end{equation}
\subsubsection{Application to BFs}\label{sec:approx_BF}
From Eqs.~(\ref{eq:bloch-x}) and (\ref{eqap:fn_phi_n}),
\begin{equation}\label{eqap:psi_approx}
	\begin{split}
	e^{ikl}e^{ikm\Delta x }c_{m,k}^{n}&=\psi_{k,p}^n \\
	&\simeq N^{-1/2}\psi_{k}^n (x_p).\\
	\end{split}
\end{equation}
And hence,
\begin{equation}\label{eq:PWE-substitution-2}
	\begin{split}
		\vert \psi^n_k \rangle
		= N^{-1/2}\sum_{n,k}  \psi^n_k(x_p) \vert x_p\rangle + O(\sigma^2_0/N^2). \\
	\end{split}
\end{equation}
It also applies to Wannier functions and in the same way:
\begin{equation}
	\vert W_l^s \rangle =N^{-1/2} \sum_p W_l^s(x_p)\vert x_p \rangle + O(\sigma^2_0/N^2).
\end{equation}
\subsubsection{Application to Potential Energy Matrix Elements}\label{sec:potential-elem}
If the potential energy as a function of $x$ is piece-wise  quadratic polynomial
over a few calculation cells, the matrix elements of operator $V(\hat x)$ is
calculated by utilizing  Eq.~(\ref{eqap:xhat_approx}):
\begin{equation}\label{apeq:v-diagonal}
	\begin{split}
		\langle x_{p_1} \vert V(\hat x) \vert  x_{p_2} \rangle &=
		\langle x_{p_1} \vert \lbrace V(\hat x) \vert x_{p_2} \rangle  \rbrace\\
		& =		\langle x_{p_1} \vert \lbrace V(x_{p_2}) \vert x_{p_2} \rangle \rbrace +O(\sigma^2_0/N^2)\\
		& = V(x_{p_2}) \delta_{p_1, {p_2}}+O(\sigma^2_0/N^2).
	\end{split}
\end{equation}
The approximation above is, in effect, equivalent of:
\begin{equation}\label{key}
	V(\hat x) \rightarrow V(\tilde x).
\end{equation}
\subsubsection{$\delta$-function Potential}
When  $V(x)$ is given as a $\delta$-function such as:
\begin{equation}\label{key}
	\begin{split}
		V (x,a) &= V_0\delta (x-a),\\
	\end{split}
\end{equation}
preserving  the character of the $\delta$-function as an  \textit{impulse},
rather than the direct projection of the delta function onto the composite band
space,  is prioritized.

The formal potential operator is calculated as follows:
\begin{equation}\label{key}
  \begin{split}
  \hat V(a) &= \int\!\!\! dx\vert x \rangle V_0\delta(x\!\!-\!\!a)\langle x \vert \\
   &=V_0\vert a\rangle\langle a\vert,
  \end{split}
\end{equation}
and hence,
\begin{equation}\label{key}
  \begin{array}{r}
    \langle f \vert \hat V(a) \vert g \rangle = V_0f(a)g(a)\quad  \left( \forall \vert f \rangle, \vert g \rangle \in \mathcal H \right).
  \end{array}
\end{equation}

In $\mathcal H_X$, on the other hand,
\begin{equation}\label{key}
  \begin{array}{r}
  \begin{aligned}
    \langle f\vert x_{p}\rangle\langle x_{p}\vert g\rangle&=f_{p}g_{p}\\
    &=\frac{1}{N}f(x_{p})f(x_{p})\quad  \left( \forall \vert f \rangle, \vert g \rangle \in \mathcal H_X \right).
  \end{aligned}
\end{array}
\end{equation}
Thus, the potential energy  operator is discretized to be:
\begin{equation}\label{key}
  \begin{array}{c}
    V_0\vert a\rangle\langle a\vert \rightarrow NV_0\vert x_{p}\rangle\langle x_{p}\vert,\\
    x_p = a,\\
   p= a/\Delta x.
  \end{array}
\end{equation}
Eventually, the numerical expression of the $\delta$-function potential becomes as follows:
\begin{equation}\label{ape: disc-potential}
	\begin{split}
		\langle x_p \vert\hat V(a)  \vert x_{p_1} \rangle =  V_0 N\delta_{p, p_1}.
	\end{split}
\end{equation}

  Note that if the direct projection of $\hat V(a) $ on to $\mathcal H_X$ is
carried out, it will  result in a possibly broader potential energy operator as
described in Sec.~\ref{sec:scdm}:
  \begin{equation}
    \begin{split}
    \sum_{p,q} \vert x_p \rangle \langle x_p \vert \hat V(a)\vert x_q \rangle
    \langle x_q \vert
    =\sum_{p,q}\vert x_p\rangle \xi_{p}(a)\xi_{q}(a)\langle x_q \vert.
    \end{split}
    \end{equation}
\subsection{Momentum Eigenvector}
In a similar way described in Sec.~\ref{app:xi_solution}, the momentum eigenvectors are obtained:
\begin{equation}
\begin{array}{c}
\tilde{P} \vert k \rangle = P(k)\vert k \rangle, \\
\vert k \rangle = \frac{1}{\sqrt N } \sum_n e^{ikn} \vert n \rangle, \\
P(k) = \sum_r P_r e^{ikr},
\end{array}
\end{equation}
where $\tilde P$ is the momentum operator in $\mathcal{H_S}$,  $\vert k \rangle $
is the momentum eigenvector with the wave number $k$, and $P(k)$ is the
eigenvalue. It is easily checked the above satisfies the following eigenvalue equations:
\begin{equation}\label{eq:Pr}
\begin{aligned}
\sum_m P_{m-n} e^{ikn}= P(k) e^{ikm}. \\
\end{aligned}
\end{equation}
 \section{MLWF and PPE Calculations}
 \subsection{Eigenvalue Equations for MLWFs }\label{app:wannier-eigeen}
 An MLWF satisfies the following equations\cite{Marzari1997}\cite{kivelson1982}:
 \begin{equation}\label{eqap:btow_lambda}
 	\begin{split}
 		\tilde{x}  \vert W \rangle &= x \vert W \rangle. \\
 	\end{split}
 \end{equation}
Since the number of energy bands involved is $n_c$ and the total unit cell
number is $L$, Eq.~(\ref{eqap:btow_lambda}) has $n_c \times L$ eigenvalues and
eigenvectors, if there is no degeneracy:
 \begin{equation}\label{eqap:btow_s_l}
 	\begin{split}
 		\tilde{x}  \vert W^s_l \rangle &= x^s_l \vert W^s_l  \rangle. \\
 	\end{split}
 \end{equation}
 An MLWF  is  expanded by the primal WFs pertaining to cell $l$ and energy band $n$, $\{\vert \tilde W^n_{l} \rangle \}$:
 \begin{equation}\label{eqap:btow_w}
 	\begin{split}
 		\vert W^s_{l} \rangle
   &= \sum_{l_1,n} U_{l,l_1}^{s,n} \vert \tilde W^n_{l_1} \rangle, \\
 	\end{split}
 \end{equation}
 with
  \begin{equation}
 	\begin{split}
  \quad \vert \tilde W^n_l \rangle= L^{-1/2}\sum_{k}e^{-ikl}\vert \psi_k^n \rangle.
 	\end{split}
 \end{equation}
 Substituting the above into Eq.~(\ref{eqap:btow_s_l}), the eigenvalue equations are obtained:
 \begin{equation}\label{eqap:btow_w_U}
 	 \begin{array}{l}
 	\sum_{l_2, n_2} X^{n_1,n_2}_{l_1, l_2} U_{l,l_2}^{s,n_2}= x^s_l U_{l,l_1}^{s,n_1},\\
   	X^{n_1,n_2}_{l_1, l_2}= \langle \tilde W^{n_1}_{l_1} \vert \tilde x \vert \tilde W^{n_2}_{l_2} \rangle.
   	 \end{array}
 \end{equation}
 From the translational invariance of system, we have:
 \begin{equation}\label{key}
 \begin{split}
 x^s_{l+1}&=x^s_l + 1.\\
 \end{split}
 \end{equation}
Since each series has corresponding eigenvalue for each cell number $l$, one
unit cell hosts $n_c$ different \textit{maximally} localized Wannier functions
and they are the candidates for the truly maximally localized Wannier
function.

In the same manner, an eigenvalue equation for MLWFs can also be obtained using the BFs:
\begin{equation}\label{eqap:btow_b_U}
  \begin{array}{l}
    \sum_{k_2, n_2} X^{n_1,n_2}_{k_1, k_2} U_{l,k_2}^{s,n_2}= x^s_l U_{l,k_1}^{s,n_1},\\
    X^{n_1,n_2}_{k_1, k_2}= \langle \psi^{n_1}_{k_1} \vert \tilde x \vert \psi^{n_2}_{k_2} \rangle.
  \end{array}
\end{equation}
The above equation is equivalent of Eq.~(\ref{eq:wannier-k-eq}) (see Eq.~(\ref{apeq:x_elem_0}))
and useful, when the gauge of the BFs is
dealt with. However, it is not suited to  solving
Eq.~(\ref{eqap:btow_s_l}), since the position operator matrix
is not diagonal dominant. In the present paper,
therefore,  MLWFs are obtained  by solving Eq.~(\ref{eqap:btow_w_U}) with the GSL
library\cite{gsl}.
\subsection{Supplement for Eq.~(\ref{eq:MLWF(x)})  }\label{app:suppl_0}
By using  the translational operator by $l$,  $\hat T(l)$:
\begin{equation}\label{eqap:bloch-MLWF-translation}
	\begin{split}
		\vert W_{l}^s \rangle &= \hat T(l)\vert W_{0}^s \rangle \\
		&= \sum_{n,k} w^{s,n}_{0,k}T(l)\vert \psi_k^n \rangle \\
		&= \sum_{n,k} w^{s,n}_{0,k}e^{-ik l}\vert \psi_k^n \rangle. \\
	\end{split}
\end{equation}
Therefore, we have:
\begin{equation}
\begin{split}
\langle x_p\vert W^s_{l}\rangle
&= \sum_{n,k} w^{s,n}_{k}e^{-ik l}\langle x_p \vert \psi_k^n \rangle \\
&=\sum_{k}w^{s,n}_{k}e^{-ikl}N^{-1/2}\psi^n_{k}(x_p)\\
&=\sum_{k}w^{s,n}_{k}N^{-1/2}\psi^n_{k}(x_p-l)\\
&=N^{-1/2}W^s_0(x_p-l)=N^{-1/2}W^s_l(x_p).
\end{split}
\end{equation}
\subsection{Orthogonalization of Translational Vectors } \label{app:orthcond}
The inner product of a vector $\vert W_l \rangle \in \mathrm{Span} \{\vert \psi_k \rangle \}$ and its translation is calculated as follows\cite{Daub10-5}:
\begin{equation} \label{key}
	\begin{split}\langle W_{l}\vert W_{l+m}\rangle & =\langle W_{l}\vert e^{-i\hat{p}m}\vert W_{l}\rangle\\
		& =\sum_{k}\langle W_{0}\vert \psi_k\rangle\langle \psi_k\vert W_{0}\rangle e^{ikl}e^{-ikl}e^{-ikm}\\
		& =\sum_{k}\vert w_{k}\vert^{2}e^{-ikm},
	\end{split}
\end{equation}
where
\begin{equation}
  w_k=  \langle  \psi_k  \vert W_0 \rangle.
\end{equation}
Therefore if the following is satisfied:
\begin{equation} \label{key}
	\vert w_k \vert^{2}=  1/L,
\end{equation}
the equation below holds:
\begin{equation} \label{key}
	\langle W_0 \vert e^{-i\hat{p}m} \vert W_0 \rangle  =  \delta_{m, 0 }.
\end{equation}
Thus,
$ \left \{ \  \ e^{im \hat{p}}  \vert W_0  \rangle  \  \  \right \}$ becomes an orthonormal set.
\subsection{Supplement for Sec.~\ref{sec:PPE-approx}}
\subsubsection{Supplement for Eq.~(\ref{eq:ape-max-is-MLWF-max})}\label{app:why x_p is maximum p}
Let  $\vert x^C_p\rangle$  be the closest vector to  $\vert W^s_{l}\rangle$, then,
\begin{equation}\label{eq:ape-max-is-MLWF-max_1}
    \begin{split}
        x^C_p(x)&=N^{1/2}\langle x^C \vert x^C_p\rangle\\
        &=N^{1/2}\langle  x^C \vert \sum_{s_1,l_1}\vert W^{s_1}_{l_1}\rangle\langle W^{s_1}_{l_1}\vert x^C_p\rangle\\
        &=N^{-1/2}\sum_{s_1,l_1}W^{s_1}_{l_1}(x)\overline{W}^{s_1}_{l_1}(x_p)\\
        &=N^{-1/2}W^s_{l}(x)\overline{W}^s_{l}(x_p)\\
        &\quad +N^{-1/2}\!\!\!\sum_{(s_1,l_1)\ne(s,l)} \!\!\!\!\!\!\!\!W^{s_1}_{l_1}(x)\overline{W}^{s_1}_{l_1}(x_p).\\
    \end{split}
\end{equation}
As $\vert x^C_p\rangle$ is closest to  $\vert W_l^s\rangle$,
\begin{equation}\label{eq:ape-MLWF-orthogonal}
    \begin{split}
        \langle  x^C_p \vert W^{s_1}_{l_1} \rangle = 0,
        \  \left((s_1,l_1)\ne(s,l) \right).
    \end{split}
\end{equation}
Therefore, we have Eq.~(\ref{eq:ape-max-is-MLWF-max}):
\[
x^C_p(x) \simeq N^{-1/2}W^s_{l}(x)\overline{W}^s_{l}(x_p).
\]
\subsubsection{Supplement for Eq.~(\ref{eq:phi=sqrt xp2})}\label{app:intra}
 From Eq.~(\ref{eq:w=x_p})
    \begin{equation}\label{key}
        \begin{split}
            \langle \psi^n_k  \vert x_p \rangle &=\langle \psi^n_k  \vert x^C_p \rangle \\
            &\simeq \langle \psi^n_k\vert
            \left\{\sqrt{\left| \langle x^C_p \vert x^C_p \rangle \right| }
             \vert W^{s}_{l} \rangle \right\}\\
            &=\sqrt{\left| \langle x^C_p \vert x^C_p \rangle \right| } e^{-ikl}w^{s,n}_{k},
        \end{split}
    \end{equation}
    by using  Eq.~(\ref{eq:MLWF-orthonormal-delta}):
    \begin{equation}\label{eqap:xcxc}
        \begin{split}
            \Phi_k &= \sqrt {\sum_n \vert \langle \psi^n_k \vert x^C_p \rangle \vert^2 }\\
            &\simeq  \sqrt{\left| \langle x^C_p \vert x^C_p \rangle \right| } \sqrt {\sum_n \vert w^{s,n}_{k}  \vert^2 }\\
            &=\sqrt{\left| \langle x^C_p \vert x^C_p \rangle \right| }.
        \end{split}
    \end{equation}
    \subsection{Contrasting MLWF and PPE}\label{contrast}
    For  $ x, y \in \mathbb R$, the position eigenvectors in continuous real coordinate system satisfy the following:
    \begin{equation}
        \begin{split}
            \hat x \vert y \rangle &= y \vert y \rangle, \\
            \langle x  \vert f \rangle &= f(x),\\
            \langle x \vert y \rangle &= \delta (x-y),\\
        \end{split}
    \end{equation}
    while the MLWFs and PPEs satisfy part of the above and they supplement each other:
    \begin{equation}\label{eq:porpMLWF}
        \begin{split}
            \tilde x^C \vert W^s_l \rangle &= (x_{0s} + l)\vert W^s_l \rangle, \\
            \langle  W^s_l  \vert f \rangle &\ne f(x_{0s}+l),\\
            \langle  W^{s_1}_{l_1}  \vert  W^{s_2}_{l_2} \rangle &=\delta_{s_1,s_2}\delta_{l_1, l_2},\\
%        \end{split}
%    \end{equation}
%    \begin{equation}\label{eq:propPPE}
%        \begin{split}
            \tilde x^C \vert x^C_p \rangle &\ne  x_p \vert x^C_p \rangle,\\
            \langle  x^C_p \vert f \rangle &= N^{-1/2}f(x_p),\\
            \langle x^C_{p_1}  \vert x^C_{p_2} \rangle &= \xi_{p_2}^C(x_{p_1}) \ne \delta_{p_1,p_2}.
        \end{split}
    \end{equation}
    Whether $\hat x$ should first be projected onto $\mathcal H_C$ or
    $\vert x_p \rangle$  should be directly projected is a matter of
    choice between the properties summarized in the upper and lower half
    of Eq.~({\ref{eq:porpMLWF}), which may depend on the necessity of the
        analysis.
\begin{widetext}
\subsection{Supplement for Eq.~(\ref{eq:wannier-k-eq})  }\label{app:supp}
By multiplying $\langle\psi_{k_{1}}^{n_{1}}\vert$ to Eq.~(\ref{eq:btow})  from left
and summing the term over $k_{2},n_{2}$, we get:
 \begin{equation}\label{apeq:x_elem_0}
 \begin{split}
\sum_{k_{2},n_{2}}\langle\psi_{k_{1}}^{n_{1}}\vert\hat{x} w_{k_{2}}^{n_{2}}\vert\psi_{k_{2}}^{n_{2}}\rangle
&=\!\left(\frac{1}{\sqrt{L}}\right)^{2}\!\!\!\!\!\sum_{k_{2},{n_{2}}m,l}\!\!(l+m\Delta x)e^{i(l+m\Delta x)(k_{2}-k_{1})}\bar{c}_{m,k_{1}}^{n_{1}}c_{m,k_{2}}^{n_{2}}w_{k_{2}}^{n_{2}} \\
&= \frac{{1}}{L} \sum_{k_{2},n_{2},m}
\left\{
 \frac{{1}}{i}\frac{\partial}{\partial\lambda}
\left(
L\delta_{k_{2},k_{1}-\lambda}e^{im\Delta x (k_{2}-k_{1})}\bar{c}_{m,k_{1}}^{n_{1}}c_{m,k_{2}}^{n_{2}}w_{k_{2}}^{n_{2}}
\right)
+L\delta_{k_{2},k_{1}}m\Delta xe^{im\Delta x (k_{2}-k_{1})}\bar{c}_{m,k_{1}}^{n_{1}}c_{m,k_{2}}^{n_{2}}w_{k_{2}}^{n_{2}}
\right\}
\\
 &=
 \sum_{m,n_{2}}
  \left\{
  \frac{1}{i}
 \frac{\partial}{\partial\lambda}
 \left(
 e^{-im\Delta x\lambda}\bar{c}_{m,k_{1}}^{n_{1}}c_{m,k_{1}-\lambda}^{n_{2}}w_{k_{1}-\lambda}^{n_{2}}
 \right)
 +m\Delta x\bar{c}_{m,k_{1}}^{n_{1}}c_{m,k_{1}}^{n_{2}}w_{k_{1}}^{n_{2}}
 \right\}
 \\
&= -\sum_{m,n_{2}}
 \left\{m\Delta x\bar{c}_{m,k_{1}}^{n_{1}}c_{m,k_{1}}^{n_{2}}w_{k_{1}}^{n_{2}}-\frac{1}{i}\bar{c}_{m,k_{1}}^{n_{1}}\frac{\partial c_{m,k_{1}}^{n_{2}}}{\partial k_{1}}w_{k_{1}}^{n_{2}}
 -\frac{1}{i} \bar{c}_{m,k_{1}}^{n_{1}}c_{m,k_{1}}^{n_{2}}\frac{\partial w_{k_{1}}^{n_{2}}}{\partial k_{1}} + m\Delta x\bar{c}_{m,k_{1}}^{n_{1}}c_{m,k_{2}}^{n_{2}}w_{k_{2}}^{n_{2}}  \right\}
 \\
 &= -\frac{1}{i}\sum_{m,n_{2}}\left\{ \bar{c}_{m,k_{1}}^{n_{1}}\frac{\partial c_{m,k_{1}}^{n_{2}}}{\partial k_{1}}w_{k_{1}}^{n_{2}}+ \bar{c}_{m,k_{1}}^{n_{1}}c_{m,k_{1}}^{n_{2}}\frac{\partial w_{k_{1}}^{n_{2}}}{\partial k_{1}} \right\}.
\end{split}
\end{equation}
\end{widetext}
Therefore:
\begin{equation}\label{eqap:wannier-k-eq}
    \begin{split}
        \frac{\partial w_{k}^{n_{1}}}{\partial k}&=-\sum_{m,n_{2}}
        \bar{c}_{m,k_{1}}^{n_{1}}
        \frac{\partial c_{m,k_{1}}^{n_{2}}}{\partial k_{1}} w^{n_2}_{k_1}
        -ix_a w_{k}^{n_{1}}\\
        &=\sum_{m,n_{2}}\frac{\partial\bar{c}_{m,k}^{n_{1}}}{\partial k}
        c_{m,k}^{n_{2}}w_{k}^{n_{2}}-ix_a w_{k}^{n_{1}},
    \end{split}
\end{equation}
where the following are utilized:
\begin{equation}
  \begin{array}{c}
    \sum_{m} \bar{c}_{m,k}^{n_1}c_{m,k}^{n_2}=\delta_{n_1, n_2},\\
    \frac{\partial}{\partial k}\sum_{m}\bar{c}_{m,k}^{n_{1}}c_{m,k}^{n_{2}}=\frac{\partial}{\partial k}\sum_{m}\bar{c}_{m,k}^{n_{2}}c_{m,k}^{n_{1}}=0,
  \end{array}
\end{equation}
and
\begin{equation}
  \begin{split}
    \sum_{k_{2}}le^{il(k_{2}-k_{1})}f(k_{2})
    &=\lim_{\lambda=0}\frac{1}{i}\frac{\partial}{\partial\lambda}
    \sum_{k_{1}}e^{il(k_{2}-k_{1}+\lambda)}f(k_{2})  \\
    &=\lim_{\lambda=0}\frac{L}{i}\frac{\partial}{\partial\lambda}
    \sum_{k_{1}}\delta_{k_{2},k_{1}-\lambda}f(k_{2})  \\
    &=\lim_{\lambda=0}\frac{L}{i}\frac{\partial}{\partial\lambda}
    f(k_{1}-\lambda) \\
    &=-\frac{L}{i}\frac{\partial}{\partial k_{1}}  f(k_{1}).
  \end{split}
\end{equation}

\bibliography{xeigen_paper}

\begin{thebibliography}{62}%
\makeatletter
\providecommand \@ifxundefined [1]{%
 \@ifx{#1\undefined}
}%
\providecommand \@ifnum [1]{%
 \ifnum #1\expandafter \@firstoftwo
 \else \expandafter \@secondoftwo
 \fi
}%
\providecommand \@ifx [1]{%
 \ifx #1\expandafter \@firstoftwo
 \else \expandafter \@secondoftwo
 \fi
}%
\providecommand \natexlab [1]{#1}%
\providecommand \enquote  [1]{``#1''}%
\providecommand \bibnamefont  [1]{#1}%
\providecommand \bibfnamefont [1]{#1}%
\providecommand \citenamefont [1]{#1}%
\providecommand \href@noop [0]{\@secondoftwo}%
\providecommand \href [0]{\begingroup \@sanitize@url \@href}%
\providecommand \@href[1]{\@@startlink{#1}\@@href}%
\providecommand \@@href[1]{\endgroup#1\@@endlink}%
\providecommand \@sanitize@url [0]{\catcode `\\12\catcode `\$12\catcode
  `\&12\catcode `\#12\catcode `\^12\catcode `\_12\catcode `\%12\relax}%
\providecommand \@@startlink[1]{}%
\providecommand \@@endlink[0]{}%
\providecommand \url  [0]{\begingroup\@sanitize@url \@url }%
\providecommand \@url [1]{\endgroup\@href {#1}{\urlprefix }}%
\providecommand \urlprefix  [0]{URL }%
\providecommand \Eprint [0]{\href }%
\providecommand \doibase [0]{https://doi.org/}%
\providecommand \selectlanguage [0]{\@gobble}%
\providecommand \bibinfo  [0]{\@secondoftwo}%
\providecommand \bibfield  [0]{\@secondoftwo}%
\providecommand \translation [1]{[#1]}%
\providecommand \BibitemOpen [0]{}%
\providecommand \bibitemStop [0]{}%
\providecommand \bibitemNoStop [0]{.\EOS\space}%
\providecommand \EOS [0]{\spacefactor3000\relax}%
\providecommand \BibitemShut  [1]{\csname bibitem#1\endcsname}%
\let\auto@bib@innerbib\@empty
%</preamble>
\bibitem [{\citenamefont {Wannier}(1937)}]{PhysRev.52.191}%
  \BibitemOpen
  \bibfield  {author} {\bibinfo {author} {\bibfnamefont {G.~H.}\ \bibnamefont
  {Wannier}},\ }\bibfield  {title} {\bibinfo {title} {The structure of
  electronic excitation levels in insulating crystals},\ }\href
  {https://doi.org/10.1103/PhysRev.52.191} {\bibfield  {journal} {\bibinfo
  {journal} {Phys. Rev.}\ }\textbf {\bibinfo {volume} {52}},\ \bibinfo {pages}
  {191} (\bibinfo {year} {1937})}\BibitemShut {NoStop}%
\bibitem [{\citenamefont {Vanderbilt}\ and\ \citenamefont
  {King-Smith}(1993)}]{PhysRevB.48.4442}%
  \BibitemOpen
  \bibfield  {author} {\bibinfo {author} {\bibfnamefont {D.}~\bibnamefont
  {Vanderbilt}}\ and\ \bibinfo {author} {\bibfnamefont {R.~D.}\ \bibnamefont
  {King-Smith}},\ }\bibfield  {title} {\bibinfo {title} {Electric polarization
  as a bulk quantity and its relation to surface charge},\ }\href
  {https://doi.org/10.1103/PhysRevB.48.4442} {\bibfield  {journal} {\bibinfo
  {journal} {Phys. Rev. B}\ }\textbf {\bibinfo {volume} {48}},\ \bibinfo
  {pages} {4442} (\bibinfo {year} {1993})}\BibitemShut {NoStop}%
\bibitem [{\citenamefont {Resta}(1994)}]{RevModPhys.66.899}%
  \BibitemOpen
  \bibfield  {author} {\bibinfo {author} {\bibfnamefont {R.}~\bibnamefont
  {Resta}},\ }\bibfield  {title} {\bibinfo {title} {Macroscopic polarization in
  crystalline dielectrics: the geometric phase approach},\ }\href
  {https://doi.org/10.1103/RevModPhys.66.899} {\bibfield  {journal} {\bibinfo
  {journal} {Rev. Mod. Phys.}\ }\textbf {\bibinfo {volume} {66}},\ \bibinfo
  {pages} {899} (\bibinfo {year} {1994})}\BibitemShut {NoStop}%
\bibitem [{\citenamefont {Wu}\ \emph {et~al.}(2006)\citenamefont {Wu},
  \citenamefont {Di\'eguez}, \citenamefont {Rabe},\ and\ \citenamefont
  {Vanderbilt}}]{PhysRevLett.97.107602}%
  \BibitemOpen
  \bibfield  {author} {\bibinfo {author} {\bibfnamefont {X.}~\bibnamefont
  {Wu}}, \bibinfo {author} {\bibfnamefont {O.}~\bibnamefont {Di\'eguez}},
  \bibinfo {author} {\bibfnamefont {K.~M.}\ \bibnamefont {Rabe}},\ and\
  \bibinfo {author} {\bibfnamefont {D.}~\bibnamefont {Vanderbilt}},\ }\bibfield
   {title} {\bibinfo {title} {Wannier-based definition of layer polarizations
  in perovskite superlattices},\ }\href
  {https://doi.org/10.1103/PhysRevLett.97.107602} {\bibfield  {journal}
  {\bibinfo  {journal} {Phys. Rev. Lett.}\ }\textbf {\bibinfo {volume} {97}},\
  \bibinfo {pages} {107602} (\bibinfo {year} {2006})}\BibitemShut {NoStop}%
\bibitem [{\citenamefont {King-Smith}\ and\ \citenamefont
  {Vanderbilt}(1993)}]{PhysRevB.47.1651}%
  \BibitemOpen
  \bibfield  {author} {\bibinfo {author} {\bibfnamefont {R.~D.}\ \bibnamefont
  {King-Smith}}\ and\ \bibinfo {author} {\bibfnamefont {D.}~\bibnamefont
  {Vanderbilt}},\ }\bibfield  {title} {\bibinfo {title} {Theory of polarization
  of crystalline solids},\ }\href {https://doi.org/10.1103/PhysRevB.47.1651}
  {\bibfield  {journal} {\bibinfo  {journal} {Phys. Rev. B}\ }\textbf {\bibinfo
  {volume} {47}},\ \bibinfo {pages} {1651} (\bibinfo {year}
  {1993})}\BibitemShut {NoStop}%
\bibitem [{\citenamefont {Mustafa}\ \emph {et~al.}(2015)\citenamefont
  {Mustafa}, \citenamefont {Coh}, \citenamefont {Cohen},\ and\ \citenamefont
  {Louie}}]{PhysRevB.92.165134}%
  \BibitemOpen
  \bibfield  {author} {\bibinfo {author} {\bibfnamefont {J.~I.}\ \bibnamefont
  {Mustafa}}, \bibinfo {author} {\bibfnamefont {S.}~\bibnamefont {Coh}},
  \bibinfo {author} {\bibfnamefont {M.~L.}\ \bibnamefont {Cohen}},\ and\
  \bibinfo {author} {\bibfnamefont {S.~G.}\ \bibnamefont {Louie}},\ }\bibfield
  {title} {\bibinfo {title} {Automated construction of maximally localized
  wannier functions: Optimized projection functions method},\ }\href
  {https://doi.org/10.1103/PhysRevB.92.165134} {\bibfield  {journal} {\bibinfo
  {journal} {Phys. Rev. B}\ }\textbf {\bibinfo {volume} {92}},\ \bibinfo
  {pages} {165134} (\bibinfo {year} {2015})}\BibitemShut {NoStop}%
\bibitem [{\citenamefont {Marzari}\ \emph {et~al.}(2012)\citenamefont
  {Marzari}, \citenamefont {Mostofi}, \citenamefont {Yates}, \citenamefont
  {Souza},\ and\ \citenamefont {Vanderbilt}}]{RevModPhys.84.1419}%
  \BibitemOpen
  \bibfield  {author} {\bibinfo {author} {\bibfnamefont {N.}~\bibnamefont
  {Marzari}}, \bibinfo {author} {\bibfnamefont {A.~A.}\ \bibnamefont
  {Mostofi}}, \bibinfo {author} {\bibfnamefont {J.~R.}\ \bibnamefont {Yates}},
  \bibinfo {author} {\bibfnamefont {I.}~\bibnamefont {Souza}},\ and\ \bibinfo
  {author} {\bibfnamefont {D.}~\bibnamefont {Vanderbilt}},\ }\bibfield  {title}
  {\bibinfo {title} {Maximally localized wannier functions: Theory and
  applications},\ }\href {https://doi.org/10.1103/RevModPhys.84.1419}
  {\bibfield  {journal} {\bibinfo  {journal} {Rev. Mod. Phys.}\ }\textbf
  {\bibinfo {volume} {84}},\ \bibinfo {pages} {1419} (\bibinfo {year}
  {2012})}\BibitemShut {NoStop}%
\bibitem [{\citenamefont {Canc\`es}\ \emph {et~al.}(2017)\citenamefont
  {Canc\`es}, \citenamefont {Levitt}, \citenamefont {Panati},\ and\
  \citenamefont {Stoltz}}]{PhysRevB.95.075114}%
  \BibitemOpen
  \bibfield  {author} {\bibinfo {author} {\bibfnamefont {E.}~\bibnamefont
  {Canc\`es}}, \bibinfo {author} {\bibfnamefont {A.}~\bibnamefont {Levitt}},
  \bibinfo {author} {\bibfnamefont {G.}~\bibnamefont {Panati}},\ and\ \bibinfo
  {author} {\bibfnamefont {G.}~\bibnamefont {Stoltz}},\ }\bibfield  {title}
  {\bibinfo {title} {Robust determination of maximally localized wannier
  functions},\ }\href {https://doi.org/10.1103/PhysRevB.95.075114} {\bibfield
  {journal} {\bibinfo  {journal} {Phys. Rev. B}\ }\textbf {\bibinfo {volume}
  {95}},\ \bibinfo {pages} {075114} (\bibinfo {year} {2017})}\BibitemShut
  {NoStop}%
\bibitem [{\citenamefont {Alexandradinata}\ \emph {et~al.}(2014)\citenamefont
  {Alexandradinata}, \citenamefont {Dai},\ and\ \citenamefont
  {Bernevig}}]{PhysRevB.89.155114}%
  \BibitemOpen
  \bibfield  {author} {\bibinfo {author} {\bibfnamefont {A.}~\bibnamefont
  {Alexandradinata}}, \bibinfo {author} {\bibfnamefont {X.}~\bibnamefont
  {Dai}},\ and\ \bibinfo {author} {\bibfnamefont {B.~A.}\ \bibnamefont
  {Bernevig}},\ }\bibfield  {title} {\bibinfo {title} {Wilson-loop
  characterization of inversion-symmetric topological insulators},\ }\href
  {https://doi.org/10.1103/PhysRevB.89.155114} {\bibfield  {journal} {\bibinfo
  {journal} {Phys. Rev. B}\ }\textbf {\bibinfo {volume} {89}},\ \bibinfo
  {pages} {155114} (\bibinfo {year} {2014})}\BibitemShut {NoStop}%
\bibitem [{\citenamefont {Cloizeaux}(1963)}]{PhysRev.129.554}%
  \BibitemOpen
  \bibfield  {author} {\bibinfo {author} {\bibfnamefont {J.~D.}\ \bibnamefont
  {Cloizeaux}},\ }\bibfield  {title} {\bibinfo {title} {Orthogonal orbitals and
  generalized wannier functions},\ }\href
  {https://doi.org/10.1103/PhysRev.129.554} {\bibfield  {journal} {\bibinfo
  {journal} {Phys. Rev.}\ }\textbf {\bibinfo {volume} {129}},\ \bibinfo {pages}
  {554} (\bibinfo {year} {1963})}\BibitemShut {NoStop}%
\bibitem [{\citenamefont {Koumpouras}\ and\ \citenamefont
  {Larsson}(2020)}]{first-guess}%
  \BibitemOpen
  \bibfield  {author} {\bibinfo {author} {\bibfnamefont {K.}~\bibnamefont
  {Koumpouras}}\ and\ \bibinfo {author} {\bibfnamefont {J.~A.}\ \bibnamefont
  {Larsson}},\ }\bibfield  {title} {\bibinfo {title} {Distinguishing between
  chemical bonding and physical binding using electron localization function
  (elf)},\ }\href@noop {} {\bibfield  {journal} {\bibinfo  {journal} {Journal
  of Physics: Condensed Matter}\ }\textbf {\bibinfo {volume} {32}},\ \bibinfo
  {pages} {315502} (\bibinfo {year} {2020})}\BibitemShut {NoStop}%
\bibitem [{\citenamefont {Smirnov}\ and\ \citenamefont
  {Usvyat}(2001)}]{PhysRevB.64.245108}%
  \BibitemOpen
  \bibfield  {author} {\bibinfo {author} {\bibfnamefont {V.~P.}\ \bibnamefont
  {Smirnov}}\ and\ \bibinfo {author} {\bibfnamefont {D.~E.}\ \bibnamefont
  {Usvyat}},\ }\bibfield  {title} {\bibinfo {title} {Variational method for the
  generation of localized wannier functions on the basis of bloch functions},\
  }\href {https://doi.org/10.1103/PhysRevB.64.245108} {\bibfield  {journal}
  {\bibinfo  {journal} {Phys. Rev. B}\ }\textbf {\bibinfo {volume} {64}},\
  \bibinfo {pages} {245108} (\bibinfo {year} {2001})}\BibitemShut {NoStop}%
\bibitem [{\citenamefont {Lopez}\ \emph {et~al.}(2012)\citenamefont {Lopez},
  \citenamefont {Vanderbilt}, \citenamefont {Thonhauser},\ and\ \citenamefont
  {Souza}}]{PhysRevB.85.014435}%
  \BibitemOpen
  \bibfield  {author} {\bibinfo {author} {\bibfnamefont {M.~G.}\ \bibnamefont
  {Lopez}}, \bibinfo {author} {\bibfnamefont {D.}~\bibnamefont {Vanderbilt}},
  \bibinfo {author} {\bibfnamefont {T.}~\bibnamefont {Thonhauser}},\ and\
  \bibinfo {author} {\bibfnamefont {I.}~\bibnamefont {Souza}},\ }\bibfield
  {title} {\bibinfo {title} {Wannier-based calculation of the orbital
  magnetization in crystals},\ }\href
  {https://doi.org/10.1103/PhysRevB.85.014435} {\bibfield  {journal} {\bibinfo
  {journal} {Phys. Rev. B}\ }\textbf {\bibinfo {volume} {85}},\ \bibinfo
  {pages} {014435} (\bibinfo {year} {2012})}\BibitemShut {NoStop}%
\bibitem [{\citenamefont {Busch}\ \emph {et~al.}(2003)\citenamefont {Busch},
  \citenamefont {Mingaleev}, \citenamefont {Garcia-Martin}, \citenamefont
  {Schillinger},\ and\ \citenamefont {Hermann}}]{Kurt2003}%
  \BibitemOpen
  \bibfield  {author} {\bibinfo {author} {\bibfnamefont {K.}~\bibnamefont
  {Busch}}, \bibinfo {author} {\bibfnamefont {S.~F.}\ \bibnamefont
  {Mingaleev}}, \bibinfo {author} {\bibfnamefont {A.}~\bibnamefont
  {Garcia-Martin}}, \bibinfo {author} {\bibfnamefont {M.}~\bibnamefont
  {Schillinger}},\ and\ \bibinfo {author} {\bibfnamefont {D.}~\bibnamefont
  {Hermann}},\ }\bibfield  {title} {\bibinfo {title} {The wannier function
  approach to photonic crystal circuits},\ }\href
  {https://doi.org/10.1088/0953-8984/15/30/201} {\bibfield  {journal} {\bibinfo
   {journal} {Journal of Physics: Condensed Matter}\ }\textbf {\bibinfo
  {volume} {15}},\ \bibinfo {pages} {R1233} (\bibinfo {year}
  {2003})}\BibitemShut {NoStop}%
\bibitem [{\citenamefont {Busch}\ \emph {et~al.}(2011)\citenamefont {Busch},
  \citenamefont {Blum}, \citenamefont {Graham}, \citenamefont {Hermann},
  \citenamefont {Köhl}, \citenamefont {Mack},\ and\ \citenamefont
  {Wolff}}]{Busch2011}%
  \BibitemOpen
  \bibfield  {author} {\bibinfo {author} {\bibfnamefont {K.}~\bibnamefont
  {Busch}}, \bibinfo {author} {\bibfnamefont {C.}~\bibnamefont {Blum}},
  \bibinfo {author} {\bibfnamefont {A.~M.}\ \bibnamefont {Graham}}, \bibinfo
  {author} {\bibfnamefont {D.}~\bibnamefont {Hermann}}, \bibinfo {author}
  {\bibfnamefont {M.}~\bibnamefont {Köhl}}, \bibinfo {author} {\bibfnamefont
  {P.}~\bibnamefont {Mack}},\ and\ \bibinfo {author} {\bibfnamefont
  {C.}~\bibnamefont {Wolff}},\ }\bibfield  {title} {\bibinfo {title} {The
  photonic wannier function approach to photonic crystal simulations: status
  and perspectives},\ }\href {https://doi.org/10.1080/09500340.2010.526256}
  {\bibfield  {journal} {\bibinfo  {journal} {Journal of Modern Optics}\
  }\textbf {\bibinfo {volume} {58}},\ \bibinfo {pages} {365} (\bibinfo {year}
  {2011})}\BibitemShut {NoStop}%
\bibitem [{\citenamefont {Wolff}\ \emph {et~al.}(2013)\citenamefont {Wolff},
  \citenamefont {Mack},\ and\ \citenamefont {Busch}}]{PhysRevB.88.075201}%
  \BibitemOpen
  \bibfield  {author} {\bibinfo {author} {\bibfnamefont {C.}~\bibnamefont
  {Wolff}}, \bibinfo {author} {\bibfnamefont {P.}~\bibnamefont {Mack}},\ and\
  \bibinfo {author} {\bibfnamefont {K.}~\bibnamefont {Busch}},\ }\bibfield
  {title} {\bibinfo {title} {Generation of wannier functions for photonic
  crystals},\ }\href {https://doi.org/10.1103/PhysRevB.88.075201} {\bibfield
  {journal} {\bibinfo  {journal} {Phys. Rev. B}\ }\textbf {\bibinfo {volume}
  {88}},\ \bibinfo {pages} {075201} (\bibinfo {year} {2013})}\BibitemShut
  {NoStop}%
\bibitem [{\citenamefont {Liu}\ \emph {et~al.}(2022)\citenamefont {Liu},
  \citenamefont {Cao}, \citenamefont {Chen}, \citenamefont {Wang},
  \citenamefont {Yang},\ and\ \citenamefont {Zhang}}]{Liu2022}%
  \BibitemOpen
  \bibfield  {author} {\bibinfo {author} {\bibfnamefont {Y.}~\bibnamefont
  {Liu}}, \bibinfo {author} {\bibfnamefont {W.}~\bibnamefont {Cao}}, \bibinfo
  {author} {\bibfnamefont {W.}~\bibnamefont {Chen}}, \bibinfo {author}
  {\bibfnamefont {H.}~\bibnamefont {Wang}}, \bibinfo {author} {\bibfnamefont
  {L.}~\bibnamefont {Yang}},\ and\ \bibinfo {author} {\bibfnamefont
  {X.}~\bibnamefont {Zhang}},\ }\bibfield  {title} {\bibinfo {title} {Fully
  integrated topological electronics},\ }\href
  {https://doi.org/10.1038/s41598-022-17010-8} {\bibfield  {journal} {\bibinfo
  {journal} {Scientific Reports}\ }\textbf {\bibinfo {volume} {12}},\ \bibinfo
  {pages} {13410} (\bibinfo {year} {2022})}\BibitemShut {NoStop}%
\bibitem [{\citenamefont {Pham}\ \emph {et~al.}(2016)\citenamefont {Pham} \emph
  {et~al.}}]{Pham2016}%
  \BibitemOpen
  \bibfield  {author} {\bibinfo {author} {\bibfnamefont {V.~H.}\ \bibnamefont
  {Pham}} \emph {et~al.},\ }\bibfield  {title} {\bibinfo {title} {Progress in
  the research and development of photonic structure devices},\ }\href
  {https://doi.org/10.1088/2043-6262/7/1/015003} {\bibfield  {journal}
  {\bibinfo  {journal} {Adv. Nat. Sci: Nanosci. Nanotechnol.}\ }\textbf
  {\bibinfo {volume} {7}},\ \bibinfo {pages} {015003} (\bibinfo {year}
  {2016})}\BibitemShut {NoStop}%
\bibitem [{\citenamefont {Takeda}\ \emph {et~al.}(2006)\citenamefont {Takeda},
  \citenamefont {Chutinan},\ and\ \citenamefont {John}}]{PhysRevB.74.195116}%
  \BibitemOpen
  \bibfield  {author} {\bibinfo {author} {\bibfnamefont {H.}~\bibnamefont
  {Takeda}}, \bibinfo {author} {\bibfnamefont {A.}~\bibnamefont {Chutinan}},\
  and\ \bibinfo {author} {\bibfnamefont {S.}~\bibnamefont {John}},\ }\bibfield
  {title} {\bibinfo {title} {Localized light orbitals: Basis states for
  three-dimensional photonic crystal microscale circuits},\ }\href
  {https://doi.org/10.1103/PhysRevB.74.195116} {\bibfield  {journal} {\bibinfo
  {journal} {Phys. Rev. B}\ }\textbf {\bibinfo {volume} {74}},\ \bibinfo
  {pages} {195116} (\bibinfo {year} {2006})}\BibitemShut {NoStop}%
\bibitem [{\citenamefont {Kohn}(1959)}]{PhysRev.115.809}%
  \BibitemOpen
  \bibfield  {author} {\bibinfo {author} {\bibfnamefont {W.}~\bibnamefont
  {Kohn}},\ }\bibfield  {title} {\bibinfo {title} {Analytic properties of bloch
  waves and wannier functions},\ }\href
  {https://doi.org/10.1103/PhysRev.115.809} {\bibfield  {journal} {\bibinfo
  {journal} {Phys. Rev.}\ }\textbf {\bibinfo {volume} {115}},\ \bibinfo {pages}
  {809} (\bibinfo {year} {1959})}\BibitemShut {NoStop}%
\bibitem [{\citenamefont {Cloizeaux}(1964{\natexlab{a}})}]{PhysRev.135.A685}%
  \BibitemOpen
  \bibfield  {author} {\bibinfo {author} {\bibfnamefont {J.~D.}\ \bibnamefont
  {Cloizeaux}},\ }\bibfield  {title} {\bibinfo {title} {Energy bands and
  projection operators in a crystal: Analytic and asymptotic properties},\
  }\href {https://doi.org/10.1103/PhysRev.135.A685} {\bibfield  {journal}
  {\bibinfo  {journal} {Phys. Rev.}\ }\textbf {\bibinfo {volume} {135}},\
  \bibinfo {pages} {A685} (\bibinfo {year} {1964}{\natexlab{a}})}\BibitemShut
  {NoStop}%
\bibitem [{\citenamefont {Kohn}\ and\ \citenamefont
  {Onffroy}(1973)}]{PhysRevB.8.2485}%
  \BibitemOpen
  \bibfield  {author} {\bibinfo {author} {\bibfnamefont {W.}~\bibnamefont
  {Kohn}}\ and\ \bibinfo {author} {\bibfnamefont {J.~R.}\ \bibnamefont
  {Onffroy}},\ }\bibfield  {title} {\bibinfo {title} {Wannier functions in a
  simple nonperiodic system},\ }\href {https://doi.org/10.1103/PhysRevB.8.2485}
  {\bibfield  {journal} {\bibinfo  {journal} {Phys. Rev. B}\ }\textbf {\bibinfo
  {volume} {8}},\ \bibinfo {pages} {2485} (\bibinfo {year} {1973})}\BibitemShut
  {NoStop}%
\bibitem [{\citenamefont {Nenciu}\ and\ \citenamefont
  {Nenciu}(1993)}]{PhysRevB.47.10112}%
  \BibitemOpen
  \bibfield  {author} {\bibinfo {author} {\bibfnamefont {A.}~\bibnamefont
  {Nenciu}}\ and\ \bibinfo {author} {\bibfnamefont {G.}~\bibnamefont
  {Nenciu}},\ }\bibfield  {title} {\bibinfo {title} {Existence of exponentially
  localized wannier functions for nonperiodic systems},\ }\href
  {https://doi.org/10.1103/PhysRevB.47.10112} {\bibfield  {journal} {\bibinfo
  {journal} {Phys. Rev. B}\ }\textbf {\bibinfo {volume} {47}},\ \bibinfo
  {pages} {10112} (\bibinfo {year} {1993})}\BibitemShut {NoStop}%
\bibitem [{\citenamefont {Marzari}\ and\ \citenamefont
  {Vanderbilt}(1997)}]{Marzari1997}%
  \BibitemOpen
  \bibfield  {author} {\bibinfo {author} {\bibfnamefont {N.}~\bibnamefont
  {Marzari}}\ and\ \bibinfo {author} {\bibfnamefont {D.}~\bibnamefont
  {Vanderbilt}},\ }\bibfield  {title} {\bibinfo {title} {Maximally localized
  generalized wannier functions for composite energy bands},\ }\href
  {https://doi.org/10.1103/PhysRevB.56.12847} {\bibfield  {journal} {\bibinfo
  {journal} {Phys. Rev. B}\ }\textbf {\bibinfo {volume} {56}},\ \bibinfo
  {pages} {12847} (\bibinfo {year} {1997})}\BibitemShut {NoStop}%
\bibitem [{\citenamefont {Kivelson}(1982)}]{kivelson1982}%
  \BibitemOpen
  \bibfield  {author} {\bibinfo {author} {\bibfnamefont {S.}~\bibnamefont
  {Kivelson}},\ }\bibfield  {title} {\bibinfo {title} {Wannier functions in
  one-dimensional disordered systems: Application to fractionally charged
  solitons},\ }\href {https://doi.org/10.1103/PhysRevB.26.4269} {\bibfield
  {journal} {\bibinfo  {journal} {Phys. Rev. B}\ }\textbf {\bibinfo {volume}
  {26}},\ \bibinfo {pages} {4269} (\bibinfo {year} {1982})}\BibitemShut
  {NoStop}%
\bibitem [{\citenamefont {Damle}\ \emph {et~al.}(2015)\citenamefont {Damle},
  \citenamefont {Lin},\ and\ \citenamefont {Ying}}]{ct500985f}%
  \BibitemOpen
  \bibfield  {author} {\bibinfo {author} {\bibfnamefont {A.}~\bibnamefont
  {Damle}}, \bibinfo {author} {\bibfnamefont {L.}~\bibnamefont {Lin}},\ and\
  \bibinfo {author} {\bibfnamefont {L.}~\bibnamefont {Ying}},\ }\bibfield
  {title} {\bibinfo {title} {Compressed representation of kohn–sham orbitals
  via selected columns of the density matrix},\ }\href
  {https://doi.org/10.1021/ct500985f} {\bibfield  {journal} {\bibinfo
  {journal} {Journal of Chemical Theory and Computation}\ }\textbf {\bibinfo
  {volume} {11}},\ \bibinfo {pages} {1463} (\bibinfo {year} {2015})},\ \bibinfo
  {note} {pMID: 26574357}\BibitemShut {NoStop}%
\bibitem [{\citenamefont {Damle}\ \emph {et~al.}(2016)\citenamefont {Damle},
  \citenamefont {Lin},\ and\ \citenamefont {Ying}}]{damle2016scdmk}%
  \BibitemOpen
  \bibfield  {author} {\bibinfo {author} {\bibfnamefont {A.}~\bibnamefont
  {Damle}}, \bibinfo {author} {\bibfnamefont {L.}~\bibnamefont {Lin}},\ and\
  \bibinfo {author} {\bibfnamefont {L.}~\bibnamefont {Ying}},\ }\href@noop {}
  {\bibinfo {title} {Scdm-k: Localized orbitals for solids via selected columns
  of the density matrix}} (\bibinfo {year} {2016}),\ \Eprint
  {https://arxiv.org/abs/1507.03354} {arXiv:1507.03354 [physics.comp-ph]}
  \BibitemShut {NoStop}%
\bibitem [{\citenamefont {Vitale}\ \emph {et~al.}(2020)\citenamefont {Vitale},
  \citenamefont {Pizzi}, \citenamefont {Marrazzo}, \citenamefont {Yates},
  \citenamefont {Marzari},\ and\ \citenamefont {Mostofi}}]{Vitale2020}%
  \BibitemOpen
  \bibfield  {author} {\bibinfo {author} {\bibfnamefont {V.}~\bibnamefont
  {Vitale}}, \bibinfo {author} {\bibfnamefont {G.}~\bibnamefont {Pizzi}},
  \bibinfo {author} {\bibfnamefont {A.}~\bibnamefont {Marrazzo}}, \bibinfo
  {author} {\bibfnamefont {J.~R.}\ \bibnamefont {Yates}}, \bibinfo {author}
  {\bibfnamefont {N.}~\bibnamefont {Marzari}},\ and\ \bibinfo {author}
  {\bibfnamefont {A.~A.}\ \bibnamefont {Mostofi}},\ }\bibfield  {title}
  {\bibinfo {title} {Automated high-throughput wannierisation},\ }\href
  {https://doi.org/10.1038/s41524-020-0312-y} {\bibfield  {journal} {\bibinfo
  {journal} {npj Computational Materials}\ }\textbf {\bibinfo {volume} {6}},\
  \bibinfo {pages} {66} (\bibinfo {year} {2020})}\BibitemShut {NoStop}%
\bibitem [{\citenamefont {Pizzi}\ \emph {et~al.}(2020)\citenamefont {Pizzi}
  \emph {et~al.}}]{Pizzi2020}%
  \BibitemOpen
  \bibfield  {author} {\bibinfo {author} {\bibfnamefont {G.}~\bibnamefont
  {Pizzi}} \emph {et~al.},\ }\bibfield  {title} {\bibinfo {title} {Wannier90 as
  a community code: new features and applications},\ }\href
  {https://doi.org/10.1088/1361-648X/ab51ff} {\bibfield  {journal} {\bibinfo
  {journal} {Journal of Physics: Condensed Matter}\ }\textbf {\bibinfo {volume}
  {32}},\ \bibinfo {pages} {165902} (\bibinfo {year} {2020})}\BibitemShut
  {NoStop}%
\bibitem [{\citenamefont {Freimuth}\ \emph {et~al.}(2008)\citenamefont
  {Freimuth}, \citenamefont {Mokrousov}, \citenamefont {Wortmann},
  \citenamefont {Heinze},\ and\ \citenamefont
  {Bl\"ugel}}]{freimuth2008publisher}%
  \BibitemOpen
  \bibfield  {author} {\bibinfo {author} {\bibfnamefont {F.}~\bibnamefont
  {Freimuth}}, \bibinfo {author} {\bibfnamefont {Y.}~\bibnamefont {Mokrousov}},
  \bibinfo {author} {\bibfnamefont {D.}~\bibnamefont {Wortmann}}, \bibinfo
  {author} {\bibfnamefont {S.}~\bibnamefont {Heinze}},\ and\ \bibinfo {author}
  {\bibfnamefont {S.}~\bibnamefont {Bl\"ugel}},\ }\bibfield  {title} {\bibinfo
  {title} {Maximally localized wannier functions within the flapw formalism},\
  }\href {https://doi.org/10.1103/PhysRevB.78.035120} {\bibfield  {journal}
  {\bibinfo  {journal} {Phys. Rev. B}\ }\textbf {\bibinfo {volume} {78}},\
  \bibinfo {pages} {035120} (\bibinfo {year} {2008})}\BibitemShut {NoStop}%
\bibitem [{\citenamefont {Stubbs}\ \emph {et~al.}(2021)\citenamefont {Stubbs},
  \citenamefont {Watson},\ and\ \citenamefont {Lu}}]{PhysRevB.103.075125}%
  \BibitemOpen
  \bibfield  {author} {\bibinfo {author} {\bibfnamefont {K.~D.}\ \bibnamefont
  {Stubbs}}, \bibinfo {author} {\bibfnamefont {A.~B.}\ \bibnamefont {Watson}},\
  and\ \bibinfo {author} {\bibfnamefont {J.}~\bibnamefont {Lu}},\ }\bibfield
  {title} {\bibinfo {title} {Iterated projected position algorithm for
  constructing exponentially localized generalized wannier functions for
  periodic and nonperiodic insulators in two dimensions and higher},\ }\href
  {https://doi.org/10.1103/PhysRevB.103.075125} {\bibfield  {journal} {\bibinfo
   {journal} {Phys. Rev. B}\ }\textbf {\bibinfo {volume} {103}},\ \bibinfo
  {pages} {075125} (\bibinfo {year} {2021})}\BibitemShut {NoStop}%
\bibitem [{\citenamefont {Daubechies}(1992{\natexlab{a}})}]{Daub10}%
  \BibitemOpen
  \bibfield  {author} {\bibinfo {author} {\bibfnamefont {I.}~\bibnamefont
  {Daubechies}},\ }\href {https://doi.org/10.1137/1.9781611970104} {\emph
  {\bibinfo {title} {Ten Lectures on Wavelets}}}\ (\bibinfo  {publisher}
  {Philadelphia, PA: Society for Industrial and Applied Mathematics},\ \bibinfo
  {year} {1992})\BibitemShut {NoStop}%
\bibitem [{\citenamefont {Battle}(1999)}]{battle1999wavelets}%
  \BibitemOpen
  \bibfield  {author} {\bibinfo {author} {\bibfnamefont {G.}~\bibnamefont
  {Battle}},\ }\href {https://books.google.co.jp/books?id=1qQwK0-\_Pz4C} {\emph
  {\bibinfo {title} {Wavelets and Renormalization}}},\ Approximations and
  Decomposition Series\ (\bibinfo  {publisher} {World Scientific},\ \bibinfo
  {year} {1999})\BibitemShut {NoStop}%
\bibitem [{\citenamefont {Evenbly}\ and\ \citenamefont
  {White}(2016)}]{PhysRevLett.116.140403}%
  \BibitemOpen
  \bibfield  {author} {\bibinfo {author} {\bibfnamefont {G.}~\bibnamefont
  {Evenbly}}\ and\ \bibinfo {author} {\bibfnamefont {S.~R.}\ \bibnamefont
  {White}},\ }\bibfield  {title} {\bibinfo {title} {Entanglement
  renormalization and wavelets},\ }\href
  {https://doi.org/10.1103/PhysRevLett.116.140403} {\bibfield  {journal}
  {\bibinfo  {journal} {Phys. Rev. Lett.}\ }\textbf {\bibinfo {volume} {116}},\
  \bibinfo {pages} {140403} (\bibinfo {year} {2016})}\BibitemShut {NoStop}%
\bibitem [{\citenamefont {Parzen}(1953)}]{parzen1953}%
  \BibitemOpen
  \bibfield  {author} {\bibinfo {author} {\bibfnamefont {G.}~\bibnamefont
  {Parzen}},\ }\bibfield  {title} {\bibinfo {title} {Electronic energy bands in
  metals},\ }\href {https://doi.org/10.1103/PhysRev.89.237} {\bibfield
  {journal} {\bibinfo  {journal} {Phys. Rev.}\ }\textbf {\bibinfo {volume}
  {89}},\ \bibinfo {pages} {237} (\bibinfo {year} {1953})}\BibitemShut
  {NoStop}%
\bibitem [{\citenamefont {Hong}\ \emph {et~al.}(2004)\citenamefont {Hong},
  \citenamefont {Wang},\ and\ \citenamefont {Gardner}}]{hong2004-9}%
  \BibitemOpen
  \bibfield  {author} {\bibinfo {author} {\bibfnamefont {D.}~\bibnamefont
  {Hong}}, \bibinfo {author} {\bibfnamefont {J.}~\bibnamefont {Wang}},\ and\
  \bibinfo {author} {\bibfnamefont {R.}~\bibnamefont {Gardner}},\ }\bibinfo
  {title} {Real analysis with an introduction to wavelets and applications}\
  (\bibinfo  {publisher} {Elsevier Science},\ \bibinfo {year} {2004})\
  Chap.~\bibinfo {chapter} {9}\BibitemShut {NoStop}%
\bibitem [{\citenamefont {Clow}\ and\ \citenamefont
  {Johnson}(2003)}]{clow2003}%
  \BibitemOpen
  \bibfield  {author} {\bibinfo {author} {\bibfnamefont {S.~D.}\ \bibnamefont
  {Clow}}\ and\ \bibinfo {author} {\bibfnamefont {B.~R.}\ \bibnamefont
  {Johnson}},\ }\bibfield  {title} {\bibinfo {title} {Wavelet-basis calculation
  of wannier functions},\ }\href {https://doi.org/10.1103/PhysRevB.68.235107}
  {\bibfield  {journal} {\bibinfo  {journal} {Phys. Rev. B}\ }\textbf {\bibinfo
  {volume} {68}},\ \bibinfo {pages} {235107} (\bibinfo {year}
  {2003})}\BibitemShut {NoStop}%
\bibitem [{\citenamefont {Keinert}(2003)}]{keinert2003}%
  \BibitemOpen
  \bibfield  {author} {\bibinfo {author} {\bibfnamefont {F.}~\bibnamefont
  {Keinert}},\ }\href {https://doi.org/10.1201/9780203011591} {\emph {\bibinfo
  {title} {Wavelets and Multiwavelets}}}\ (\bibinfo  {publisher} {Chapman and
  Hall/CRC},\ \bibinfo {year} {2003})\BibitemShut {NoStop}%
\bibitem [{\citenamefont {Yan}\ and\ \citenamefont
  {Wang}(2006)}]{PhysRevB.74.224303}%
  \BibitemOpen
  \bibfield  {author} {\bibinfo {author} {\bibfnamefont {Z.-Z.}\ \bibnamefont
  {Yan}}\ and\ \bibinfo {author} {\bibfnamefont {Y.-S.}\ \bibnamefont {Wang}},\
  }\bibfield  {title} {\bibinfo {title} {Wavelet-based method for calculating
  elastic band gaps of two-dimensional phononic crystals},\ }\href
  {https://doi.org/10.1103/PhysRevB.74.224303} {\bibfield  {journal} {\bibinfo
  {journal} {Phys. Rev. B}\ }\textbf {\bibinfo {volume} {74}},\ \bibinfo
  {pages} {224303} (\bibinfo {year} {2006})}\BibitemShut {NoStop}%
\bibitem [{\citenamefont {Checoury}\ and\ \citenamefont
  {Lourtioz}(2006)}]{CHECOURY2006360}%
  \BibitemOpen
  \bibfield  {author} {\bibinfo {author} {\bibfnamefont {X.}~\bibnamefont
  {Checoury}}\ and\ \bibinfo {author} {\bibfnamefont {J.-M.}\ \bibnamefont
  {Lourtioz}},\ }\bibfield  {title} {\bibinfo {title} {Wavelet method for
  computing band diagrams of 2d photonic crystals},\ }\href
  {https://doi.org/https://doi.org/10.1016/j.optcom.2005.08.027} {\bibfield
  {journal} {\bibinfo  {journal} {Optics Communications}\ }\textbf {\bibinfo
  {volume} {259}},\ \bibinfo {pages} {360} (\bibinfo {year}
  {2006})}\BibitemShut {NoStop}%
\bibitem [{\citenamefont {Daubechies}(1992{\natexlab{b}})}]{Daub10-6}%
  \BibitemOpen
  \bibfield  {author} {\bibinfo {author} {\bibfnamefont {I.}~\bibnamefont
  {Daubechies}},\ }\bibinfo {title} {Ten lectures on wavelets}\ (\bibinfo
  {publisher} {Society for Industrial and Applied Mathematics},\ \bibinfo
  {year} {1992})\ Chap.~\bibinfo {chapter} {6}\BibitemShut {NoStop}%
\bibitem [{\citenamefont {Daubechies}(1992{\natexlab{c}})}]{Daub10-5}%
  \BibitemOpen
  \bibfield  {author} {\bibinfo {author} {\bibfnamefont {I.}~\bibnamefont
  {Daubechies}},\ }\bibinfo {title} {Ten lectures on wavelets}\ (\bibinfo
  {publisher} {Society for Industrial and Applied Mathematics},\ \bibinfo
  {year} {1992})\ Chap.~\bibinfo {chapter} {5}\BibitemShut {NoStop}%
\bibitem [{\citenamefont {Galassi}\ \emph {et~al.}()\citenamefont {Galassi}
  \emph {et~al.}}]{gsl}%
  \BibitemOpen
  \bibfield  {author} {\bibinfo {author} {\bibfnamefont {M.}~\bibnamefont
  {Galassi}} \emph {et~al.},\ }\href@noop {} {\bibinfo {title} {{GNU}
  scientific library reference manual}},\ \bibinfo {howpublished}
  {\url{http://www.gnu.org/software/gsl/}}\BibitemShut {NoStop}%
\bibitem [{\citenamefont {Daubechies}(1992{\natexlab{d}})}]{Daub10-8}%
  \BibitemOpen
  \bibfield  {author} {\bibinfo {author} {\bibfnamefont {I.}~\bibnamefont
  {Daubechies}},\ }\bibinfo {title} {Ten lectures on wavelets}\ (\bibinfo
  {publisher} {Society for Industrial and Applied Mathematics},\ \bibinfo
  {year} {1992})\ Chap.~\bibinfo {chapter} {8}\BibitemShut {NoStop}%
\bibitem [{\citenamefont {Daubechies}(1992{\natexlab{e}})}]{Daub10-7}%
  \BibitemOpen
  \bibfield  {author} {\bibinfo {author} {\bibfnamefont {I.}~\bibnamefont
  {Daubechies}},\ }\bibinfo {title} {Ten lectures on wavelets}\ (\bibinfo
  {publisher} {Society for Industrial and Applied Mathematics},\ \bibinfo
  {year} {1992})\ Chap.~\bibinfo {chapter} {7}\BibitemShut {NoStop}%
\bibitem [{\citenamefont {Daubechies}\ and\ \citenamefont
  {Lagarias}(1991)}]{Daub-Lag-1}%
  \BibitemOpen
  \bibfield  {author} {\bibinfo {author} {\bibfnamefont {I.}~\bibnamefont
  {Daubechies}}\ and\ \bibinfo {author} {\bibfnamefont {J.~C.}\ \bibnamefont
  {Lagarias}},\ }\bibfield  {title} {\bibinfo {title} {Two-scale difference
  equations. i. existence and global regularity of solutions},\ }\href
  {https://doi.org/10.1137/0522089} {\bibfield  {journal} {\bibinfo  {journal}
  {SIAM Journal on Mathematical Analysis}\ }\textbf {\bibinfo {volume} {22}},\
  \bibinfo {pages} {1388} (\bibinfo {year} {1991})}\BibitemShut {NoStop}%
\bibitem [{\citenamefont {Daubechies}(1988)}]{Daub1988}%
  \BibitemOpen
  \bibfield  {author} {\bibinfo {author} {\bibfnamefont {I.}~\bibnamefont
  {Daubechies}},\ }\bibfield  {title} {\bibinfo {title} {Orthonormal bases of
  compactly supported wavelets},\ }\href
  {https://doi.org/10.1002/cpa.3160410705} {\bibfield  {journal} {\bibinfo
  {journal} {Commun. Pure Appl. Math.}\ }\textbf {\bibinfo {volume} {41}},\
  \bibinfo {pages} {909} (\bibinfo {year} {1988})}\BibitemShut {NoStop}%
\bibitem [{\citenamefont {Cloizeaux}(1964{\natexlab{b}})}]{PhysRev.135.A698}%
  \BibitemOpen
  \bibfield  {author} {\bibinfo {author} {\bibfnamefont {J.~D.}\ \bibnamefont
  {Cloizeaux}},\ }\bibfield  {title} {\bibinfo {title} {Analytical properties
  of $n$-dimensional energy bands and wannier functions},\ }\href
  {https://doi.org/10.1103/PhysRev.135.A698} {\bibfield  {journal} {\bibinfo
  {journal} {Phys. Rev.}\ }\textbf {\bibinfo {volume} {135}},\ \bibinfo {pages}
  {A698} (\bibinfo {year} {1964}{\natexlab{b}})}\BibitemShut {NoStop}%
\bibitem [{\citenamefont {Prodan}\ and\ \citenamefont
  {Kohn}(2005)}]{Prodan2005}%
  \BibitemOpen
  \bibfield  {author} {\bibinfo {author} {\bibfnamefont {E.}~\bibnamefont
  {Prodan}}\ and\ \bibinfo {author} {\bibfnamefont {W.}~\bibnamefont {Kohn}},\
  }\bibfield  {title} {\bibinfo {title} {Nearsightedness of electronic
  matter},\ }\href {https://doi.org/10.1073/pnas.0505436102} {\bibfield
  {journal} {\bibinfo  {journal} {Proc. Natl. Acad. Sci. U.S.A.}\ }\textbf
  {\bibinfo {volume} {102}},\ \bibinfo {pages} {11635} (\bibinfo {year}
  {2005})}\BibitemShut {NoStop}%
\bibitem [{\citenamefont {Benzi}\ \emph {et~al.}(2013)\citenamefont {Benzi},
  \citenamefont {Boito},\ and\ \citenamefont {Razouk}}]{Benzi2013}%
  \BibitemOpen
  \bibfield  {author} {\bibinfo {author} {\bibfnamefont {M.}~\bibnamefont
  {Benzi}}, \bibinfo {author} {\bibfnamefont {P.}~\bibnamefont {Boito}},\ and\
  \bibinfo {author} {\bibfnamefont {N.}~\bibnamefont {Razouk}},\ }\bibfield
  {title} {\bibinfo {title} {Decay properties of spectral projectors with
  applications to electronic structure},\ }\href
  {https://doi.org/10.1137/100814019} {\bibfield  {journal} {\bibinfo
  {journal} {SIAM Review}\ }\textbf {\bibinfo {volume} {55}},\ \bibinfo {pages}
  {3} (\bibinfo {year} {2013})}\BibitemShut {NoStop}%
\bibitem [{\citenamefont {Gupta}\ and\ \citenamefont
  {Bradlyn}(2022)}]{Gupta2022}%
  \BibitemOpen
  \bibfield  {author} {\bibinfo {author} {\bibfnamefont {V.}~\bibnamefont
  {Gupta}}\ and\ \bibinfo {author} {\bibfnamefont {B.}~\bibnamefont
  {Bradlyn}},\ }\bibfield  {title} {\bibinfo {title} {Wannier-function methods
  for topological modes in one-dimensional photonic crystals},\ }\href
  {https://doi.org/10.1103/PhysRevA.105.053521} {\bibfield  {journal} {\bibinfo
   {journal} {Phys. Rev. A}\ }\textbf {\bibinfo {volume} {105}},\ \bibinfo
  {pages} {053521} (\bibinfo {year} {2022})}\BibitemShut {NoStop}%
\bibitem [{\citenamefont {G.Grosso}(2000)}]{Grosso2000}%
  \BibitemOpen
  \bibfield  {author} {\bibinfo {author} {\bibnamefont {G.Grosso}},\
  }\href@noop {} {\emph {\bibinfo {title} {Solid State Physics}}},\ CBMS-NSF
  Regional Conf. Series in Appl. Math.\ (\bibinfo  {publisher} {Academic
  Press},\ \bibinfo {year} {2000})\BibitemShut {NoStop}%
\bibitem [{\citenamefont {Johnston}(2019)}]{johnston2019}%
  \BibitemOpen
  \bibfield  {author} {\bibinfo {author} {\bibfnamefont {D.~C.}\ \bibnamefont
  {Johnston}},\ }\bibfield  {title} {\bibinfo {title} {Attractive kronig-penney
  band structures and wave functions},\ }\href@noop {} {\bibfield  {journal}
  {\bibinfo  {journal} {arXiv preprint arXiv:1905.12084}\ } (\bibinfo {year}
  {2019})}\BibitemShut {NoStop}%
\bibitem [{\citenamefont {Vellasco-Gomes}\ \emph {et~al.}(2020)\citenamefont
  {Vellasco-Gomes}, \citenamefont {{de Figueiredo Camargo}},\ and\
  \citenamefont {Bruno-Alfonso}}]{VELLASCO2020}%
  \BibitemOpen
  \bibfield  {author} {\bibinfo {author} {\bibfnamefont {A.}~\bibnamefont
  {Vellasco-Gomes}}, \bibinfo {author} {\bibfnamefont {R.}~\bibnamefont {{de
  Figueiredo Camargo}}},\ and\ \bibinfo {author} {\bibfnamefont
  {A.}~\bibnamefont {Bruno-Alfonso}},\ }\bibfield  {title} {\bibinfo {title}
  {Energy bands and wannier functions of the fractional kronig-penney model},\
  }\href {https://doi.org/https://doi.org/10.1016/j.amc.2020.125266} {\bibfield
   {journal} {\bibinfo  {journal} {Applied Mathematics and Computation}\
  }\textbf {\bibinfo {volume} {380}},\ \bibinfo {pages} {125266} (\bibinfo
  {year} {2020})}\BibitemShut {NoStop}%
\bibitem [{\citenamefont {Romano}\ \emph {et~al.}(2010)\citenamefont {Romano},
  \citenamefont {Nacbar},\ and\ \citenamefont {Bruno-Alfonso}}]{Romano2010}%
  \BibitemOpen
  \bibfield  {author} {\bibinfo {author} {\bibfnamefont {M.~C.}\ \bibnamefont
  {Romano}}, \bibinfo {author} {\bibfnamefont {D.~R.}\ \bibnamefont {Nacbar}},\
  and\ \bibinfo {author} {\bibfnamefont {A.}~\bibnamefont {Bruno-Alfonso}},\
  }\bibfield  {title} {\bibinfo {title} {Wannier functions of a one-dimensional
  photonic crystal with inversion symmetry},\ }\href
  {https://doi.org/10.1088/0953-4075/43/21/215403} {\bibfield  {journal}
  {\bibinfo  {journal} {J. Phys. B}\ }\textbf {\bibinfo {volume} {43}},\
  \bibinfo {pages} {215403} (\bibinfo {year} {2010})}\BibitemShut {NoStop}%
\bibitem [{\citenamefont {Wang}\ \emph {et~al.}(2014)\citenamefont {Wang},
  \citenamefont {Lazar}, \citenamefont {Park}, \citenamefont {Millis},\ and\
  \citenamefont {Marianetti}}]{Wang2014}%
  \BibitemOpen
  \bibfield  {author} {\bibinfo {author} {\bibfnamefont {R.}~\bibnamefont
  {Wang}}, \bibinfo {author} {\bibfnamefont {E.~A.}\ \bibnamefont {Lazar}},
  \bibinfo {author} {\bibfnamefont {H.}~\bibnamefont {Park}}, \bibinfo {author}
  {\bibfnamefont {A.~J.}\ \bibnamefont {Millis}},\ and\ \bibinfo {author}
  {\bibfnamefont {C.~A.}\ \bibnamefont {Marianetti}},\ }\bibfield  {title}
  {\bibinfo {title} {Selectively localized wannier functions},\ }\href
  {https://doi.org/10.1103/PhysRevB.90.165125} {\bibfield  {journal} {\bibinfo
  {journal} {Phys. Rev. B}\ }\textbf {\bibinfo {volume} {90}},\ \bibinfo
  {pages} {165125} (\bibinfo {year} {2014})}\BibitemShut {NoStop}%
\bibitem [{\citenamefont {Cohen}\ and\ \citenamefont
  {Schlenker}(1993)}]{Cohen1993}%
  \BibitemOpen
  \bibfield  {author} {\bibinfo {author} {\bibfnamefont {A.}~\bibnamefont
  {Cohen}}\ and\ \bibinfo {author} {\bibfnamefont {J.-M.}\ \bibnamefont
  {Schlenker}},\ }\bibfield  {title} {\bibinfo {title} {Compactly supported
  bidimensional wavelet bases with hexagonal symmetry},\ }\href
  {https://doi.org/10.1007/BF01198004} {\bibfield  {journal} {\bibinfo
  {journal} {Constructive Approximation}\ }\textbf {\bibinfo {volume} {9}},\
  \bibinfo {pages} {209} (\bibinfo {year} {1993})}\BibitemShut {NoStop}%
\bibitem [{\citenamefont {Cuyt}\ \emph {et~al.}(2008)\citenamefont {Cuyt},
  \citenamefont {Backeljauw}, \citenamefont {Petersen}, \citenamefont
  {Bonan-Hamada}, \citenamefont {Verdonk}, \citenamefont {Waadeland},\ and\
  \citenamefont {Jones}}]{cuyt2008handbook}%
  \BibitemOpen
  \bibfield  {author} {\bibinfo {author} {\bibfnamefont {A.}~\bibnamefont
  {Cuyt}}, \bibinfo {author} {\bibfnamefont {F.}~\bibnamefont {Backeljauw}},
  \bibinfo {author} {\bibfnamefont {V.}~\bibnamefont {Petersen}}, \bibinfo
  {author} {\bibfnamefont {C.}~\bibnamefont {Bonan-Hamada}}, \bibinfo {author}
  {\bibfnamefont {B.}~\bibnamefont {Verdonk}}, \bibinfo {author} {\bibfnamefont
  {H.}~\bibnamefont {Waadeland}},\ and\ \bibinfo {author} {\bibfnamefont
  {W.}~\bibnamefont {Jones}},\ }\href
  {https://books.google.co.jp/books?id=DQtpJaEs4NIC} {\emph {\bibinfo {title}
  {Handbook of Continued Fractions for Special Functions}}},\ SpringerLink:
  Springer e-Books\ (\bibinfo  {publisher} {Springer Netherlands},\ \bibinfo
  {year} {2008})\ p.\ \bibinfo {pages} {344}\BibitemShut {NoStop}%
\bibitem [{\citenamefont {Ahlfers}(1979)}]{ahlfers1966complex}%
  \BibitemOpen
  \bibfield  {author} {\bibinfo {author} {\bibfnamefont {L.~V.}\ \bibnamefont
  {Ahlfers}},\ }\href@noop {} {\emph {\bibinfo {title} {Complex Analysis; an
  Introduction to the Theory of Analytic Functions of One Complex Variable}}},\
  \bibinfo {edition} {3rd}\ ed.\ (\bibinfo  {publisher} {McGraw-Hill},\
  \bibinfo {year} {1979})\ Chap.~\bibinfo {chapter} {2}\BibitemShut {NoStop}%
\bibitem [{\citenamefont {Pinsky}(2008)}]{pinsky2008introduction}%
  \BibitemOpen
  \bibfield  {author} {\bibinfo {author} {\bibfnamefont {M.}~\bibnamefont
  {Pinsky}},\ }\href {https://books.google.co.jp/books?id=PyISCgAAQBAJ} {\emph
  {\bibinfo {title} {Introduction to Fourier Analysis and Wavelets}}},\
  Graduate studies in mathematics\ (\bibinfo  {publisher} {American
  Mathematical Society},\ \bibinfo {year} {2008})\ Chap.~\bibinfo {chapter}
  {1}\BibitemShut {NoStop}%
\bibitem [{\citenamefont {Abramowitz}\ and\ \citenamefont
  {Stegun}(1964)}]{AbramowitzStegun64}%
  \BibitemOpen
  \bibfield  {author} {\bibinfo {author} {\bibfnamefont {M.}~\bibnamefont
  {Abramowitz}}\ and\ \bibinfo {author} {\bibfnamefont {I.~A.}\ \bibnamefont
  {Stegun}},\ }\href@noop {} {\emph {\bibinfo {title} {Handbook of Mathematical
  Functions with Formulas, Graphs, and Mathematical Tables}}}\ (\bibinfo
  {publisher} {Dover Publications},\ \bibinfo {address} {New York},\ \bibinfo
  {year} {1964})\ p.~\bibinfo {pages} {36}\BibitemShut {NoStop}%
\bibitem [{\citenamefont {Daubechies}(1992{\natexlab{f}})}]{Daub10-2}%
  \BibitemOpen
  \bibfield  {author} {\bibinfo {author} {\bibfnamefont {I.}~\bibnamefont
  {Daubechies}},\ }\bibinfo {title} {Ten lectures on wavelets}\ (\bibinfo
  {publisher} {Society for Industrial and Applied Mathematics},\ \bibinfo
  {year} {1992})\ Chap.~\bibinfo {chapter} {2}\BibitemShut {NoStop}%
\end{thebibliography}%
\end{document}